\documentclass[epj]{svjour}
\usepackage{graphics}
\usepackage{rotate}
\newcommand{\lsim}{\raisebox{-0.13cm}{~\shortstack{$<$ \\[-0.07cm] $\sim$}}~} 
\newcommand{\gsim}{\raisebox{-0.13cm}{~\shortstack{$>$ \\[-0.07cm] $\sim$}}~} 
\newcommand{\beq}{\begin{eqnarray}} 
\newcommand{\eeq}{\end{eqnarray}} 
\newcommand{\tb}{\tan \beta}

\begin{document}
\title{Implications of the Higgs discovery for the MSSM\thanks{Review to appear in a 
special issue of EPJC. Extended version of talks given at 
various recent conferences.}}
\author{Abdelhak Djouadi\inst{1}% etc
% \thanks is optional - remove next line if no needed
%\thanks{\emph{Present address: TH Unit, CERN, Gen\`eve, Switzerland.}}%
}                     % Do not remove
\institute{ Laboratoire de Physique Th\'eorique, U. Paris--Sud and  CNRS,  F--91405
Orsay, France.
}
\date{}%Received: date / Revised version: date}
\abstract{
The implications of the discovery of the Higgs boson at the LHC
with a mass of approximately 125 GeV are summarised in the context of 
the minimal supersymmetric extension of the Standard Model, the MSSM. 
Discussed are the implications from the measured mass and production/decay 
rates of the observed particle and from the constraints in the search for 
the heavier Higgs states at the LHC. 
 % end of PACS codes
} %end of abstract
\maketitle
\section{Introduction}
\label{intro}

The ATLAS and CMS historical discovery 
of a particle with a mass of approximately 125 GeV \cite{discovery1,discovery2} and 
pro\-per\-ties that are compatible with those of a scalar Higgs boson 
\cite{Higgs,HHG,Review1,Review2}
has far reaching consequences not only for the Standard Model (SM) of the electroweak 
and strong interactions, but also for new physics models beyond it. This is 
particularly true for supersymmetric theories (SUSY) \cite{SUSY} that are widely considered 
to be the most attractive extensions of the SM as they naturally protect the 
Higgs mass against large radiative corrections and stabilise the hierarchy 
between the electroweak and Planck scales, besides of allowing for the unification 
of the three gauge coupling constants and providing a good candidate for the 
dark matter in the universe, the lightest SUSY particle. 

In the minimal supersymmetric extension of the SM (MSSM), two Higgs doublet 
fields $H_u$ and $H_d$ are required to break the electroweak symmetry, leading 
to a physical spectrum with five Higgs particles: two CP--even $h$ and $H$, a 
CP--odd $A$ and two charged $H^\pm$ states \cite{HHG,Review2}. Two parameters 
are needed to describe the MSSM Higgs sector at the tree level: one Higgs mass, 
which is generally taken to be that of the pseudoscalar boson $M_A$, and the 
ratio of vacuum expectation values of the two Higgs fields, $\tan\beta =v_d/v_u$, 
expected to lie in the range $1\! \lsim \! \tb \! \lsim \! 60$. The masses of 
the CP--even $h,H$ and the charged $H^\pm$ states, as well as the mixing angle 
$\alpha$ in the CP--even sector are uniquely defined in terms of 
these two inputs at tree-level, but this nice property is spoiled at higher orders
\cite{CR-1loop,CR-eff,CR-2loop,CR-3loop,adkps,Sven,Carena}.

At high $M_A$ values, $M_A \! \gg \! M_Z$, one is in the so--called decoupling
regime \cite{decoup} in which the neutral CP--even state $h$ is light and has almost exactly the
properties of the SM Higgs boson,  i.e. its couplings to fermions and gauge bosons
are the same as the standard Higgs,  while the  other CP--even $H$ and the charged
$H^\pm$ bosons become heavy and mass degenerate with the $A$ state, $M_H \! \approx
\! M_{H^\pm} \! \approx \! M_A$, and decouple from the massive gauge bosons. In this regime, 
the MSSM Higgs sector thus looks almost exactly as the one of  the SM with its unique 
Higgs boson.

There is, however,   one major difference between the two cases: while in the
SM  the Higgs mass is essentially a free parameter (and should simply be smaller
than about 1 TeV in order to insure unitarity in the high--energy scattering of 
massive gauge bosons), the lightest MSSM CP--even Higgs particle mass is bounded
from above and, depending on the SUSY parameters that enter the important quantum
corrections, is restricted to $M_h^{\rm max} 
\! \approx \! 90$--130 GeV. The lower value comes from experimental constraints, 
in particular Higgs searches at LEP \cite{PDG,LEP}, while the upper bound assumes
a SUSY breaking scale that is not too high, $M_S\! \lsim\! {\cal O}$ (1 TeV), in order
to avoid too much fine-tuning in the model.  Hence, the requirement that the 
MSSM $h$ boson coincides with the one observed at the LHC, i.e. with $M_h \approx 125$ 
GeV and almost SM--like couplings as the LHC data seem to indicate, would place very 
strong constraints on the MSSM 
parameters, in particular the SUSY scale $M_S$, through their contributions to the 
radiative corrections to the Higgs sector. This comes in addition to the limits
that have been obtained from the search of the heavier Higgs states at the LHC, as
well as from the negative search for supersymmetric particles. 

In this review, we summarise the implications of the available LHC Higgs results 
for the MSSM Higgs sector. We first discuss the consequences of the $M_h$ measured value
for the various unconstrained (with the many free parameters defined at the weak scale) 
and constrained (with parameters obeying some universal boundary conditions
at the high scale) versions of the MSSM. We then discuss the impact of the measured 
production and decay rates of the observed particle on the various Higgs couplings 
and, hence, the MSSM parameters. The impact of the negative search of the heavy 
$H,A$ and $H^\pm$ states is summarized. 
An outlook is given in a concluding section.

%\newpage

\section{Implications of the Higgs mass value} 

\subsection{The Higgs masses in the MSSM}

In the MSSM, the tree--level masses of the CP--even $h$ and $H$  bosons depend
only on $M_A$ and $\tb$. However, many parameters of the MSSM such as the 
masses of the third generation stop and sbottom squarks $m_{\tilde t_i},
m_{\tilde b_i}$  and their trilinear couplings $A_{t}, A_b$ enter $M_h$ and $M_H$ 
through quantum corrections.  In the basis $(H_d,H_u)$, the CP--even Higgs  mass
matrix  can be written in full generality as
\beq
{\cal M}^2&=&M_{Z}^2
\left(
\begin{array}{cc}
  c^2_\beta & -s_\beta c_\beta \\
 -s_\beta c_\beta & s^2_\beta \\
\end{array}
\right)
+M_{A}^2
\left(
\begin{array}{cc}
 s^2_\beta & -s_\beta c_\beta \\
 -s_\beta c_\beta& c^2_\beta \\
\end{array}
\right) \nonumber \\
&+&
\left(
\begin{array}{cc}
 \Delta {\cal M}_{11}^2~~ &  \Delta {\cal M}_{12}^2 \\
 \Delta {\cal M}_{12}^2~~ &\Delta {\cal M}_{22}^2 \\
\end{array}
\right)
\label{mass-matrix}
\eeq
where we use the short--hand notation $s_\beta \equiv \sin\beta$ etc$\dots$  and
introduce the radiative corrections  by a general $2\times 2$ matrix  
$\Delta {\cal M}_{ij}^2$. One can then easily derive the neutral CP even Higgs 
boson masses and the mixing angle $\alpha$ that diagonalises the $h$ and $H$ 
states, $H= \cos\alpha H_d^0 + \sin\alpha H_u^0$  and $h=-\sin\alpha H_d^0 + \cos
\alpha H_u^0$: 
\begin{eqnarray}
\hspace{-1.0cm}
M_{h/H}^2&=&\frac{1}{2} \big( M_{A}^2+M_{Z}^2+ \Delta {\cal M}_{11}^2+ 
\Delta {\cal M}_{22}^2  \mp  N \big) \\
\hspace{-1.0cm}
\tan \alpha&=&\frac{2\Delta {\cal M}_{12}^2 - (M_{A}^2 + M_{Z}^2) s_{\beta}}
{ \Delta {\cal M}_{11}^2 -  \Delta {\cal M}_{22}^2 + (M_{Z}^2-M_{A}^2)
c_{2\beta} + N
}
\end{eqnarray}
\vspace*{-5mm}
\begin{eqnarray}
N &=& \sqrt{M_{A}^4 + M_{Z}^4 - 2 M_{A}^2 M_{Z}^2 c_{4\beta} + C} \nonumber \\
C &=&  4 \Delta {\cal M}_{12}^4\! + \!( \Delta {\cal M}_{11}^2 \!- \! 
\Delta {\cal M}_{22}^2)^2 \!- \! 
 2 (M_{A}^2 \! - \! M_{Z}^2)\\ &\times & ( \Delta {\cal M}_{11}^2 \! - \! \Delta M_{22}^2) 
 c_{2\beta} \!   - \!
 4 (M_{A}^2 \! + \!M_{Z}^2)  \Delta {\cal M}_{12}^2 s_{2\beta} \nonumber
\end{eqnarray}

The by far leading one--loop radiative corrections  to the 
mass matrix of eq.~(\ref{mass-matrix}) are controlled by the top Yukawa coupling, 
$\lambda_t =  m_t/v \sin\beta$ with $v=246$ GeV, which appears with the fourth power. 
One obtains a very simple analytical expression for the radiative correction 
matrix $\Delta {\cal M}_{ij}^2$ if only 
this contribution is taken into account \cite{CR-1loop}
\beq
\label{higgscorr}
\Delta {\cal M}_{11}^2& \sim & \Delta {\cal M}_{12}^2 \sim 0 \ , \\
\Delta {\cal M}_{22}^2& \sim & \epsilon \! = \! \frac{3 \bar{m}_t^4}{2\pi^2 v^2\sin^
2\beta} \left[ \log \frac{M_S^2}{\bar{m}_t^2} \!+ \! \frac{X_t^2}{M_S^2} \left( 1 \! -
\! \frac{X_t^2}{12M_S^2} \right) \right] \nonumber
%-\frac{3\,\bar{m}_b^4}{2\,\pi^2 \,v^2}\, \frac{X_b^4}{12\,M_S^4}\;\;,
\eeq
where $M_S$ is the geometric average of the two stop masses $M_S =\sqrt{
m_{\tilde{t}_1}m_{\tilde{t}_2}} $ defined to be the SUSY--breaking scale 
and $X_{t}$ is the stop mixing parameter given by $X_t\!= \! A_t\! - \! 
\mu/\tb$ with $\mu$ the higgsino mass parameter; $\bar{m}_t$ is the 
running ${\rm \overline{MS}}$ top quark mass to account for the leading 
two--loop QCD  corrections in a renormalisation--group improved approach
(some refinements can be include as well). 

Other soft SUSY--breaking parameters, in particular $\mu$ and $A_b$
(and in general the corrections controlled by the bottom Yukawa coupling 
$\lambda_b\! = \! m_b/v \cos\beta$ which at large value of $\mu
\tb$ become relevant) as well as the gaugino mass parameters $M_{1,2,3}$, 
provide a small but non--negligible correction to $\Delta {\cal M}_{ij}^2$ and can 
thus also have an impact on the loop corrections 
\cite{CR-eff,adkps,Sven,Carena}. 

The maximal value of the $h$  mass, $M_h^{\rm max}$ is 
given in the leading one--loop approximation above by 
\begin{eqnarray}
M_h^2 \stackrel{M_A \gg M_Z} \to  M_Z^2 \cos^2 2 \beta +
\Delta {\cal M}_{22}^2
\end{eqnarray}
and is obtained for the  choice of parameters \cite{adkps,Sven,Carena}:

\noindent -- a decoupling regime with heavy $A$ states, 
$M_A\! \sim \mathcal{O}$(TeV);

\noindent -- large values of the parameter $\tb$, $\tb \gsim 10$;

\noindent -- heavy stops, i.e. large $M_S$ values and we choose in general $M_S\! 
\leq\!3$ TeV to avoid a too large fine-tuning \cite{FT,NSUSY}; 

\noindent -- a stop trilinear coupling $X_t=\sqrt{6}M_S$, the so--called ma\-xi\-mal 
mixing scenario that maximizes the stop loops \cite{Mh-max}.

If the parameters are optimized as above, the maximal $h$ mass value can reach the 
level of $M_h^{\rm max} \approx 130$ GeV.   

An important aspect is that in the decoupling regime $M_A \! \gg \! M_Z$, the heavier
CP--even  and the charged Higgs states become almost  degenerate in mass with the 
CP--odd state, $M_H \approx  M_{H^\pm}  \approx M_A$, 
while the mixing angle $\alpha$ becomes close to  $\alpha \approx \frac{\pi}{2} 
-\beta$ making the couplings of the light $h$ state to fermions and massive gauge 
bosons SM--like, and decoupling the $H,H^\pm$ from the weak bosons as is
the case for the state $A$ by virtue of CP invariance. 
 
In this section, we discuss the implications of the measured mass value of the 
observed Higgs boson at the LHC \cite{paper1,paper2,papers-mass,Stop2,H=observed}
that we identify with the lightest state $h$ of
the MSSM. We consider the phenomenological MSSM  \cite{pMSSM} in which the relevant 
soft SUSY parameters are allowed to vary freely (but with some restrictions) 
and  constrained MSSM scenarios such as the minimal supergravity
(mSU\-GRA) \cite{mSUGRA}, gauge mediated (GMSB) \cite{GMSB}  and  anomaly mediated
(AMSB) \cite{AMSB} supersymmetry breaking models (for a review, see again 
Ref.~\cite{Review2}). We also discuss the implications of
such an $M_h$ value for scenarios in which  the supersymmetric spectrum is extremely
heavy, the so--called split SUSY \cite{split} or high--scale SUSY models 
\cite{high-scale}. Finally, a new parametrisation of the Higgs sector which 
uses of the  information $M_h\!=\!125$ GeV, is discussed \cite{Habemus}.

\subsection{Implications for the phenomenological MSSM}

In an unconstrained MSSM, there is a large number of soft SUSY-breaking parameters, 
${\cal O}(100)$, but analyses can be performed in the so--called ``phenomenological 
MSSM" (pMSSM)~\cite{pMSSM}, in which CP conservation, flavour diagonal sfermion mass 
and coupling matrices and universality of the first and second sfermion generations 
are imposed. The pMSSM involves then 22 free parameters in addition to those of the
SM: besides $\tb$ and $M_A$, these are the higgsino mass $\mu$,  the three gaugino 
masses $M_{1,2,3}$, the diagonal left-- and right--handed sfermion mass parameters 
$m_{ {\tilde f}_{L,R}}$ and the trilinear sfermion couplings $A_f$. 

\begin{figure*}[!t]
\vspace*{-3cm}
\begin{center}
\resizebox{0.45\textwidth}{!}{\includegraphics{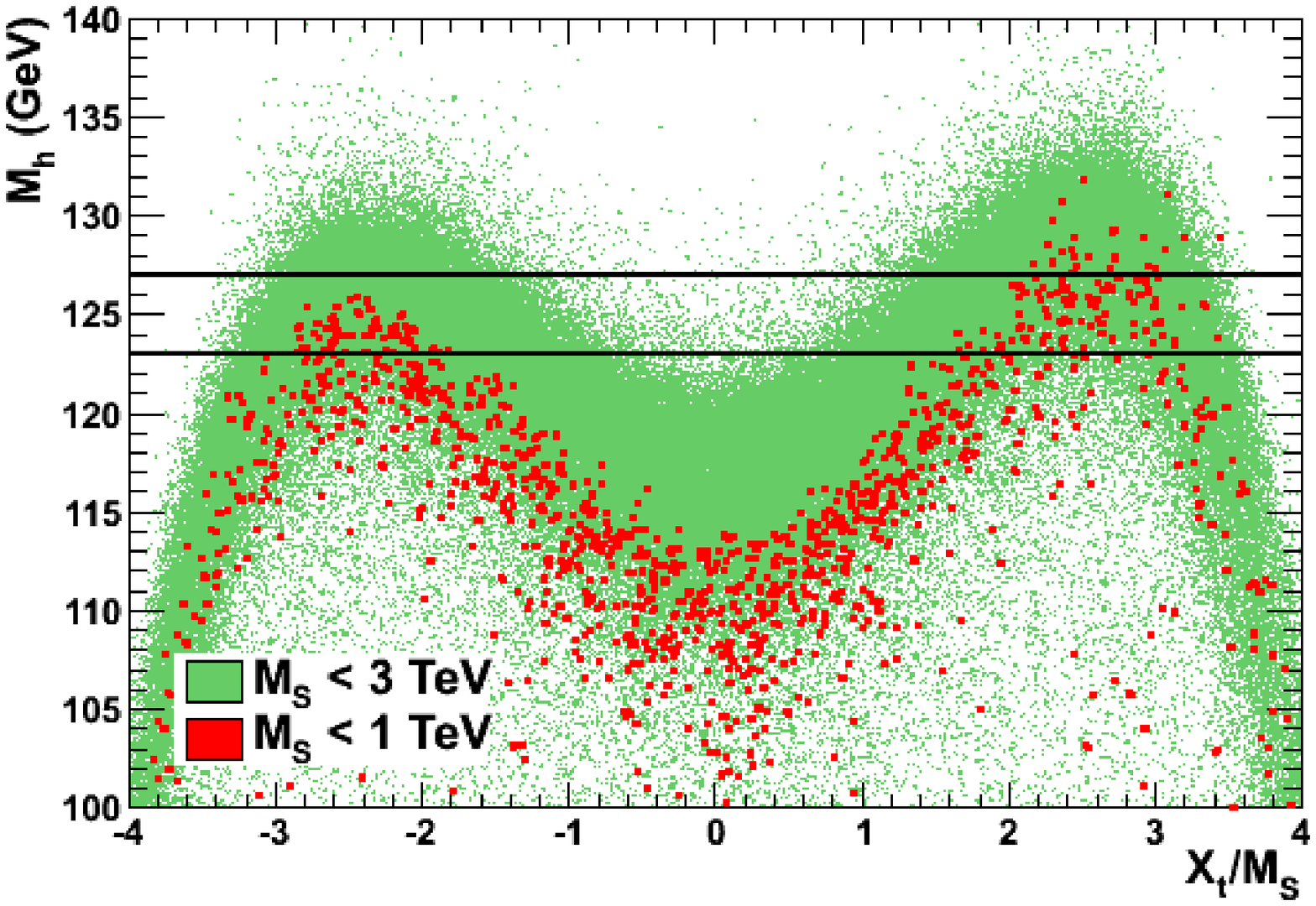} }%\hspace*{-5mm}
\resizebox{0.45\textwidth}{!}{\includegraphics{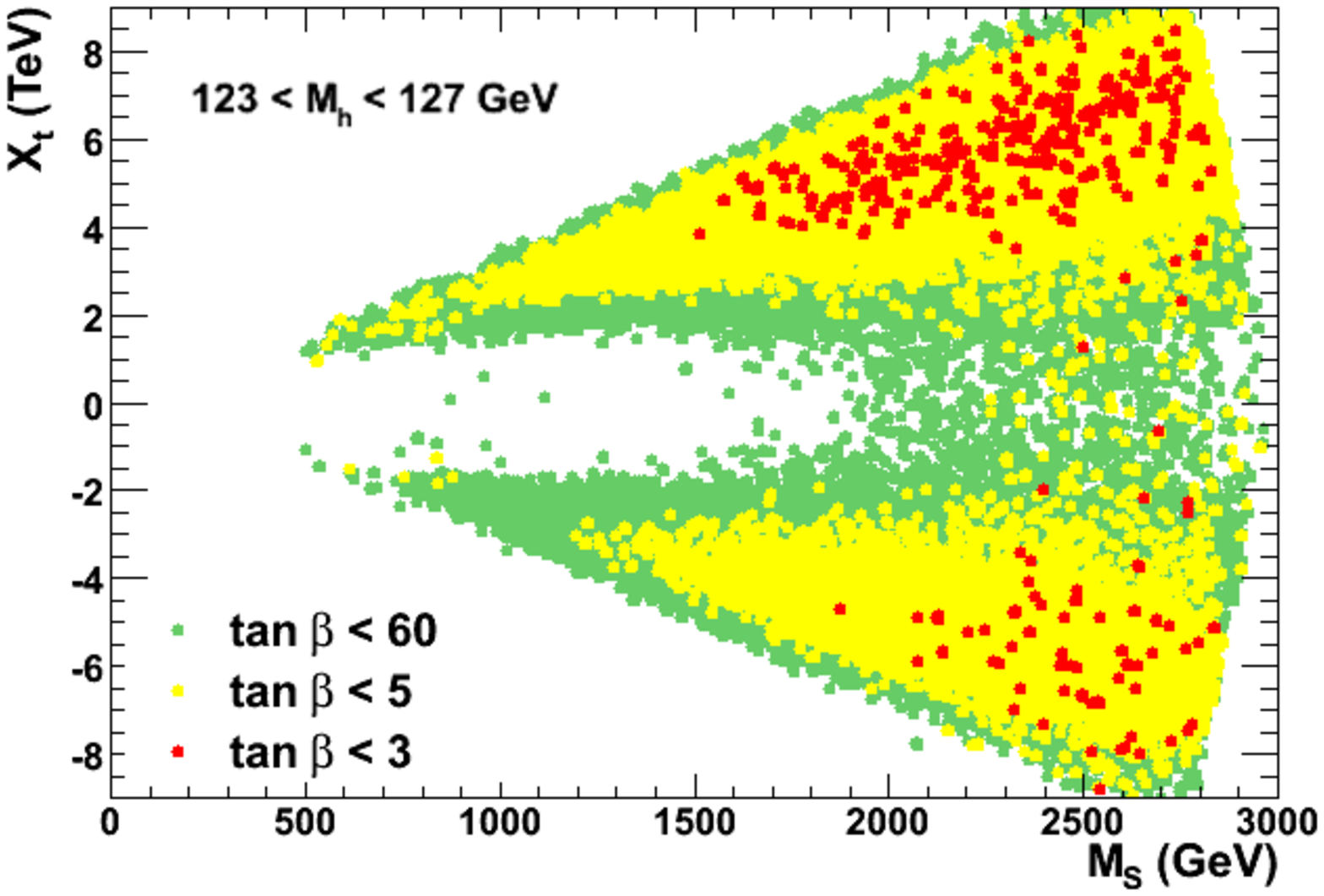} }
\end{center}
\vspace*{-3.7cm}
\caption{The maximal value of the $h$ boson mass as a function 
of $X_t/M_S$ in the pMSSM when  all other soft SUSY--breaking parameters 
and $\tb$ are scanned  (left)
and the contours for the Higgs mass range 123 $<M_h<$127 GeV in the $[M_S,X_t]$ plane for some
selected  range of $\tan\beta$ values (right); from Ref.~\cite{paper1}.}
\label{Fig:pMSSM}
\vspace*{-4mm}
\end{figure*}

As discussed above, an estimate of the upper bound  on $M_h$ can be obtained by 
including the corrections that involve only the parameters $M_S$ and $X_t$.
However, to be more precise, one could scan the full pMSSM 22 parameter space
in order to include the subleading corrections. To do so, one can use RGE programs
such as {\tt Suspect} \cite{Suspect}   which calculate the Higgs and superparticle 
spectrum in the MSSM including the most  up-to-date information \cite{adkps}. 

To obtain the value $M_h^{\rm max}$ with the full radiative corrections, a large scan 
of the pMSSM parameters in an uncorrelated way was performed \cite{paper1,paper2} 
in the  domains:
\begin{eqnarray}
\vspace*{-5mm}
1\leq \tb \leq 60 \, , \  50~{\rm GeV} \leq M_A \leq 3~{\rm TeV}\, , \nonumber
\\ -9~{\rm TeV} \leq A_t, A_b, A_\tau  \leq 9~{\rm TeV} \; ,~~~~  \nonumber \\
~~~50~{\rm GeV} \leq m_{\tilde f_L}, m_{\tilde f_R}, M_3 
\leq 3~{\rm TeV}\, ,~~~ \nonumber \\ 
50~{\rm GeV} \leq M_1, M_2, |\mu|  
\leq 1.5~{\rm TeV}.
\label{scan-range}
\vspace*{-5mm}
\end{eqnarray}
\vspace*{-5mm}

The results are shown in Fig.~\ref{Fig:pMSSM} where, in the  left--hand side,  the
obtained maximal value $M_h^{\rm max}$ is displayed as a
function of the ratio of  parameters $X_t/M_S$. The resulting values are
confronted to the mass range $123~{\rm GeV} \leq M_h \leq 127~{\rm GeV}$
when  the parametric uncertainties from the
SM inputs such 	as the top quark mass and the  theoretical uncertainties in
the determination of $M_h$ are included\footnote{This uncertainty is obtained by comparing
the outputs of {\tt SuSpect} and {\tt FeynHiggs} \cite{feynhiggs} which
use different schemes for the radiative corrections: while the former uses the
$\overline{\rm DR}$ scheme, the latter uses the on--shell scheme; the difference in
the obtained $M_h$ amounts to $\approx \pm 2$--3 GeV in general. To this, one has 
to add an uncertainty of $\pm 1$ GeV from the top quark mass measurement
at the Tevatron, $m_t=173 \pm 1$ GeV \cite{mt-TeV}. Note that it is not entirely clear whether 
this mass is indeed the pole quark mass. A more rigorous
determination of the pole mass from the measured top--pair cross section at the
Tevatron gives a lower value with a larger uncertainty, $m_t^{\rm pole}\!=171 
\!\pm \!3$ GeV \cite{Alekhin}.}. 
 
For $M_S \! \lsim \! 1$ TeV, only the scenarios with $X_t/M_S$ values close to maximal  
mixing $X_t/M_S \approx \sqrt 6$ survive. The no--mixing scenario $X_t \approx 0$ 
is ruled out for $M_S
\lsim 3$ TeV, while the typical mixing scenario, $X_t \approx M_S$, needs large $M_S$ and moderate
to large $\tan\beta$ values.  From the scan, one obtains a maximum $M_h^{\rm max}$=136, 126 and 123 GeV with maximal, typical and zero mixing, respectively.

What are the implications for the mass of the lightest stop state $\tilde t_1$? This is
illustrated in the right--hand side of Fig.~\ref{Fig:pMSSM} where,  shown are the contours 
in the $[M_S,X_t]$ plane in which one obtains $123 \! < \! M_h \!  < \! 
127$~GeV from the pMSSM scan; the regions in which $\tan\beta 
\lsim 3, 5$ and 60 are highlighted. One sees again that a large part of the parameter
space is excluded  if the Higgs mass constraint is imposed. In particular,  
large $M_S$ values, in general corresponding to large $m_{\tilde t_1}$ are favored.
However, as $M_S\! = \! \sqrt { m_{\tilde t_1} m_{\tilde t_2}}$, the possibility that 
$m_{\tilde t_1}$ is of the order of a few 100
GeV is still allowed, provided that stop mixing (leading to a significant
$m_{\tilde t_1}, m_{\tilde t_2}$ splitting) is large \cite{paper2,Stop2}. 

Masses above 1 TeV for the scalar partners of light quarks and for the gluinos
are also required by the direct searches of SUSY particles at the LHC 
\cite{LHC-SUSY,Craig}, confirming 
the need of high $M_S$ values. Nevertheless, relatively light stops 
as well as electroweak sparticles such as sleptons, charginos and neutralinos 
are still possible allowing for a ``natural SUSY" \cite{NSUSY} despite of
the value  $M_h \approx 125$ GeV. Nevertheless, the present LHC SUSY searches 
\cite{LHC-SUSY,Craig} are constraining more and more this natural scenario.
 
\subsection{Implications for constrained MSSM scenarios} 

In constrained MSSM scenarios (cMSSM),  the various soft SUSY--breaking parameters 
obey a number of universal boundary conditions at a high energy scale, thus reducing 
the number of basic input parameters to a handful. The various soft SUSY--breaking 
parameters are evolved via the MSSM renormalisation group equations down to the low  
energy scale $M_S$ where the conditions of proper electroweak symmetry breaking (EWSB)
are imposed. 

Three classes of such models have been widely
discussed in the literature. There is first
the minimal supergravity (mSUGRA) model \cite{mSUGRA} in which  
SUSY--breaking is assumed to occur in a hidden sector which communicates with the 
visible sector only via flavour-blind gravitational interactions, 
leading to universal soft breaking terms, namely a common $m_{1/2}, m_0, A_0$
values for the gaugino masses, sfermion masses and sfermion trilinear couplings. 
Then come the gauge mediated \cite{GMSB} and anomaly mediated \cite{AMSB}   
SUSY--breaking (GMSB and AMSB) scenarios  in which
SUSY--breaking is communicated to the visible sector via, respectively,
 gauge interactions and a  super-Weyl anomaly.

These models are described by $\tb$, the sign of $\mu$ and a few continuous
parameters. Besides of allowing for both signs of $\mu$, requiring  $1\leq\tan
\beta\leq 60$ and, to avoid excessive fine--tuning in the EWSB
conditions,  imposing the bound  $M_S =M_{\rm EWSB} < 3~{\rm TeV}$,  
we  adopt  the  following ranges for the input parameters of these 
scenarios: 

\begin{figure*}[!t]
\vspace*{-2.5cm}
\begin{center}
\begin{tabular}{ll}
\begin{minipage}{8cm}
\resizebox{0.95\textwidth}{!}{\includegraphics{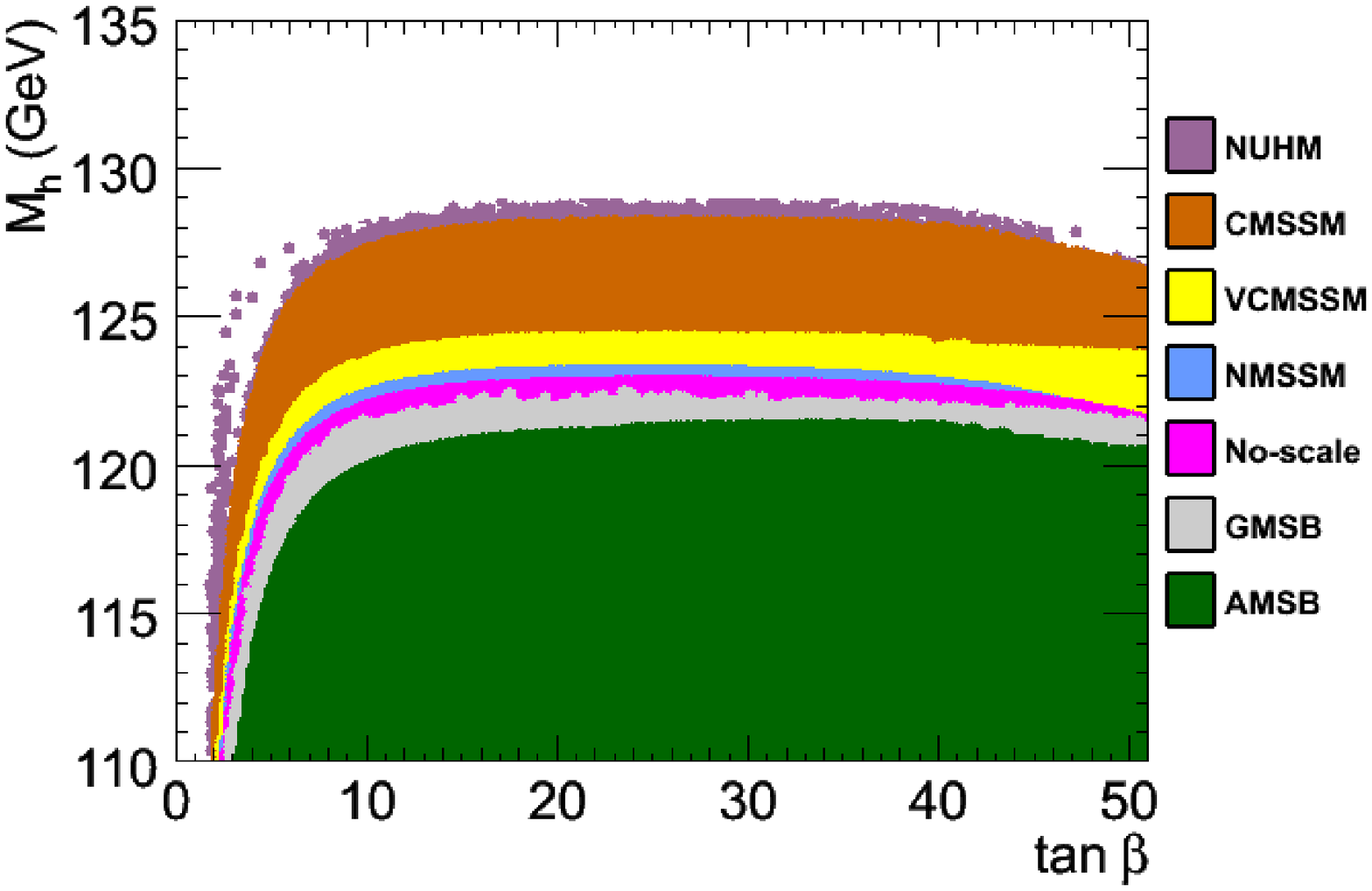}}\hspace*{-1cm} 
\end{minipage}
& \hspace*{-.2cm} 
\begin{minipage}{8cm}
\resizebox{0.95\textwidth}{!}{\includegraphics{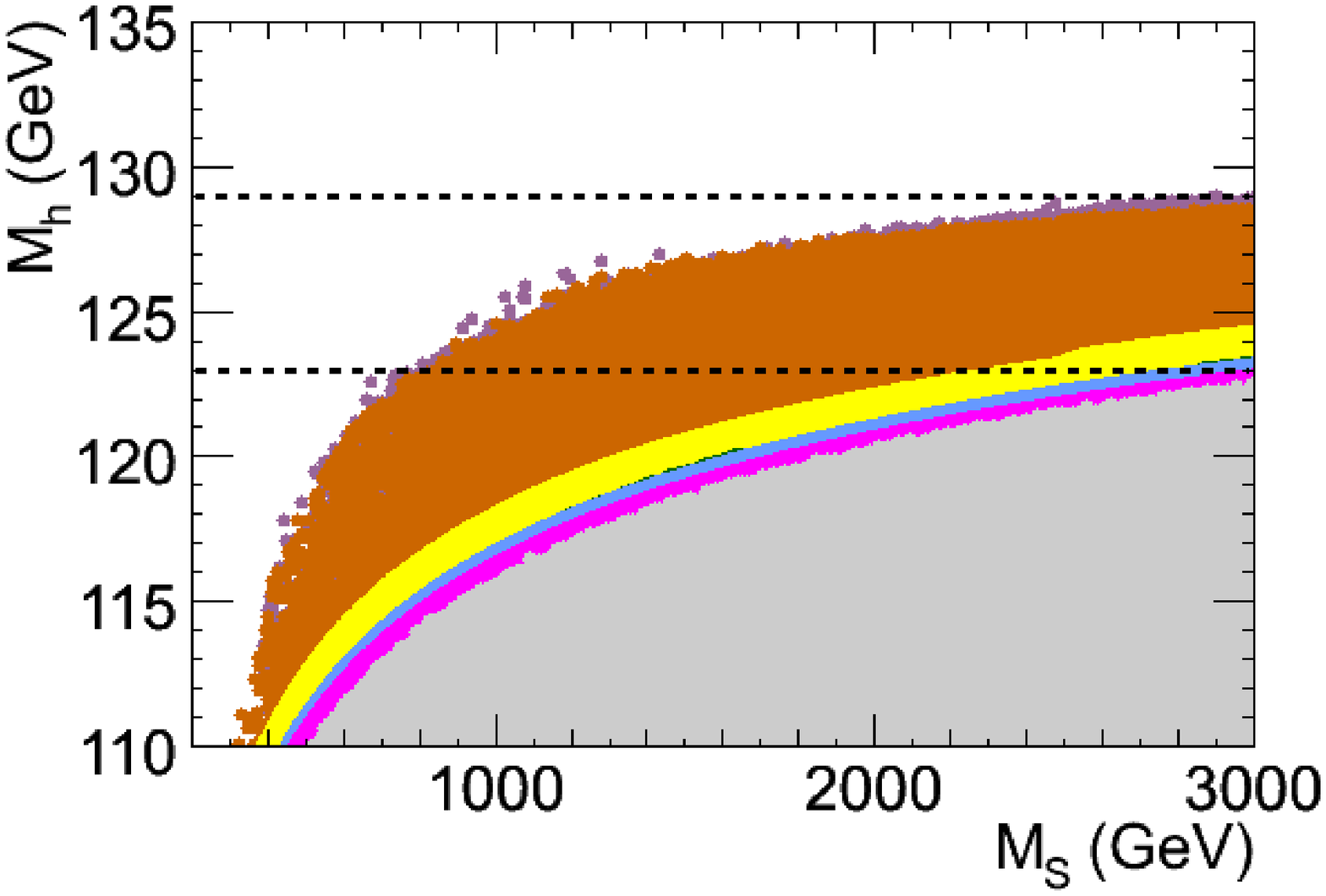}}
\end{minipage}
\end{tabular}
\end{center}
\vspace*{-3.3cm}
\caption{The maximal value of the $h$ boson mass as a function 
of $\tb$ (left) and $M_S$ (right) with a scan of all other parameters
in various constrained MSSM scenarios.  The range $123\! < \! M_h\! <\! 129$ GeV 
for the light $h$ boson mass is
highlighted. From Ref.~\cite{paper1}.}
\label{Fig:cMSSM}
\end{figure*}

\begin{table*}
\begin{center} 
\begin{tabular}{rccc} 
mSUGRA: & 50 GeV $\leq m_0 \leq$ 3 TeV, & 50 GeV $\leq m_{1/2} \leq$ 3 TeV, & 
$|A_0|\leq 9 $ TeV; \\ 
GMSB: & 10 TeV $\leq \Lambda\leq$ 1000 TeV, & 1 $\leq M_{\rm mess} / \Lambda
\leq 10^{11}$, & $N_{\rm mess} =$ 1; \\ 
AMSB: & 1 TeV $ \leq m_{3/2}\leq$ 100 TeV, & 50 GeV $\leq m_0\leq$ 3 TeV. &  
\end{tabular}
\end{center}
\vspace*{-5mm}
\end{table*}

Hence, in contrast to the pMSSM,  the various parameters which enter the radiative 
corrections to $M_h$  are not all independent in these constrained scenarios, 
as a  consequence of the relations between SUSY breaking parameters
that are set at the high--energy scale and  the requirement that electroweak
symmetry breaking is triggered radiatively for each set of input parameters.  
The additional constraints  make that it is not possible to freely 
tune the parameters that enter the Higgs sector to obtain the pMSSM maximal value of 
$M_h$. In order to obtain even a rough determination of 
$M_h^{\rm max}$  in a given constrained SUSY scenario, it is necessary to scan
through the allowed range of values for the basic input parameters.

Using again the program {\tt Suspect}, a  full scan  of these scenarios has been 
performed in Ref.~\cite{paper1} and the results for $M_h^{\rm max}$ are shown in the 
left-hand side
of Fig.~\ref{Fig:cMSSM}  as a function of $\tan\beta$, the input parameter that is
common to all models, and in the right-hand side of the figure as a function of $M_S$. 
In the adopted parameter space of the models and with the
central values of the SM inputs, the obtained upper $h$ mass value  is 
$M_h^{\rm max} \approx $ 121 GeV in the AMSB scenario, i.e. much less that 125
GeV, while in the  GMSB scenario one has $M_h^{\rm max} \approx $ 122 GeV
(these values are obtained for $\tan\beta \approx 20$). Thus,
clearly, these two scenarios are disfavoured if the lightest $h$
particle has indeed a mass in the range 123--127~GeV and $M_S \lsim 3$ TeV. In 
mSUGRA, one obtains $M_h^{\rm max} \! = \! 128$ GeV and, thus, some
parameter space would still survive the $M_h$ constraint.

The upper bound on $M_h$ in these scenarios  can be qualitatively understood by
considering in each model the allowed values of the trilinear coupling $A_t$,
which essentially determines the stop mixing parameter $X_t$ and thus the value
of $M_h$ for a given scale $M_S$. In GMSB, one has $A_t\approx 0$ at relatively
high scales and its magnitude does not significantly increase in the evolution
down to the scale  $M_S$; this implies that we are almost in the no--mixing
scenario which  gives a low value of $M_h^{\rm max}$ as can be seen from
Fig.~\ref{Fig:pMSSM}. In  AMSB, one has a non-zero $A_t$ that is fully predicted
at any renormalisation scale in terms of the Yukawa and gauge couplings;
however, the ratio $A_t/M_S$ with $M_S$ determined  from the overall SUSY
breaking scale $m_{3/2}$ turns out to be rather small, implying again that we
are close to the no--mixing scenario.  Finally, in the mSUGRA model, since we
have allowed $A_t$ to vary in a wide range as $|A_0|\leq  9$ TeV,  one can get a
large $A_t/M_S$ ratio which leads to a heavier Higgs particle. However, one
cannot easily reach $A_t$ values such that $X_t/M_S \approx \sqrt 6$ so that we
are not in the maximal--mixing scenario  and  the higher upper bound on $M_h$ in
the pMSSM  cannot be reached.

In the case of mSUGRA, one can study several interesting special 
cases : the no-scale scenario  with $m_0\! \approx \! A_0 \! \approx \! 0$ \cite{no-scale}, 
the scenario $m_0 \! \approx \! 0$ and $A_0 \! \approx \! -\! \frac14 m_{1/2}$ which
approximately corresponds  to the constrained next-to--MSSM (cNMSSM) \cite{cNMSSM},
 $A_0\! \approx\! -m_0$ which corresponds to a very constrained 
MSSM (VCMSSM) \cite{VCMSSM}, and a non--universal Higgs mass model 
(NUHM) \cite{NUHM} in which the  soft
SUSY--breaking scalar mass terms are  
different for the sfermions and for the
two Higgs  doublet fields.

In two particular cases, namely 
the ``no--scale" and the  ``approximate cNMSSM" scenarios, the upper bound on
$M_h$ is much lower than in the more general mSUGRA case and, in fact, barely
reaches $M_h \approx 123$ GeV. The main reason is that these
scenarios  involve small values of $A_0$ at the GUT scale, $A_0 \approx 0$ for
no--scale and $A_0 \approx -\frac14 m_{1/2}$ for the cNMSSM which lead to
$A_t$ values at the weak scale that are too low to generate a significant stop
mixing and, hence, one is again close to the no--mixing scenario. Thus, only a
very small fraction of the parameter space of these two sub--classes of the
mSUGRA model survive
if we impose 123 $< M_h <$ 127 GeV. These models should thus have a very
heavy sfermion spectrum as a value $M_S \gsim 3$ TeV is required to increase $M_h^{\rm
max}$. In the VCMSSM case, the value $M_h \simeq 125$ GeV can be reached as $|A_0|$
can be large for large $m_0$, $A_0 \approx -m_0$, allowing for typical
mixing.

Finally, since the NUHM  is more general than  mSUGRA as we have two more free
parameters, the $[\tan\beta, M_h]$ area shown in  Fig.~\ref{Fig:cMSSM} is larger
than in mSUGRA. However, since we are in the decoupling regime and the
value of $M_A$ does not matter much (as long as it a larger than a few hundred
GeV) and  the key weak--scale parameters entering the determination of $M_h$,
i.e. $\tan \beta, M_S$ and $A_t$ are approximately the same in both models, one
obtains a bound $M_h^{\rm max}$ that is only slightly higher in NUHM compared to
the mSUGRA case. 

In these constrained scenarios and, in particular in the general 
mSUGRA model, most of the scanned points giving the appropriate Higgs mass correspond
to the decoupling regime of the MSSM Higgs sector and, hence, to an $h$ boson
with a SM--Higgs cross section and branching ratios. Furthermore,  as the resulting 
SUSY spectrum for $M_h\!=\!125\!\pm \!2$ GeV is rather heavy in these
scenarios (easily evading the LHC limits from direct sparticle searches
\cite{LHC-SUSY}), one obtains 
very small contributions to observables like the
anomalous muon magnetic moment  $(g-2)_\mu$ and to 
$B$--physics observables such as the rates BR($B_s\! \to \! \mu^+\mu^-$)
or BR$(b\! \to \! s\gamma)$ \cite{B-physics}. Hence, the resulting spectrum complies with all 
currently available constraints. In addition, as will be discussed later,
the correct cosmological density for the LSP neutralino required by recent measurements 
\cite{DM-review} can be easily satisfied. The $M_h$ value provides thus
a unique constraint in this decoupling regime.

\subsection{Split and high--scale SUSY models} 

In the preceding discussion, we have always assumed that the SUSY--breaking
scale is  relatively low, $M_S \lsim 3$ TeV, which implies a natural SUSY
scenario \cite{NSUSY} with supersymmetric and heavier Higgs particles that
could be observed at the LHC. 
However, as already mentioned,  this choice is mainly
dictated by fine--tuning considerations which are a rather subjective matter as
there is no compelling criterion to quantify the acceptable amount of tuning.
One could well abandon the SUSY solution to the hierarchy problem and 
have a very high $M_S$ which implies  that, except for
the lightest $h$ boson, no other scalar particle is accessible at the LHC or at
any foreseen collider.

This  argument has been advocated to construct the so--called split SUSY
scenario \cite{split} in which the soft SUSY--breaking mass terms for all the
scalars of the theory, except for one Higgs doublet, are extremely large, i.e. 
their common value $M_S$ is such that $M_S \gg 1$ TeV (such a situation occurs
e.g. in some string motivated models \cite{String}.). Instead,
the mass parameters for the spin--$\frac12$ particles,  the gauginos and the
higgsinos, are left in the vicinity of the EWSB scale,  allowing for a solution
to the dark matter  problem and a successful gauge coupling unification, the two
other SUSY virtues. The split SUSY models are much more predictive than the usual
pMSSM as only a handful parameters are needed to describe the low energy theory.
Besides the common value $M_S$ of the soft SUSY-breaking sfermion and one Higgs
mass parameters, the basic inputs are essentially the three gaugino masses
$M_{1,2,3}$ (which can be unified to a common value at $M_{\rm GUT}$ as in
mSUGRA), the higgsino parameter $\mu$ and $\tan\beta$. The  trilinear couplings
$A_f$, which are expected to have values close to the EWSB scale set by the 
gaugino/higgsino masses that are much smaller than $M_S$, will play a  negligible role. 

Concerning the Higgs sector, the main feature of split SUSY is that at the 
high scale $M_S$,  the boundary condition on the quartic Higgs coupling  
is determined by SUSY:
\beq
\label{boundlam}
\lambda(M_S) = \frac{1}{4}\left[ g^2(M_S)+g^{\prime 2}(M_S)
\right] \,\cos^22\beta~. 
\eeq
where $g$ and $g'$ are the SU(2) and U(1) gauge couplings. Here, $\tan\beta$ is
not a parameter of the low-energy effective theory as it enters only the boundary
condition above and cannot be interpreted as the ratio of the two Higgs vevs.  

If the scalars are very heavy, they will lead to radiative corrections in the
Higgs sector that are significantly  enhanced by large logarithms, $\log( 
M_S/M_{\rm EWSB})$ where $M_{\rm EWSB} \approx |\mu|, M_{2}$. In order to have 
reliable predictions, one has to properly
decouple the heavy states from the low-energy theory and resum the large
logarithmic corrections; in addition, the radiative corrections due to the
gauginos and the higgsinos  have to be implemented.  Following the early work of
Ref.~\cite{split}, a comprehensive study of the split SUSY spectrum has  been
performed in Ref.~\cite{bds}.
All the features of the model have been implemented in the  code {\tt
SuSpect} \cite{Suspect} upon which the analysis  presented 
in Ref.~\cite{paper1} and summarised here is based. 

One can adopt an even more radical attitude than in split SUSY and
assume that the gauginos and higgsinos are also very heavy, with a mass close to
the scale $M_S$; this is the case in the so--called high--scale SUSY model
\cite{high-scale}.  Here, one abandons  not only the SUSY solution to the fine-tuning
problem but also the solution to the dark matter problem by means of  the  LSP
and the successful unification of the gauge couplings.  However, there
will still be a trace of SUSY at low energy: the matching of the SUSY and
low--energy theories is indeed encoded in the Higgs quartic coupling  $\lambda$
of eq.~(\ref{boundlam}). Hence, even if broken at very high scales, SUSY
would still lead to a ``light" Higgs whose mass will give information
on $M_S$ and $\tan\beta$. 

\begin{figure*}[!t]
\vspace*{-.1cm}
\begin{center}
\hspace*{-2mm}
\mbox{
\resizebox{0.4\textwidth}{!}{\includegraphics{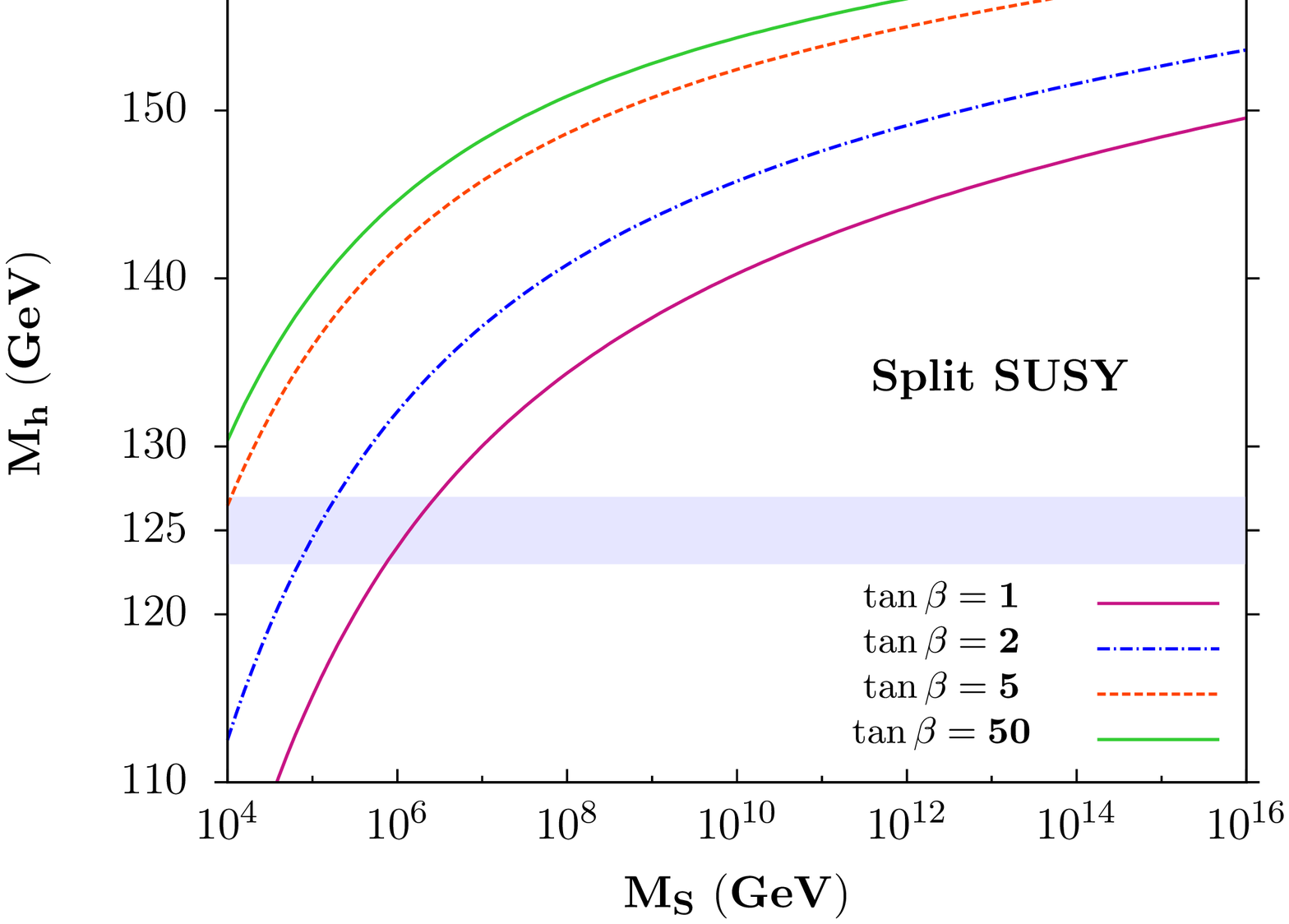}}\hspace{-.3mm} 
\resizebox{0.4\textwidth}{!}{\includegraphics{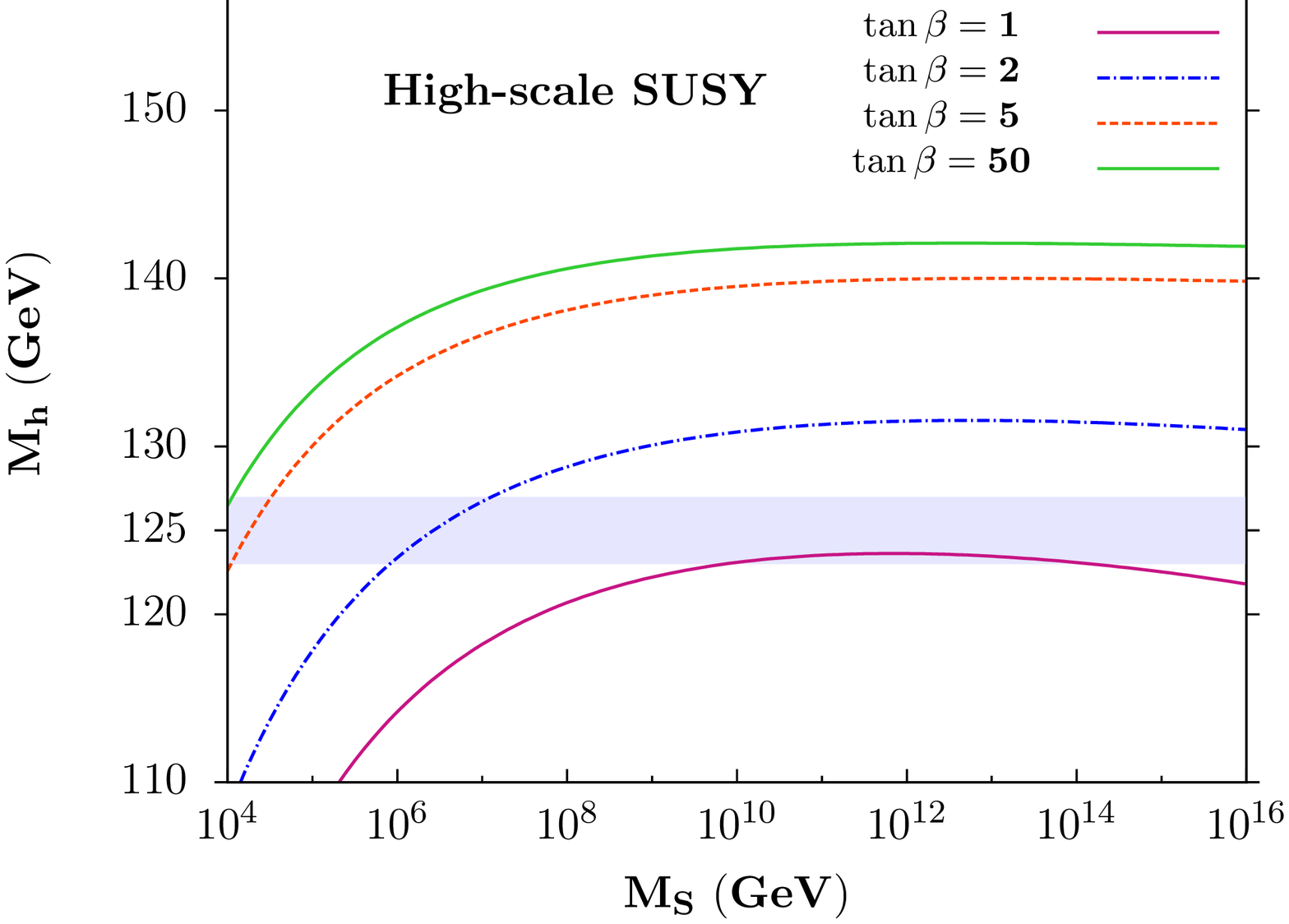}}
}
\end{center}
\vspace*{-5.cm}
\caption[]{The value of $h$ boson mass as a function of the SUSY scale $M_S$ for 
several values of $\tan\beta=1,2,5,50$ in the split--SUSY (left) and high--scale 
SUSY (right) scenarios. From Ref.~\cite{paper1}.}
\vspace*{-4mm}
\label{Fig:hMSSM}
\end{figure*}

The treatment of the Higgs sector of the high--scale SUSY scenario is similar to
that of split SUSY: one simply needs to decouple the gauginos and higgsinos
from the low energy spectrum (in particular remove their contributions to the
renormalisation group evolution of the gauge and Yukawa couplings and to the
radiative corrections to $M_h$) and set their masses to $M_S$. The version of the 
program {\tt Suspect} which handles the 
split SUSY case can be adapted to also cover the $M_1 \! \approx \! M_2 \!
\approx \! M_3 \! \approx \! |\mu| \! \approx \! M_S$ case. 

Using this tool, a scan in the $[\tan\beta, M_S]$ plane  has been performed to 
determine the value of $M_h$ in the
split SUSY and high--scale SUSY scenarios; in the former case,  
$M_{\rm EWSB} \approx \sqrt{|M_2\mu| } \approx 246$ GeV was chosen for the low scale. 
The results are shown in Fig.~\ref{Fig:hMSSM} where $M_h$ is displayed as a 
function of $M_S$ for selected values of $\tan\beta$ in both split (left plot) 
and high--scale (right plot) SUSY.

As expected, the maximal $M_h$ values are obtained at high $\tan\beta$ and $M_S$
values and, at the scale $M_S \approx 10^{16}$ GeV at which the couplings $g$
and $g'$ approximately unify in the split SUSY scenario, one obtains $M_h
\approx 160$ GeV for the higher $\tan\beta=50$ value. Not included is the 
error bands in the  SM inputs that would
lead to an uncertainty of about 2 GeV on $M_h$,  which is now mainly due to the 1 GeV
uncertainty on $m_t$. In addition, the zero--mixing
scenario was assumed as the parameter $A_t$ is expected to be much smaller than $M_S$; this
approximation might not be valid for $M_S$ values below 10 TeV and a maximal
mixing $A_t/M_S = \sqrt 6$ would increase the Higgs mass value by up to 10 GeV
at $M_S ={\cal O} (1~{\rm TeV})$ as was discussed earlier for the pMSSM. In the
high--scale SUSY scenario, one obtains a value  $M_h\approx 142$ GeV  (with
again an uncertainty of approximately 2 GeV from the top mass) for high
$\tan\beta$ values and at the unification scale $M_S \approx 10^{14}$ GeV  
\cite{high-scale}. Much smaller $M_h$ values, in the 120 GeV range, can be
obtained for lower scales and $\tan\beta$. 

Hence, the requirement that the Higgs mass is in the range 123 $\lsim M_h \lsim$ 127 GeV
imposes strong constraints on the parameters of these two models. For this
mass range, very large scales are needed for $\tan\beta\approx 1$ in the 
high--scale SUSY scenario, while scales not too far from $M_S\!
\approx \! 10^{4}~{\rm GeV}$ are required at $\tan\beta\! \gg \! 1$ in both the split
and high--scale scenarios. In this case,  SUSY should manifest itself at scales 
much below $M_{\rm GUT}$ if  $M_h\approx 125$ GeV.

\subsection{Splitting the Higgs and sfermion sectors}

In the previous high scale scenarios, the Higgs mass parameters were assumed 
to be related to the mass scale of the scalar fermions in such a way that
the masses of the heavier Higgs particles are also of the order of the
SUSY scale, $M_A 
\approx M_S$. However,  this needs  not to be true in general and one can, 
for instance, have a NUHM--like scenario where the Higgs masses are decoupled 
from those of the sfermions. If one is primarily concerned with the MSSM Higgs 
sector, one may be rather conservative and allow any value for $M_A$ 
irrespective of the SUSY--breaking scale $M_S$. This is the quite 
``model--independent" approach that has been advocated in Refs.~\cite{paper4,slim}: 
take $M_A$ as a free parameter of the pMSSM, with values ranging from ${\cal O}( 
100$ GeV) up to ${\cal O}(M_S)$, but make no restriction on $M_S$ which can 
be set to any value, even very high.

\begin{figure}[!h]
\begin{center}
\vspace*{-6.9cm}
\hspace*{-2cm}
\resizebox{0.7\textwidth}{!}{\includegraphics{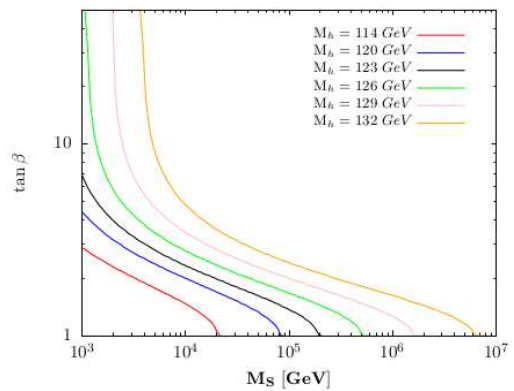}}
\end{center}
\vspace*{-6.7cm}
\caption[]{Contours for fixed values $M_h=120, 123, 126,129$ and 132 GeV
in the $[\tb, M_S]$ plane in the decoupling limit  $M_A \gg M_Z$; the ``LEP2 
contour" for $M_h=114$ GeV is also shown.}
\label{Fig:mass}
\vspace*{-.6cm}
\end{figure}

An important consequence of this possibility is that it reopens the low
$\tan\beta$ region, $\tan\beta \! \lsim \! 3$, that was long thought to be forbidden
if one requires a SUSY scale $M_S \! \lsim \! 1$ TeV, as a result of the limit $M_h 
\! \gsim \! 114$ GeV from the negative search of a SM--like Higgs boson at LEP \cite{LEP}. 
If the  SUSY scale is large enough, these small $\tb$ values would
become viable again. To estimate the required magnitude of $M_S$, 
one can still use  {\tt  Suspect} in which 
the possibility $M_S \gg 1$ TeV is implemented \cite{bds} with the
full set of radiative corrections up to two--loops included. In  Fig.~\ref{Fig:mass}, displayed 
are the contours in the plane $[\tb, M_S]$ for  fixed mass values $M_h=120$--132 GeV 
of the observed Higgs state (these include a 3 GeV theoretical 
uncertainty and also a 3 GeV uncertainty on the top quark mass \cite{Alekhin}
that is 
conservatively added linearly in the extreme cases).  The maximal mixing $X_t\! =\! \sqrt 
6 M_S$ scenario is assumed with 1 TeV gaugino/higgsino mass parameters.

One observes that values of $\tb \approx 1 $ are possible
and allow for an acceptable $M_h$ value provided the scale $M_S$ is large
enough. For instance,  while one can accommodate a scale $M_S \approx 1$ TeV  
with $\tb \approx 5$, a large scale $M_S \approx 20$ TeV is required to obtain
$\tb \approx 2$; to reach the limit $\tb=1$, an order of magnitude increase of
$M_S$ will be required. Outside the decoupling regime, the  obtained $M_S$ for a
given $M_h$ value will be of course larger. For completeness, also shown is the
contour for the LEP2 limit $M_h=114$ GeV which illustrates the fact that 
 $\tb \approx 1$ is still allowed provided that $M_S \gsim
20$ TeV. 

\subsection{A new parametrisation of the Higgs sector}

It was pointed out in Refs.~\cite{paper4,R1} that when the measured value of the $h$ 
boson mass $M_h\!=\!125$ GeV is taken into account, the MSSM Higgs  sector with  
solely the dominant radiative corrections included, can be again described with 
only two free parameters such as $\tb$  and $M_A$ as it was the case at 
tree--level. In other words,  the dominant radiative corrections that involve
the SUSY parameters  are fixed by the value of $M_h$. This observation leads 
to a rather simple parametrisation of the MSSM Higgs sector. 

More specifically, let us assume that in the $2\times 2$ matrix for the radiative 
corrections to the CP--even Higgs mass matrix eq.~(\ref{mass-matrix}), only the 
leading $\Delta{\cal M}^{2}_{22}$ entry of eq.~(\ref{higgscorr}) that involves 
the by far dominant stop--top sector contribution is taken into account;
this is the so--called $\epsilon$ approximation and its refinements \cite{CR-eff,Carena}. 
In this $\Delta{\cal M}^{2}_{22} \gg \Delta{\cal M}^{2}_{11}, \Delta{\cal  M}^{2}_{
12}$  limit, one can simply trade $\Delta {\cal M}^{2}_{22}$ for the by now known 
$h$ mass value $M_h=125$ GeV and obtain
\begin{eqnarray}
\begin{array}{l} 
M_{H}^2 = \frac{(M_{A}^2+M_{Z}^2-M_{h}^2)(M_{Z}^2 c^{2}_{\beta}+M_{A}^2
s^{2}_{\beta}) - M_{A}^2 M_{Z}^2 c^{2}_{2\beta} } {M_{Z}^2 c^{2}_{\beta}+M_{A}^2
s^{2}_{\beta} - M_{h}^2} \\
\ \ \  \alpha = -\arctan\left(\frac{ (M_{Z}^2+M_{A}^2) c_{\beta} s_{\beta}} {M_{Z}^2
c^{2}_{\beta}+M_{A}^2 s^{2}_{\beta} - M_{h}^2}\right)
\end{array}
\label{wide} 
\end{eqnarray}
This was called the habemus MSSM or hMSSM in Ref.~\cite{Habemus}. 

However, this interesting and simplifying feature has to been demonstrated for all 
MSSM parameters  and, in particular, one needs to prove that the impact of the 
subleading corrections $\Delta {\cal M}^{2}_{11}$ and $\Delta {\cal M}^{2}_{12}$
is small. To do so, a scan of the pMSSM parameter space using the program 
{\tt SuSpect}, in which the full two--loop radiative corrections to the Higgs sector  
are implemented, has been performed \cite{Habemus}. For a chosen  ($\tb$,$M_A$) 
input set, the soft--SUSY parameters that play an important role in the Higgs sector 
are varied in the following ranges:
$|\mu|\leq 3$~TeV,   $|A_t,A_b|\leq 3 M_S$, $1$~TeV$\leq \! M_3 \! \leq \!3$ TeV
and $0.5$~TeV$ \!\leq \!M_S \!\leq  \!3$~TeV ($\approx 3$ TeV is the scale up to
which programs such as {\tt SuSpect} are expected to be reliable). The usual GUT
relation between the  weak scale gaugino masses  $6 M_1\!= \!3 M_2 \!=
\!M_3$ has been assumed and $A_u,A_d, A_\tau\! =\! 0$ has been set 
(these  last parameters have little impact on the radiative corrections).  
The MSSM Higgs sector parameters have been computed all across the parameter space,
selecting the points which satisfy the constraint  $123 \! \leq \! M_h \!  \leq
\! 129$ GeV when uncertainties are included. For each of theses points,  the Higgs 
parameters have been compared to
those obtained in the simplified MSSM approximation,  $\Delta {\cal M}^{2}_{11}
\! = \! \Delta {\cal M}^{2}_{12} \! =\! 0$, with the lightest Higgs boson mass
as input. While the requirement that ${M}_h$ should lie in the range 123--129 GeV
has been made, $M_h$ was allowed to be different from the one obtained in  the 
``exact" case $\Delta {\cal M}^{2}_{11}, \Delta {\cal M}^{2}_{12} \neq 0$. 

\begin{figure}[!htbp]
\begin{center}
\vspace*{-3.4cm}
\mbox{
\hspace*{-1cm}
\resizebox{0.36\textwidth}{!}{\includegraphics{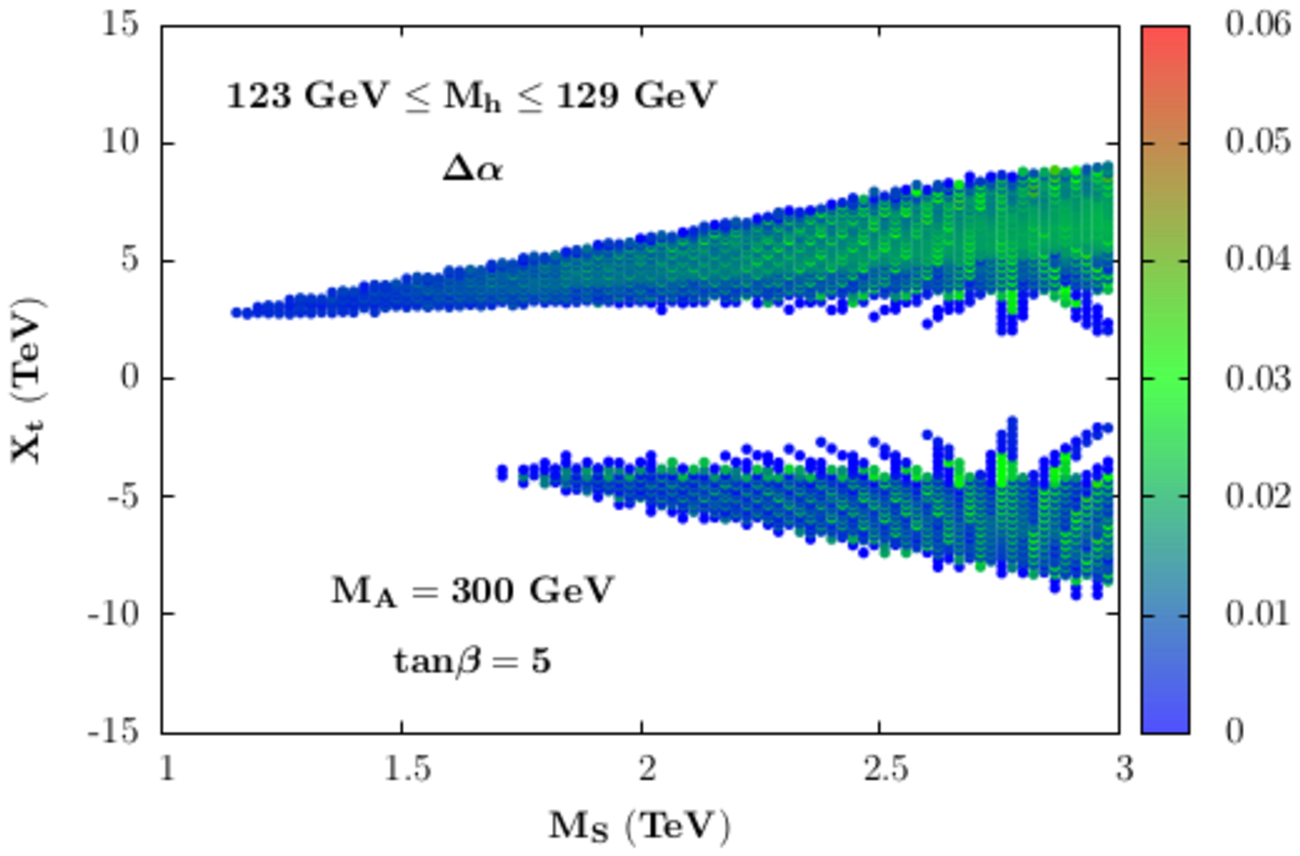}}\hspace*{-2cm} 
\resizebox{0.36\textwidth}{!}{\includegraphics{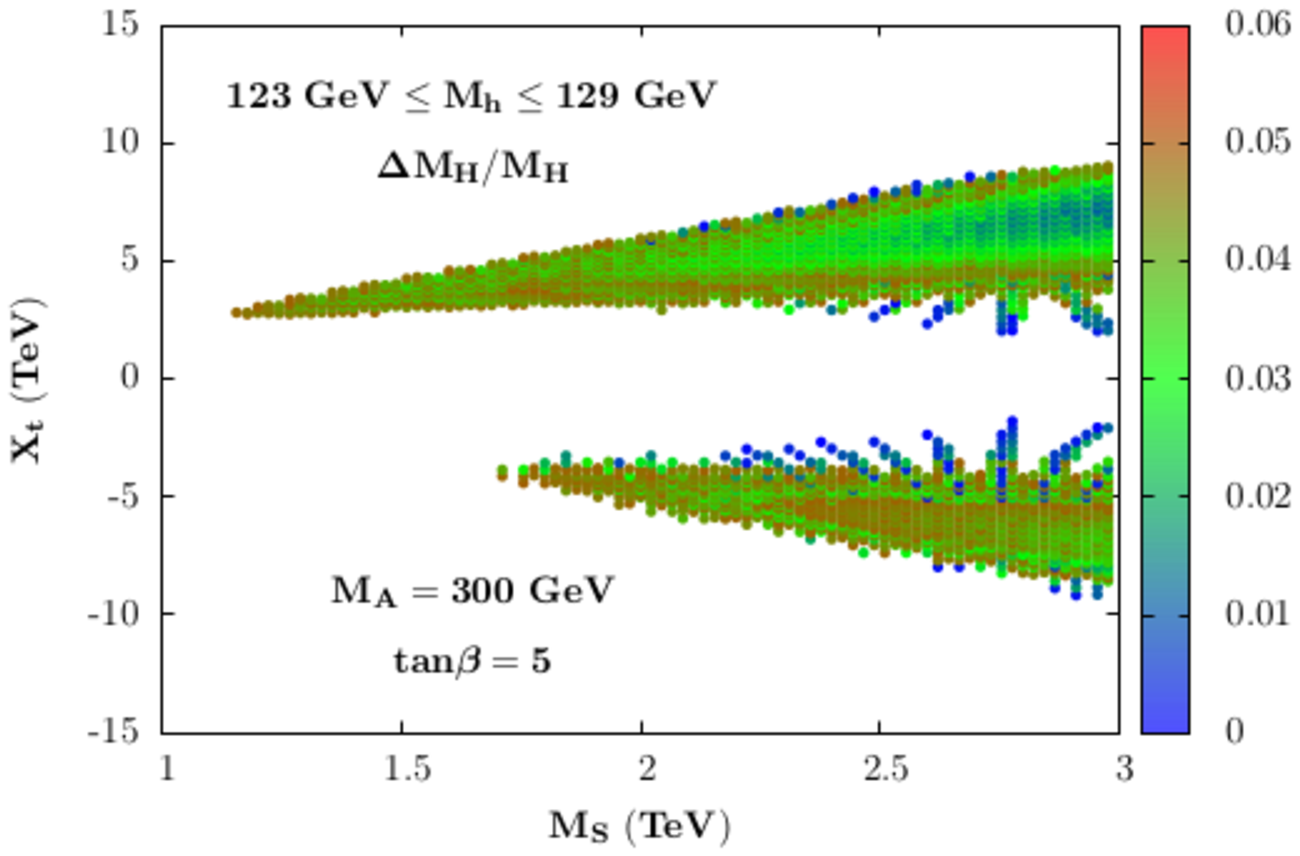}} } \\ \vspace*{-5.9cm}
\mbox{
\hspace*{-1.2cm}
\resizebox{0.36\textwidth}{!}{\includegraphics{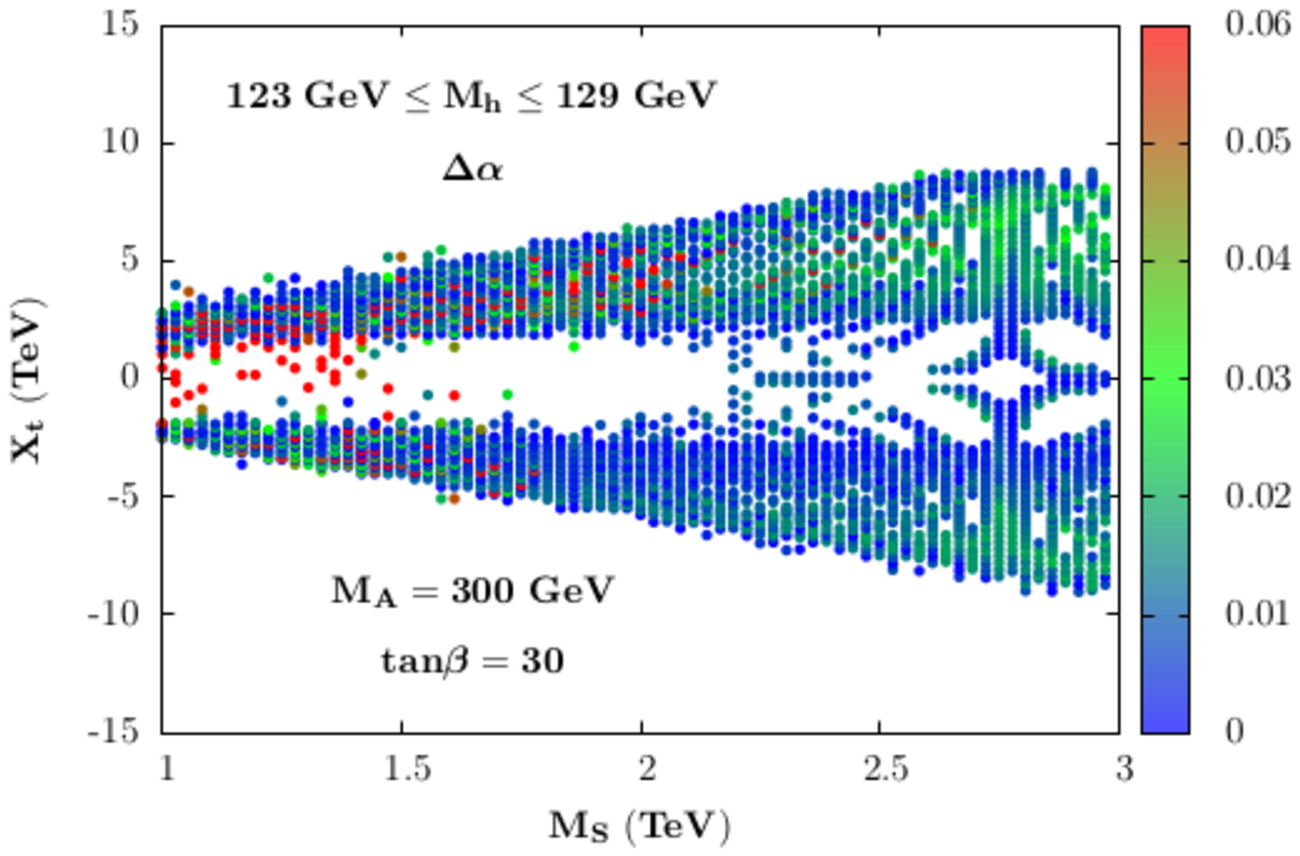}}\hspace*{-2cm}
\resizebox{0.36\textwidth}{!}{\includegraphics{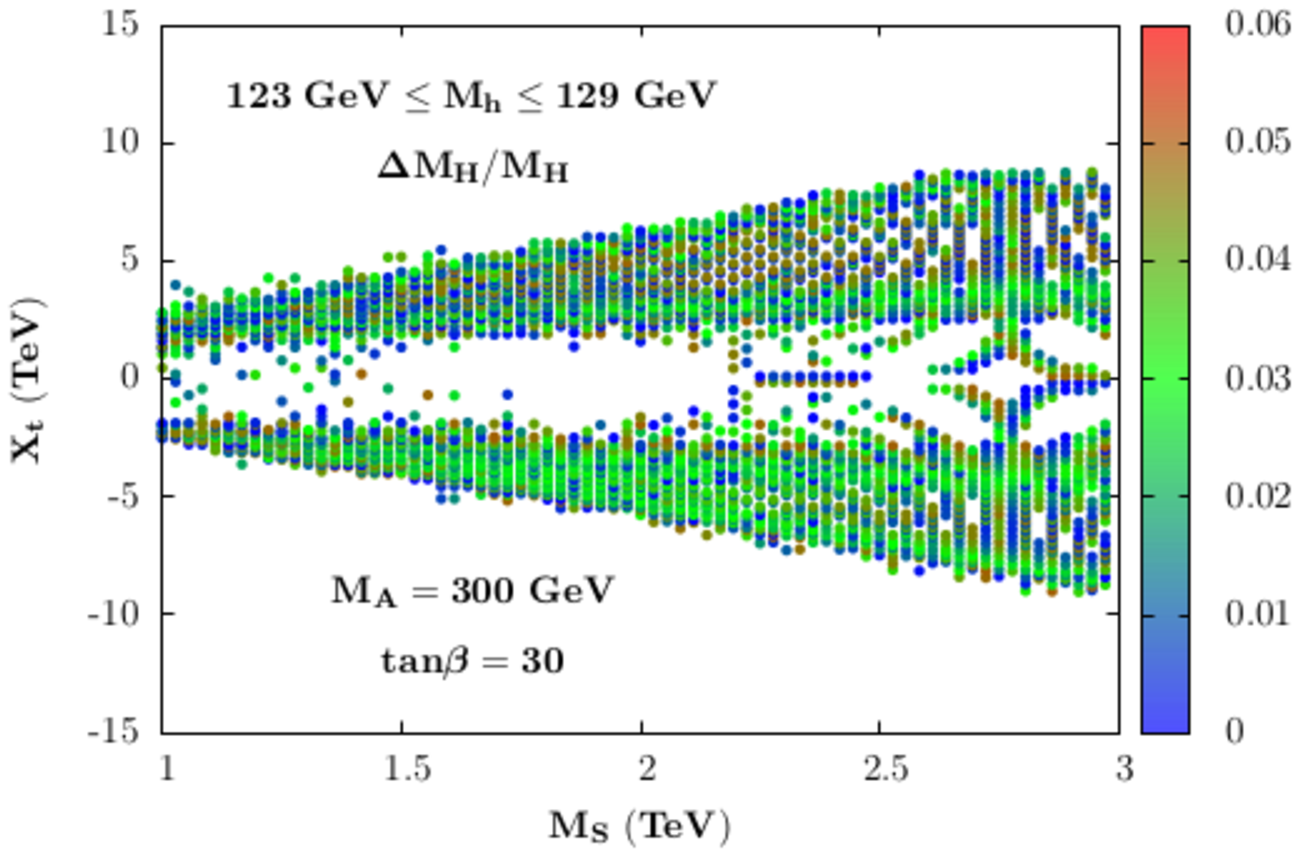}} 
} 
\vspace*{-3.4cm}
\caption{
The variation of the mass $M_H$ and the mixing angle $\alpha$
are shown as separate vertical colored scales, in
the plane $[M_{S},X_{t}]$ when the  full two loop corrections are included with
and without the subleading matrix elements $\Delta {\cal M}^{2}_{11}$ and
$\Delta {\cal M}^{2}_{12}$. $M_A\!=\!300$ GeV, $\tan\beta=5$ or
and 30 are taken and the other parameters are varied as described in the 
text~\cite{Habemus}.}
\label{approx2}
\end{center}
\vspace*{-5mm}
\end{figure}

Displayed in Fig.~\ref{approx2} are the differences between the values of the mass 
$M_H$ and the mixing angle $\alpha$ that are  obtained when the two possibilities
$ \Delta {\cal M}^{2}_{11}\!= \!\Delta {\cal M}^{2}_{12} \!= \!0$ and $\Delta {\cal
M}^{2}_{11}, \Delta {\cal M}^{2}_{12} \! \neq \! 0$ are considered. This is
shown  in the plane $[M_{S},X_{t}]$ with $X_t=A_t-\mu \cot \beta$ when all other
parameters are scanned as above. The $A$ boson mass was fixed to $M_A\!=\!300$ GeV
(a similar result was obtained for $M_A\approx 1$ TeV) 
and two representative values $\tb=5$ and $30$ are used.  The conservative approach of 
plotting only points which maximize these differences has been adopted.

In all cases, the difference between the two $M_H$ values is very small (in
fact, much smaller than the $H$ boson total decay width $\Gamma_H$), less than a few
percent, while for $\alpha$ the difference does not exceed  $\approx 0.025$ for
low values of $\tb$ but at high $\tb$ values, one can reach the level of
$\approx 0.05$ in some rare situations (large values of $\mu$,  which enhance
the $\mu \tb$ contributions). Nevertheless, at high enough $\tb$,  we are far in
the decoupling regime already for $M_A \gsim 200$ GeV and such a difference does
not significantly affect the couplings of the $h$ and $H$ bosons which,
phenomenologically, are the main  ingredients. 

Hence, even when including the full set of radiative corrections, it remains a good approximation to use eqs.~(\ref{wide}) to derive the
parameters $M_H$ and $\alpha$ in terms of the inputs $\tb, M_A$ and the measured
$M_h$ value. 

In the case of the charged Higgs boson (whose physics is described 
by $\tb, M_{H^\pm}$ and  eventually $\alpha$), the radiative
corrections to $M_{H^\pm}$ are much smaller for large enough $M_A$ and one has, at  the few
percent level in most cases (which is again smaller than the total $H^\pm$ decay width),
\beq
M_{H^\pm} \simeq \sqrt { M_A^2 + M_W^2}\, .
\eeq 

In conclusion, this approximation allows to ignore the radiative
corrections to the Higgs masses and their complicated dependence  on the MSSM
parameters and to use a simple formula to derive the other parameters of the
Higgs sector, $\alpha$,  $M_H$ as well as $M_{H^\pm}$. This considerably 
simplifies phenomenological analyses in the MSSM Higgs sector which up to now rely either on 
large scans of the parameter space (as in the previous subsections) or resort to 
benchmark scenarios in which most of the MSSM parameters are fixed (as is the 
case of Ref.~\cite{benchmarks} for instance).

\section{Implications of the Higgs production rates}

\subsection{Light Higgs decay and production at the LHC}

In many respects, the Higgs particle was born under a very lucky star as the 
mass value of $\approx 125$ GeV (although too high for a natural SUSY) allows 
to produce it at the  LHC in many redundant channels  and to detect it in a  
variety of decay modes. This allows detailed studies of the Higgs properties
as will be discussed in this section. 

\begin{figure}[hbtp]
\begin{center}
\vspace*{-.4cm}
\hspace*{-.8cm}
\resizebox{0.5\textwidth}{!}{\includegraphics{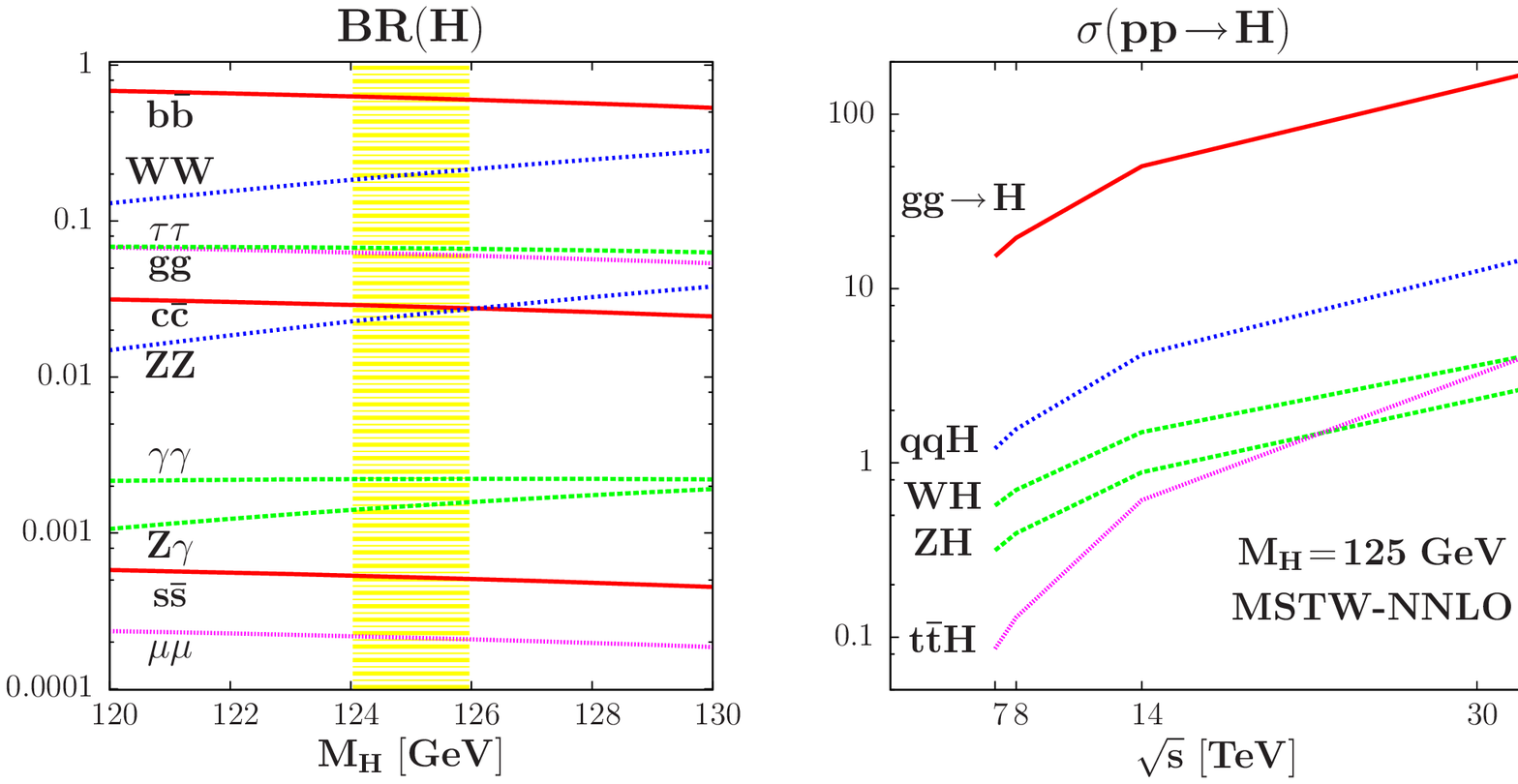}}
\vspace*{-.4cm}
\end{center}
\caption{The SM--like Higgs boson branching ratios in the mass range $120$--130
GeV (left) and its production cross sections at proton colliders as a function of  
the c.m. energy (right) \cite{ratios}.}
\label{fig:allH}
\vspace*{-.3cm}
\end{figure}

We start by summarizing the production and decay at the LHC of a light SM--like 
Higgs particle, which should correspond to the lightest MSSM $h$ boson in the 
decoupling regime.  First, for $M_h \approx 125$ GeV, the Higgs mainly decays 
into $b \bar b$ pairs but the decays into $WW^*$ and $ZZ^*$ final states, before 
allowing the gauge bosons to decay leptonically $W \! \to \! \ell \nu$ and $Z\! \to 
\! \ell \ell$ ($\ell\! =\! e,\mu$), are also significant. The $h\! \to \! \tau^+\tau^-$ 
channel (as well 
as the $gg$ and $c\bar c$ decays that are not detectable at the LHC) is also of significance, 
while the clean loop induced $h\to \gamma \gamma$ mode can be easily detected albeit 
its small rates. The very rare $h\to Z\gamma$  and even $h\to \mu^+\mu^-$ channels 
should be accessible at the LHC but only with a much larger data sample. This is illustrated 
in the left--hand side of Fig.~\ref{fig:allH} where the  decay branching fractions 
of a SM--like Higgs are displayed for the narrow mass range $M_h=120$--130 GeV 

On the other hand, many Higgs production processes have significant cross sections as 
is shown in the right--hand side of Fig.~\ref{fig:allH} where they are displayed
at a proton collider at various past, present and foreseen center of mass energies
for a 125 GeV SM--like Higgs boson; the MSTW parton densities \cite{PDF-MSTW} have been 
used. 

While the by far dominant gluon 
fusion mechanism $gg\to h$  (ggF) has extremely large rates ($\approx\! 20$ pb at 
$\sqrt s\!=\! 7$--8 TeV), the subleading channels, i.e. the vector boson fusion (VBF) 
$qq \to hqq$ and the Higgs--strahlung (HV) $q\bar q \to hV$ with $V=W,Z$ mechanisms, 
have cross sections which should allow for a study of the Higgs particle already at $\sqrt 
s\gsim 8$ TeV with the amount of integrated luminosity, $\approx 25$ fb$^{-1}$, that has 
been collected by each experiment. The Higgs--top associated process $p p\to t\bar 
t h$ (ttH) would require higher energy and luminosity. 

This pattern already allows the ATLAS and CMS experiments to observe the Higgs boson
in several channels and to measure some its couplings in a reasonably accurate way. 
The channels that have been searched are $h \! \to \! ZZ^* \! \to \! 4\ell^\pm,  
h\! \to \! WW^* \! \to \! 2\ell 2 \nu, h \! \to \! \gamma\gamma$ where the Higgs is 
mainly produced in ggF with subleading contributions from $hjj$  
in the VBF process, $h \! \to \! \tau \tau$  where the Higgs is produced in 
association with one (in ggF) and two (in VBF) jets, and finally $h\to b \bar b$ 
with the Higgs produced in the HV process. One can ignore for the moment
the additional search channels $h\! \to \! \mu \mu$ and $h \! \to \! Z\gamma$
for which the sensitivity is still too low with the  data collected so far. 

A convenient way to scrutinize the couplings of the produced $h$ boson is to consider 
their deviation from the SM expectation. One then considers for a given search 
channel the signal strength modifier $\mu$ which, with some approximation, can  be 
identified with the Higgs production cross section times  decay branching fractions 
normalized to the SM value. For the $h\! \to \! XX$ decay channel, one would have 
in the narrow width approximation, 
\beq
\mu_{XX}\vert_{\rm th} & = & \frac {\sigma( pp \to h \to XX)}{ \sigma( pp \to h
\to XX)|_{\rm SM}} \nonumber \\ &= &  \frac {\sigma( pp \to h)\times {\rm BR} (h \to XX)}{
\sigma( pp \to h)|_{\rm SM} \times {\rm BR} (h \to XX)|_{\rm SM} } .
\label{mudef}  
\eeq  
which, from the experimental side would correspond  to  
\beq 
\mu_{XX}\vert_{\rm exp} \simeq  \frac {N^{\rm ev}_{XX}}{ \epsilon \! \times\!
\sigma( pp \! \to \! h)|_{\rm SM} \! \times \! {\rm BR} (h \! \to \! XX)|_{\rm SM} \! 
\times \! {\cal L}}  \label{muEXP}  
\eeq  
where $N^{\rm ev}_{XX}$ is the measured number of events in the 
$XX$ channel, $\epsilon$ the selection efficiency and  ${\cal L}$
the  luminosity.

\begin{figure}[!h]
\vspace*{-3mm}
\begin{tabular}{ll}
\begin{minipage}{4cm}
\hspace*{-.3cm}
\resizebox{1.1\textwidth}{!}{\includegraphics{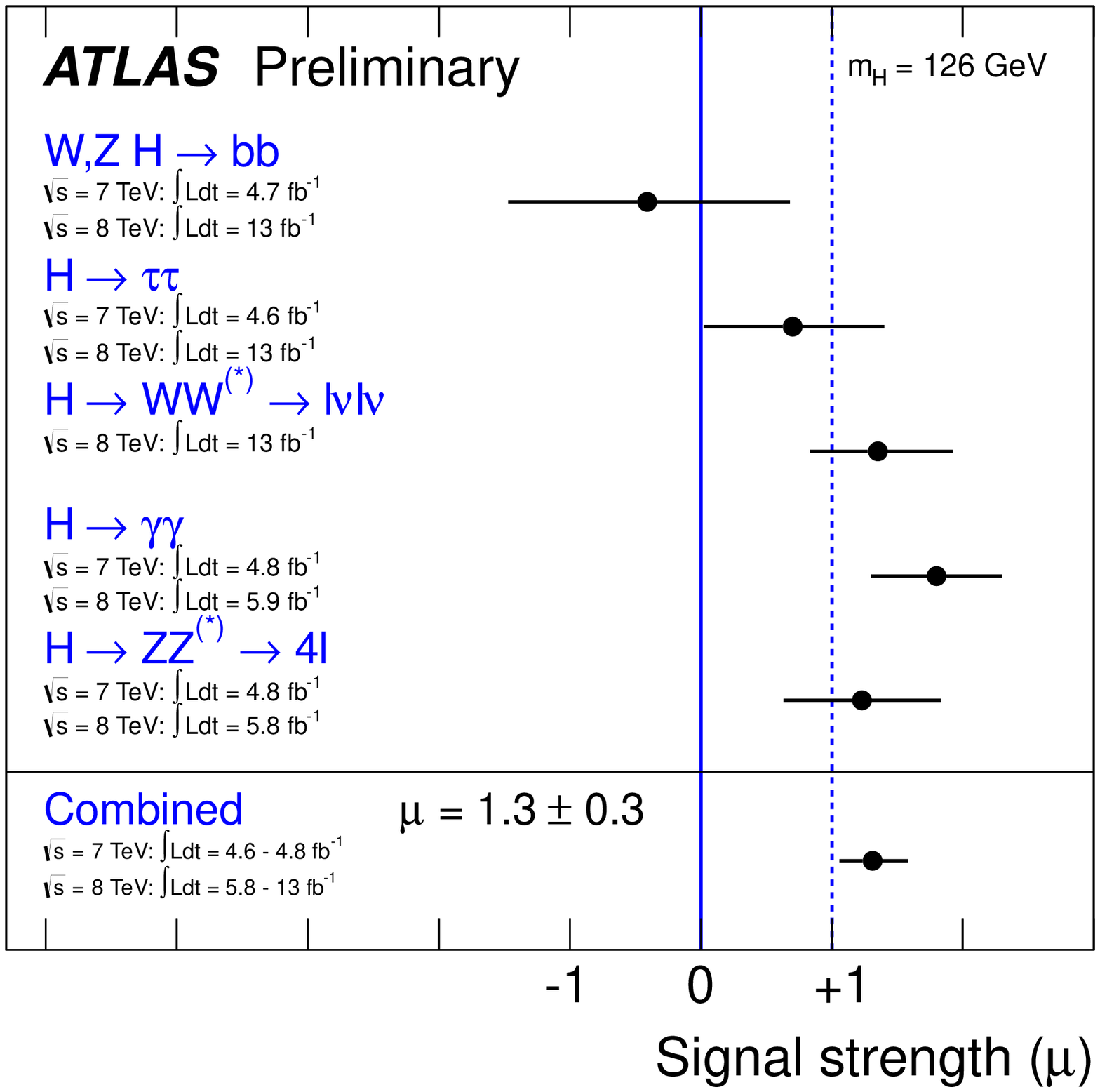}}\hspace*{-.3cm}
\end{minipage}
& \hspace*{-.4cm} 
\begin{minipage}{4cm}
\resizebox{1.1\textwidth}{!}{\includegraphics{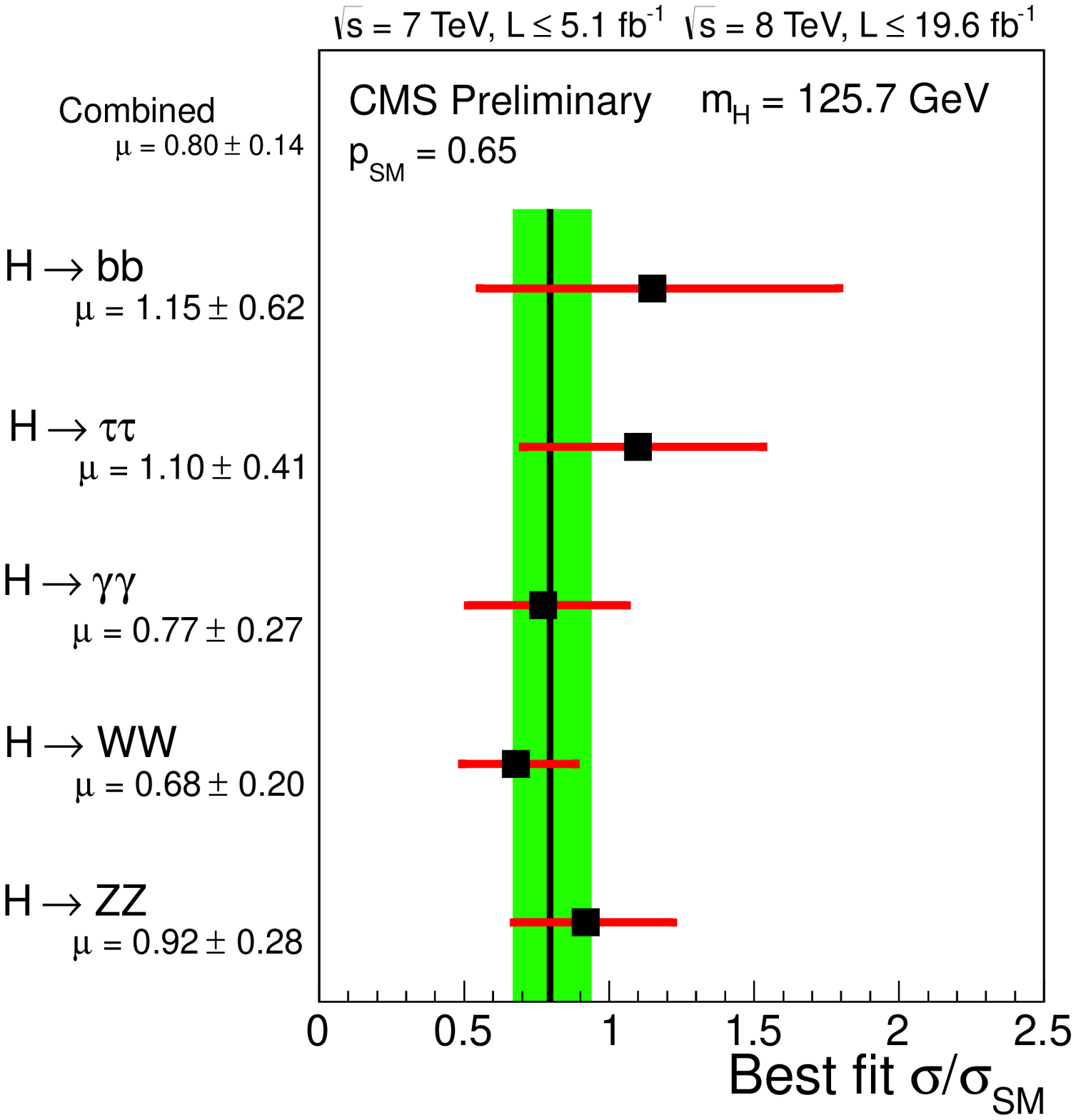}}\hspace*{-.3cm}
\end{minipage}
\end{tabular}
\vspace*{-1mm}
\caption{The signal strengths on the SM Higgs boson in the various search channels provided 
by ATLAS \cite{mu-ATLAS} and CMS \cite{mu-CMS} with the data collected so far at $\sqrt s=7$+8 TeV.
}
\label{Fig:constraints}
\end{figure}

ATLAS and CMS have provided the signal strengths for the various final states
with a luminosity of, respectively,  $\approx 5$ fb$^{-1}$ for  the 2011 run at
$\sqrt s=7$ TeV and $\approx 20$ fb$^{-1}$ for the 2012 run at $\sqrt s=8$ TeV.  
The constraints given by the two collaborations are shown in Fig.~\ref{Fig:constraints}.

When the various analyzed Higgs search channels are combined,
this leads to a global signal strength \cite{mu-ATLAS,mu-CMS}
\begin{eqnarray}  
{\rm ATLAS}: &   \mu_{\rm tot} =1.30 \pm 0.30\, \nonumber \\ 
{\rm CMS}:   &  \mu_{\rm tot}  =0.87 \pm 0.23 \,  \label{muvalues} 
\end{eqnarray}
which shows a  good agreement with the SM expectation. In fact, when the ATLAS and 
CMS values are combined, one finds a global signal strength that is very close 
to unity, implying that the observed Higgs is rather SM--like. \newpage

Hence, already  with the rather limited statistics at hand, the accuracy of the 
measurements  in eq.~(\ref{muvalues}) is reaching  the 20\% level for the 
ATLAS and CMS collaborations. This is  at the same time impressive and worrisome. 
Indeed, as mentioned earlier the main Higgs production channel is the top and 
bottom quark loop mediated gluon fusion mechanism and, at
$\sqrt s\!=\!7$ or 8 TeV, the  three other mechanisms contribute at a level below
15\% when their rates are added and  before kinematical cuts are applied.  
The majority of the signal events presently observed at the
LHC, in particular in the main search channels  $h \! \to \!  \gamma \gamma, h \! 
\to \!   ZZ^*  \to  4\ell,  h  \! \to \!  WW^*  \to 2 \ell 2\nu$
and, to a lesser extent $h \!   \to \!   \tau\tau$, thus come
from the ggF mechanism which  is known to be affected by large
theoretical uncertainties. 

As a matter of fact, although the cross section $\sigma(gg\to h)$ is known up 
next--to--next--to--leading order (NNLO) in perturbative QCD 
(and at least at NLO for the electroweak interaction) \cite{LHCXS}, there is a
significant  residual scale dependence which points to the possibility that
still higher order contributions beyond NNLO cannot be  totally excluded. In
addition, as the process is of ${\cal O}(\alpha_s^2)$ at LO and is initiated by
gluons, there are sizable uncertainties due to the  gluon parton distribution
function (PDF) and the value of the coupling $\alpha_s$. A third source of 
theoretical uncertainties, the use of an effective field theory (EFT) approach 
to calculate the radiative corrections  beyond the NLO approximation, should in 
principle also be considered \cite{BD1,BD}. In addition, large uncertainties 
arise when the $gg\! \to\! h$ cross section is broken into the jet categories
$h\! +\! 0j, h\! +\! 1j$ and $ h\! +\! 2j$ \cite{LHCXS2}. In total, 
the combined theoretical uncertainty has been estimated to be of order $\Delta^{
\rm th} \! \approx \! \pm 15\%$ by the LHC Higgs cross 
section working group  \cite{LHCXS} and it would  increase up to $\Delta^{\rm th}  
 \! \approx \pm $20\% if the EFT uncertainty is also 
included\footnote{Note that also in the VBF process,  despite the fact that the 
inclusive cross section has only a few percent combined scale and PDF uncertainty  
\cite{LHCXS}, the contamination by the  $gg\! \to\! h\!+\!2j$ channel makes the total  uncertainty in the $h\!+\!jj$ final ``VBF" sample rather large. Indeed ${\cal O}
(30\%)$ $gg\! \to \! h\!+\!2j$ events will remain even  after the specific cuts 
that select the VBF configuration have been applied, and the rate is affected by 
a much larger uncertainty than  the inclusive $gg\!\to\! h$ process, up to 50\%  when
one adds the scale and PDF uncertainties \cite{LHCXS2}.} \cite{BD}. 

Hence, the theoretical uncertainty is  already at the level of the accuracy of 
the cross section measured by the ATLAS and CMS collaborations, eq.~(\ref{muvalues}).
Another drawback of the analyses
 is that they involve strong theoretical assumptions on the
total Higgs width since some contributing decay channels  not accessible at the
LHC are assumed  to be SM--like and possible invisible Higgs decays in scenarios
beyond the SM are supposed not to occur.

In Ref.~\cite{ratios}, following earlier work \cite{Dieter}, it has been suggested to 
consider the decay ratios $D_{XX}$  defined as 
\beq
D_{XX}^{\rm p} &= &\frac {\sigma^{\rm p} ( pp \to h \to XX)}{ \sigma^{\rm p}
( pp \to h \to VV)} \\ &=&  \frac {\sigma^{\rm p} ( pp \to h)\times {\rm BR} 
(h \to XX)}{ \sigma^{\rm p}( pp \to h) \times {\rm BR} (h \to VV)} \\ &= &
\frac{\Gamma( h \to XX)}{ \Gamma ( h \to VV)}  \propto \frac{c_{X}^2} {c_{V}^2} 
\label{ratio} 
\eeq 
for a specific production process ${\rm p= ggF, VBF, VH}$ or all  (for inclusive
production) and  for a given decay channel $h\to XX$ when the reference channel
$h\to VV$ is used. In these ratios,  the cross sections $\sigma^p (pp \! \to \! h)$ 
and hence, their significant theoretical  uncertainties will cancel out, leaving 
out only the ratio of decay branching fractions  and hence of partial decay widths.
These can be obtained with the program {\tt HDECAY} \cite{HDECAY} which 
includes all higher order effects  and are affected 
by much smaller uncertainties. Thus, the total decay
width which includes contributions from channels not under control such as
possible invisible Higgs decays, do  not appear in the ratios $D_{XX}^{\rm
p}$.  Some common experimental systematical uncertainties such as the one from
the luminosity measurement and the small uncertainties in the Higgs decay 
branching ratios also cancel out. We are thus, in principle,  left with only
with the statistical uncertainty and some (non common) systematical errors.
The ratios $D_{XX}$ involve, up to  kinematical factors and known radiative
corrections, only the ratios  $\vert
c_X \vert^2/$ $\vert c_V\vert^2$ of the Higgs reduced  couplings to
the particles  $X$ and $V$ compared to the SM expectation, $c_X \equiv
g_{hXX}/g_{hXX}^{\rm SM}$. 

For the time being, three independent ratios can be considered:  $D_{\gamma \gamma},
D_{\tau \tau}$ and $D_{bb}$. $D_{\gamma \gamma}$ is the ratio of the inclusive ATLAS and CMS 
di-photon and $ZZ$ channels that are largely dominated by the ggF mechanism; $D_{\tau \tau}$
is the signal strength ratio in the $\tau\tau$ and $WW$  searches where one selects Higgs production in ggF with an associated jet or in the VBF production mechanism;  
$D_{bb}$ is the ratio of the  $h\to b\bar b$ and $h\to WW$ decays in
$hV$ production for which the sensitivities are currently too low.

In order to test the compatibility of the couplings of the $M_h=125$ GeV Higgs state 
with the SM expectation, one can perform a fit based on the $\chi_R^2$  function 
\begin{eqnarray}
\chi^2_R  &=&  \frac{[D_{\gamma \gamma}^{ggF}-\frac{\mu_{\gamma\gamma}}{\mu_{ZZ}}
\vert^{ggF}_{\rm exp}]^2} {\big [ \delta(\frac{\mu_{\gamma\gamma}}{\mu_{ZZ}})_{ggF}\big ]^2} 
+ \frac{[D_{bb}^{VH}-\frac{\mu_{bb}}{\mu_{WW}}\vert^{Vh}_{\rm exp}]^2} {\big [ 
\delta(\frac{\mu_{bb}}{\mu_{WW}})_{Vh}\big ]^2} \nonumber\\
&+& 
\frac{[D_{\tau \tau}^{ggF}-\frac{\mu_{\tau\tau}}{\mu_{WW}}\vert^{ggF}_{\rm exp}]^2}
{\big [ \delta(\frac{\mu_{\tau\tau}}{\mu_{WW}})_{ggF}\big ]^2} \!+ \! % \nonumber \\ &+& 
\frac{[D_{\tau \tau}^{\rm VBF}-\frac{\mu_{\tau\tau}}{\mu_{WW}}\vert^{\rm VBF}_{\rm exp}]^2}
{\big [ \delta(\frac{\mu_{\tau\tau}}{\mu_{WW}})_{\rm VBF}\big ]^2}  ~~~ 
%\nonumber  
\label{eq:Chi2R}
\end{eqnarray} 

The errors  $\delta(\mu_{XX}/{\mu_{VV}})$ are computed assuming no correlations between 
the different final state searches.  The uncertainties on the ratios are derived from the
individual errors that are dominated by the experimental uncertainties as one expects 
that the theoretical uncertainties largely cancel out in  the ratios
$D_{\gamma\gamma}$, $D_{bb}$ and $D_{\tau \tau}$.

For the signal strengths above, the theoretical uncertainties have to be treated 
as a bias (and not as if they were associated with a statistical distribution) and 
the fit has to be performed for the two extremal values of the  signal strengths: 
$\mu_{i} \vert_{ \rm exp} \pm \delta \mu_i/\mu_i \vert_{\rm th}$ with  the theoretical uncertainty $\delta\mu_i/ \mu_i \vert_{\rm th}$  conservatively assumed to be $\pm 20\%$ 
for both the gluon and vector boson fusion mechanisms (because of the contamination
due to $gg\to h+2j$ in the latter case) and $\approx 5\%$ for $hV$ associated production.

\subsection{Fit of the Higgs couplings and their ratios}

A large number of analyses of the Higgs couplings from the LHC data have 
been performed in the SM and its extensions and a partial list is given
in  Refs.~\cite{Fits,paper3,fit}. 

In the MSSM, the couplings of the CP--even Higgs particles $h$ and $H$
to gauge bosons and fermions, compared to the SM Higgs couplings, are 
changed by factors that involve the sine and the cosine of the mixing
angles $\beta$ and $\alpha$. Outside the decoupling regime where they reach
unity, the reduced couplings (i.e. normalized to their SM values) of the
lighter $h$ state to third generation $t,b,\tau$  fermions and gauge bosons
$V\! = \!W/Z$ are for instance given by 
\begin{eqnarray}  
c_V^0  \! = \!  \sin(\beta \!-\! \alpha)  ,   c_t^0 \!  = \!  
\cos  \alpha/ \sin\beta ,   c_b^0 \!  = \!  -\!
\sin  \alpha/ \cos\beta  \label{Eq:MSSMlaws} 
\end{eqnarray}
They thus depend not only on the two  inputs $[\tan\beta, M_A]$ as it occurs 
at tree--level but, a priori, on the entire MSSM spectrum as a result of the 
radiative corrections, in the same way as the Higgs masses.  In principle, as 
discussed earlier, knowing $\tan\beta$ and $M_A$ and fixing $M_h$ to its  measured 
value, the couplings can be determined in general. However, this is true 
when only the radiative corrections to the Higgs masses are included. 
Outside the regime in which the pseudoscalar $A$ boson and the
supersymmetric particles are very heavy, there are also direct radiative
corrections to the Higgs couplings  not contained in the mass matrix of 
eq.~(\ref{mass-matrix}) and which  can  alter  this simple picture.

First, in the case of $b$--quarks, additional one--loop vertex corrections
modify the tree--level $h b \bar b$ coupling: they grow as $ m_b \mu \tan\beta$
and can be very large at high  $\tb$. The dominant component comes from the
SUSY--QCD corrections  with sbottom--gluino loops that can be approximated by
$\Delta_b \simeq  2\alpha_s/(3\pi) \times \mu m_{\tilde{g}} \tb /{\rm max}
(m_{\tilde{g}}^2,  m_{\tilde{b}_1}^2,m_{\tilde{b}_2}^2) \label{deltab}$
\cite{deltab}. Outside the decoupling regime the $c_b$ coupling reads
\beq
c_b \approx c_b^0 \times [1- \Delta_b/(1+\Delta_b) \times (1+ \cot\alpha \cot\beta)]  
\label{cb}\!
\eeq
with $\tan\alpha \to -1/\tb$ for $M_A\! \gg\! M_Z$. A large $\Delta_b$  would 
significantly alter the dominant  $h\! \to b\bar b$ partial width and affect 
the  branching fractions of  all other decay  modes. 

In addition, the $ht\bar t$ coupling is derived indirectly from the $gg\! \to\! h$ 
production cross section and the $h\! \to\! \gamma \gamma$ decay branching ratio,
two processes that are generated by triangular loops. In the MSSM, these loops 
involve not only the top quark (and the $W$ boson in the decay $h\to \gamma \gamma$) 
but also contributions from supersymmetric particles, if not
too heavy. In the case of $gg\to h$ production, only the contributions of stops
is generally important. Including the later and working in the limit $M_h\!
\ll \! m_t, m_{\tilde{t}_{1,2} }$,  the coupling $c_t$ from the 
ggF process\footnote{In the case of the production process $gg/q\bar q \to h t\bar t$, 
it is still $c_t^0$ which should describe the $ht\bar t$ coupling, but the constraints on 
the $h$ properties from this process are presently very weak.}
 is approximated by  \cite{Stop}
\beq 
c_t \! \approx \! c_t^0 \bigg[ 1 \!+\!  \frac {m_t^2}{ 4 m_{\tilde t_1}^2 
m_{\tilde t_2}^2 } ( m_{\tilde t_1}^2 \! + \! m_{\tilde t_2}^2  \! - \! 
X_t^2) \bigg] 
\label{ct}
\eeq
which shows that indeed, $\tilde t$  contributions can be very large for
light stops and for large stop mixing.  In the  $h \to \gamma\gamma$ decay 
rate, because the $t, \tilde t$ electric charges  are the same, the $ht\bar 
t$ coupling is shifted by the same amount. If one ignores the 
usually small contributions of the other sparticles (to be discussed in
the next subsection), the $ht\bar t$ vertex 
can be simply parametrised by the effective coupling of eq.~(\ref{ct})

We note that the $h$ couplings to $\tau$ leptons and
$c$ quarks do not receive the direct corrections of eqs.~(\ref{cb}) 
and (\ref{ct}) and one should still have $c_c=c_t^0$ and 
$c_\tau= c_b^0$. However, using $c_{t,b}$ or $c_{t,b}^0$ in this case  has
almost no impact in practice as these couplings appear only in  the branching
ratios for the decays $h \to c\bar c$ and $\tau^+ \tau^-$ which are small 
and the direct  corrections should not be too large. One can thus, in a first 
approximation,  assume that $c_c=c_t$ and $c_\tau=c_b$. Another caveat is due 
to the invisible Higgs decays which are assumed to
be absent and which will be discussed later. 

Hence, because of the direct corrections,  the Higgs couplings cannot be described 
only by $\beta$ and $\alpha$ as in eq.~(\ref{Eq:MSSMlaws}). To characterize 
the Higgs particle at the LHC, it was advocated that at least three independent 
$h$ couplings should be considered, namely $c_t$, $c_b$ and $c_V=c_V^0$ \cite{Habemus}. 
One can thus define the following effective Lagrangian,     
\begin{eqnarray} 
{\cal L}_h  &\! =\! &   c_V  g_{hWW}  h  W_{\mu}^+ W^{- \mu} +  c_V  g_{hZZ}  
h Z_{\mu}^0 Z^{0 \mu} \label{Eq:LagEff}\\ &\!-\! &   
  c_t y_t  h  \bar t_L  t_R \! -\!  c_t  y_c h  \bar c_L  c_R \! -\! 
  c_b  y_b   h  \bar b_L b_R  \! - \! c_b y_\tau h  \bar \tau_L \tau_R 
 \! + \! {\rm h.c.} \nonumber 
\end{eqnarray}
where $y_{t,c,b,\tau}=m_{t,c,b,\tau}/v$ are the Yukawa couplings of the heavy
SM fermions, $g_{hWW}\! = \! 2M^2_W/v$ and $g_{hZZ}\! =\! M^2_Z/v$ the $hWW$
and $HZZ$ couplings and $v$ the SM Higgs vev.

\begin{figure}[!t] 
\vspace*{-5cm}
\begin{center}
\resizebox{0.6\textwidth}{!}{\includegraphics{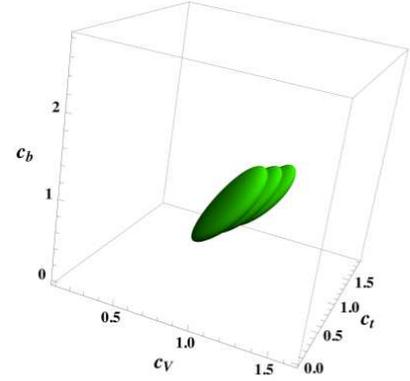}}
\end{center}
\vspace*{-5cm}
\caption{{The best-fit region at $68\%{\rm CL}$ for the Higgs signal 
strengths in the $[c_t,c_b,c_V]$ space \cite{Habemus}. The three overlapping
regions are for the central and extreme  choices of 
the theoretical prediction for the Higgs rates including uncertainties.}}
\label{fig:3D}
\vspace*{-2mm}
\end{figure}

In Ref.~\cite{Habemus}, a three--dimensional fit of the $h$ couplings was performed 
in the space $[c_t, c_b, c_V]$, assuming  $c_c\!=\!c_t$ and $c_\tau\!=\!c_b$  as discussed 
above and of course the custodial symmetry relation $c_V\!=\!c_W\!= \!c_Z$ which 
holds in supersymmetric models.  The results of this fit are presented in 
Fig.~\ref{fig:3D} for $c_t,c_b,c_V \! \geq \! 0$. The best-fit value for the 
couplings, with  the $\sqrt s=7+$8 TeV ATLAS and CMS data turns out to be   
$c_t=0.89, ~ c_b=1.01$ and $c_V=1.02$.

%\newpage

\begin{figure}[!ht]
\vspace*{-3.5cm}
\begin{center}
\mbox{\hspace*{-5mm}
\resizebox{0.5\textwidth}{!}{\includegraphics{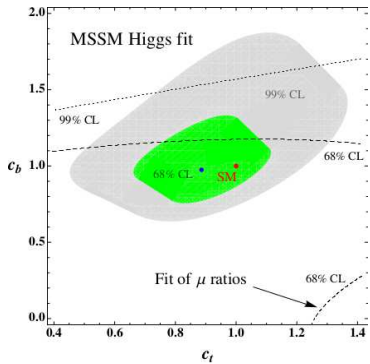}}
}
\vspace*{-4.5cm}
\caption{{Best-fit regions at $68\%$ and  $99\%{\rm
CL}$ for the Higgs signal strengths and their ratios in the  plane $[c_t,c_b]$.  
The best-fit point is indicated in blue. From Ref.~\cite{Habemus}.}}
\label{fig:2D}
\end{center}
\vspace*{-5mm}
\end{figure}

In scenarios where the direct corrections in eqs.~(\ref{cb})--(\ref{ct}) 
are not quantitatively significant (i.e. considering either not too large values 
of $\mu \tan\beta$ or high sfermion masses), one can use the MSSM relations of 
eq.~(\ref{Eq:MSSMlaws}) to reduce the number of effective parameters down to two.
This allows to perform two-parameter fits in the planes $[c_V,c_t]$, $[c_V,c_b]$ 
and $[c_t,c_b]$. As an example, the fit of the signal strengths and their ratios 
in the $[c_t,c_b]$ plane is 
displayed in Fig.~\ref{fig:2D}. In this two--dimensional case,
the best-fit point is located at $c_t=0.88$ and $c_b=0.97$, while $c_V \simeq 1$. 
Note that although  for the best--fit
point one has $c_b \lsim 1$, actually $c_b \gsim 1$ in most of the  $1\sigma$ region.

\begin{figure}[!h]
\begin{center}
\vspace*{-4.9cm}
\mbox{\hspace*{-1.5cm}
\resizebox{0.6\textwidth}{!}{\includegraphics{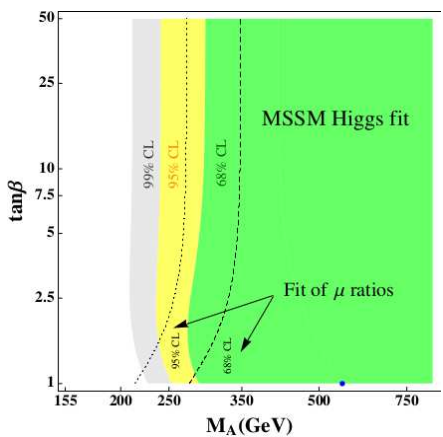}}
}
\vspace*{-5.3cm}
\caption{Best-fit regions for the signal strengths and their ratios in the 
plane $[\tan \beta, M_A]$; the best point is in  blue \cite{Habemus}.}
\label{fig:2DMA}
\vspace*{-17mm}
\end{center}
\end{figure}

Using the formulae eq.~(\ref{wide}) for the angle $ \alpha$ and using the 
input $M_h \! \approx 125$ GeV, one can make a fit in the plane $[\tan
\beta,  M_A]$. This is shown in Fig.~\ref{fig:2DMA} where
the 68\%, 95\% and  99\%CL contours from the signal strengths and their ratios
are displayed when  the theory uncertainty is taken as a bias. 
The best-fit point when the
latter uncertainty is set to zero, is obtained  for the  values
$\tan\beta\!=\! 1$  and $M_A \! = \! 557 \; {\rm GeV}$,   which implies
for the other parameters using $M_h=125$~GeV~:
$M_H= 580$~GeV, $M_{H^\pm}= 563$~GeV and $ \alpha=-0.837~{\rm rad}$ which leads to 
$\cos (\beta-\alpha) \simeq -0.05$. Such a point with $\tb\approx 1$ implies
an extremely large  value of the SUSY scale, $M_S = {\cal O}(100)$ TeV, for
$M_h\approx 125$ GeV.  One should note, however, that the $\chi^2$ value is relatively
stable all over the $1\sigma$ region.
Hence, larger values of $\tb$ (and lower values of $M_A$) could also
be accommodated reasonably well by the fit, allowing thus for not too
large $M_S$ values. In all, cases one has $M_A \gsim 200$ GeV though.

\newpage

\subsection{An excess in the $\gamma\gamma$ channel?}

In the early LHC data, a significant excess in the $h\to \gamma \gamma$ 
detection channel was observed, raising the hope that it could be the first
signal for physics beyond the SM.  This excess has unfortunately faded away 
with more statistics and with the full 25$\;$fb$^{-1}$ data collected at 
$\sqrt s\!=\!7\!+\!8$ TeV, there is now only a $\approx 2 \sigma$ excess in
ATLAS which measures $\mu_{\gamma \gamma}=1.6\pm 0.3$,  while the signal
strength measured by the CMS collaboration is $\mu_{\gamma \gamma}=
0.9\pm 0.3$ which is SM--like. Nevertheless, it would be interesting 
to briefly discuss this excess as, besides the fact that it has triggered 
a vast literature,  the $h\! \to \! \gamma \gamma$ channel is the one where 
new physics and SUSY in particular, is most likely to manifest itself.  

\begin{figure}[hbtp]
\begin{center}
\vspace*{-.3cm}
\resizebox{0.25\textwidth}{!}{\includegraphics{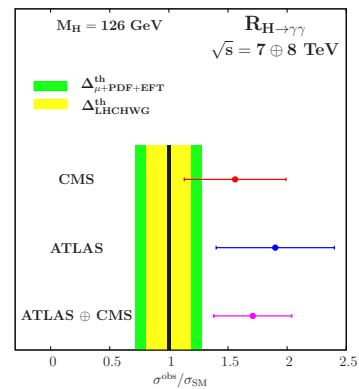}}
\vspace*{-.2cm}
\end{center}
\caption{The value of $\mu_{\gamma \gamma}$ given by the ATLAS and CMS 
collaborations with the $\approx 10$ fb$^{-1}$ data collected at $\sqrt 
s\!=7$ and 8 TeV, as well as their combination, compared to two estimates 
of the theoretical uncertainty  bands; from Ref.~\cite{excess}.}
\label{fig:TH-gamma}
\vspace*{-.3cm}
\end{figure}

First, it has been realized early that this excess, if not due to a statistical 
fluctuation, can be easily explained or reduced in the context of the SM by invoking
the large theoretical uncertainties that affect the production times decay
rate in the dominant channel, $gg\to h\to \gamma\gamma$. This is shown in 
Fig.~\ref{fig:TH-gamma}, where the ATLAS and CMS ratios $R_{\gamma\gamma}\equiv
\mu_{\gamma\gamma}$ and their combination, obtained with the $\approx 10$ fb$^{-1}$ 
data collected at $\sqrt s\!=7+$8 TeV, is compared to the theory uncertainty
bands obtained by the LHC Higgs group \cite{LHCXS} and in Ref.~\cite{BD}. 
It is clear that including the theory uncertainty as a bias helps to reduce
the discrepancy between measurement and expectation and e.g. 
 the excess reduces to $1.3\sigma$ from the original 
$\gsim 2\sigma$ value.

Ignoring this option, let us summarise the various possibilities that could
explain this excess in the context of the MSSM. Deviations of $\mu_{\gamma\gamma}$ 
from the SM value may be due to modifications of either the production cross section
or the decay branching fraction or to both. The $h$ decay branching fractions may be 
modified by a change of the $h$ total decay width. Since the dominant decay mode 
is $h \to b \bar b$, a change of the effective $hb\bar b$ coupling by the direct vertex
corrections of eq.~(\ref{cb}) outside the decoupling regime, can change all other 
Higgs rates including $h\to\gamma \gamma$. The total width can also be modified by 
additional decay channels to 
SUSY particles and the only ones that are allowed by experimental constraints
are invisible decays into the LSP that will be discussed later.

Nevertheless, these two possibilities would not only affect the $h\to\gamma \gamma$ rate
but also those of other channels such as $h\! \to \! ZZ$ where no excess has been 
observed. It is thus more appropriate to look at deviation in the $h\to \gamma \gamma$ loop
induced decay only. In the MSSM, the $h\to \gamma \gamma$ process receives 
contributions from scalar top and bottom quarks, staus and charginos
as briefly is summarised below.

\begin{figure}[hbtp]
\begin{center}
\vspace*{-1.97cm}
\hspace*{-1.cm}
\resizebox{.55\textwidth}{!}{\includegraphics{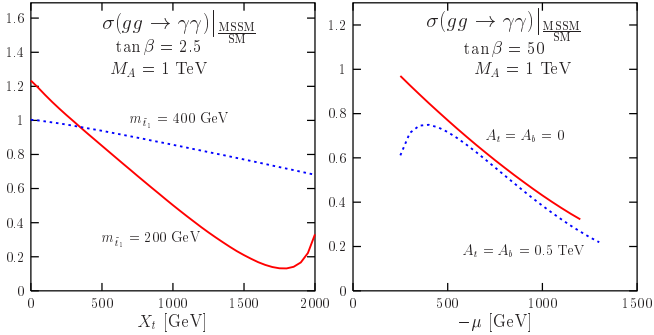}}\\[-7.9cm]
%\resizebox{.35\textwidth}{!}{\includegraphics{Pfig6a.ps}}
\end{center}
\vspace*{-83mm}
\caption{%Top: 
The deviation of $\mu_{\gamma \gamma}$ from its SM expectation 
from stop (left) and sbottom (right) contributions in various scenarios
to the $ \sigma(gg\to h)\! \times \! {\rm BR}( h \! \to \! \gamma \gamma)$ rate; 
from Ref.~\cite{Stop}. 
%Bottom: contours for the rate $\mu_{\gamma \gamma} \equiv 
%R_{\gamma \gamma}$ in the $[X_t/m_{\tilde t}, m_{\tilde t}$] plane where the stops 
%provide the `correct" $M_h \approx 125$ GeV when calculated with Suspect or FeynHiggs;
%Ref.~\cite{stop} 
}
\label{gg-stop}
\vspace*{-.3cm}
\end{figure}

-- {Light stops}: as already discussed,  the $M_h\!=\!125$ GeV constraint requires 
large $M_S \! = \! \sqrt{ m_{\tilde t_1} m_{\tilde t_2}}$ and/or $X_t$ values. 
If $m_{\tilde t_1} \!  \lsim \! 500$ GeV, one should have maximal mixing 
$X_t \! \approx \! \sqrt 6 M_S$ and, in this case, the $h \tilde t_1 \tilde t_1$ coupling
of eq.~(\ref{ct}) is large and leads to a sizeable change of the $gg\! \to \! h \! \to
\gamma \gamma$ rate; cf. Fig.~\ref{gg-stop} (left).  However, an enhancement of the 
$h\to \gamma\gamma$ rate is over-compensated by a suppression of  $\sigma(gg\to h)$
that seems not to occur. $\mu_{\gamma \gamma}$ is enhanced only in the no-mixing case,
$X_t\! \approx \! 0$, which requires very heavy stops which decouple from the amplitude
\cite{Stop2,Stop}.   

-- {Light sbottoms}:  a light $\tilde b_R$ state does not conflict with 
the $M_h$ value as its corrections to the mass are small. For $m_{\tilde b_1} \lsim$
500~GeV, it contributes to the $hgg$ loop but it  reduces the $gg\to h$
production rate; Fig.~\ref{gg-stop} (right). In turn, it has little impact on the 
$h\to \gamma\gamma$ rate because of the largely dominating $W$ loop and the small $\tilde b_1$
electric charge. For $m_{\tilde b_1}\gsim 1$ TeV, as indicated by direct LHC searches,
$\mu_{\gamma\gamma}$ is unaffected by sbottoms loops \cite{Stop}.

-- {Light staus}: they lead to the largest contributions and have received most of
the attention in the literature; see  e.g. Ref.~\cite{Htau}.  For low $m_{\tilde 
\tau_{L/R}}$ values,  a few 100~GeV, and large mixing $X_\tau= A_\tau - \mu
\tan \beta$, with $\tan \beta \approx 60$ and $|\mu|$=0.5--1~TeV, the lighter 
stau state has a mass close to the LEP2 bound, $m_{\tilde \tau_1} \approx$ 100~GeV 
and its coupling to the $h$ boson, $g_{h\tilde \tau \tilde \tau} \propto m_\tau X_\tau$, 
is huge. The $\tilde \tau_1$ contribution can hence significantly increase 
BR($h\to \gamma \gamma$), up to 50\% \cite{Htau}, but this occurs only 
for extreme choices of the parameters.

-- {Light charginos}: the $h \chi_i^+ \chi_i^-$ couplings are in general small
and are maximal when the $\chi_i^\pm$ states are almost equal higgsino--wino
mixtures.  For a mass above 100~GeV and maximal couplings to 
the $h$ boson, the $\chi^\pm_1$ contributions to the $h\to \gamma \gamma$ rate do not exceed the 
10--15\% level (with a sign being the same as the sign of $\mu$) \cite{inos}. 

Of course, different contributions can sum up resulting in more sizeable shifts. However, 
a 50\% deviation of the rate is unlikely and occurs only in extreme situations.

%\begin{figure}[!h]
%\vspace*{-5mm}
%\begin{tabular}{ll}
%\begin{minipage}{5cm}
%\hspace*{-.3cm}
%\resizebox{.8\textwidth}{!}{\includegraphics{Pfig6b.ps}}\vspace*{-2.2cm}
%\end{minipage}
%& \hspace*{-1.3cm} 
%%
%\begin{minipage}{5cm}
%\resizebox{.9\textwidth}{!}{\includegraphics{Pfig6c.ps}}\hspace*{-.3cm}
%\end{minipage}
%%
%\end{tabular}
%\vspace*{6mm}
%\caption{Left: contours for $\mu_{\gamma \gamma}$ in 
%the $[m_{\tilde \tau_L}, \mu]$] plane where light stau's enhance the $h\to \gamma \gamma$
%decay at high $\tb$ values, $\tb=60$; from Ref.~\cite{stop}. Right: the rate in the
%$[M_2, \mu]$] plane where light charginos enhance $\mu_{\gamma \gamma}$ by less
%than 5\% (blue) and 10\% (red) for $\tb=3$; from  Ref.~\cite{inos}.}
%\label{Fig:gamma}
%\vspace*{-5mm}
%\end{figure}

%The impact of stau and chargino loops is shown in Fig.~\ref{Fig:gamma}.
%Hence, if there was a $\gamma\gamma$ excess,  the enhancement in the 
%$h \to \gamma \gamma$ rate in the MSSM would have been modest in most cases and
% significant  only in extreme scenarios. 

\subsection{Invisible Higgs decays?}

\begin{figure}[h!]
\vspace*{-3.2cm}
\begin{center}
\resizebox{.37\textwidth}{!}{\includegraphics{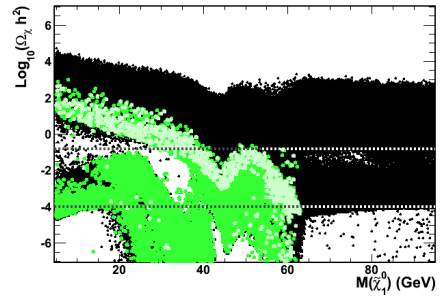}}
\end{center}
\vspace*{-3.1cm}
\caption{The neutralino relic density $\log_{10} (\Omega_{\chi} h^2)$ as a function of 
$m_{\chi_1^0}$ compatible with BR$(h\to \chi_1^0 \chi_1^0) \geq 15\%$ (green) 
and the LHC Higgs data at 90\%CL (light green). The horizontal lines show the WMAP  
constraint on $\Omega_{\chi} h^2$. From Ref.~\cite{paper3}.} 
\label{fig:DM}
\vspace*{-5mm}
\end{figure}

Invisible decays can also affect the properties of the observed $h$ particle. 
In the MSSM, because of the LEP2 constraints, the only possible invisible channel 
for the $h$ boson is  into pairs of the LSP neutralinos, $h \! \to\! \chi^0_1  
\chi^0_1$. BR$_{\rm inv}$ can be important for $m_{\tilde \chi^0_1} \! <\!$ 60GeV 
and  for not too large $M_1$ and $|\mu|$ values which make the LSP a 
higgsino--gaugino mixture with significant couplings to the $h$ state. Such an LSP 
would have the relic density $\Omega h^2$ required by the WMAP results \cite{DM-review} 
since it 
will annihilate efficiently through the s--channel exchange of the $h$ boson. 
However, BR$_{\rm inv}$ should be small in this case. This is exemplified 
in Figure~\ref{fig:DM} where $\log_{10} (\Omega_{\chi} h^2)$ is shown as a function of 
$m_{\chi_1^0}$ for the pMSSM points that satisfy the LHC Higgs constraints and 
BR$(h\! \to \! \chi_1^0 \chi_1^0)\! \geq \!15\%$. Only a small area 
in the region $30 \! \lsim \! m_{\chi_1^0} \! \lsim \! 60$ GeV fulfils these conditions. 

\begin{figure}[!h]
\vspace*{-3.8cm}
\mbox{\hspace*{-.1cm}\resizebox{.45\textwidth}{!}{\includegraphics{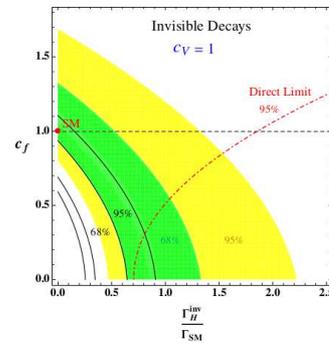}}
}
\vspace*{-3.8cm}
\caption{
$1\sigma$ and $2\sigma$ domains from $\mu_{ZZ}$ for $c_V\!=\!1$ in the 
plane $[c_f,  \Gamma_H^{\rm inv}/\Gamma_H^{\rm tot}]$ \cite{fit}. The dependence
on the theory uncertainties are shown by the black curves that 
indicate the other possible extreme domains. The direct upper limit on 
$\Gamma_H^{\rm inv}$ from direct searches at LHC for 
$c_V\!=c_f\!=\!1$~\cite{inv-ATLAS} is also shown.
\label{fig:Invfit}}
%\end{center}
\vspace*{-1mm}
\end{figure}

The invisible Higgs decay width can be constrained indirectly by a fit of the Higgs 
couplings and in particular with the signal strength $\mu_{ZZ}$ which is the most 
accurate one and has the least theoretical ambiguities. $\Gamma_H^{\rm inv}$ 
enters in the signal strength through the total width $\Gamma_H^{\rm tot}$, 
$\mu_{ZZ}\! \propto\! \Gamma (H\! \to \! ZZ)/\Gamma_H^{\rm tot}$ with $\Gamma_H^{\rm tot} 
\! = \! \Gamma_H^{\rm inv}\! + \! \Gamma_H^{\rm SM}$ and $\Gamma_H^{\rm SM}$ calculated  
with free coefficients $c_f$ and $c_V$.  The resulting $1\sigma$ or $2\sigma$ ranges
are shown in Fig.~\ref{fig:Invfit} where $c_f$ is freely varied while $c_V=1$ \cite{fit}. 
This gives $\Gamma_H^{\rm inv}/  \Gamma_H^{\rm SM} \lsim 50\%$ at the $95\%~{\rm CL}$ 
if the assumption $c_f=c_V=1$ is made. 

\begin{figure}[!h]
\begin{tabular}{ll}
\hspace*{-.5cm}
\begin{minipage}{15.5cm}
\vspace*{-2.6cm}
%\mbox{
%\resizebox{.3\textwidth}{!}{\includegraphics{ATLAS-inv.ps}}
\resizebox{.35\textwidth}{!}{\includegraphics{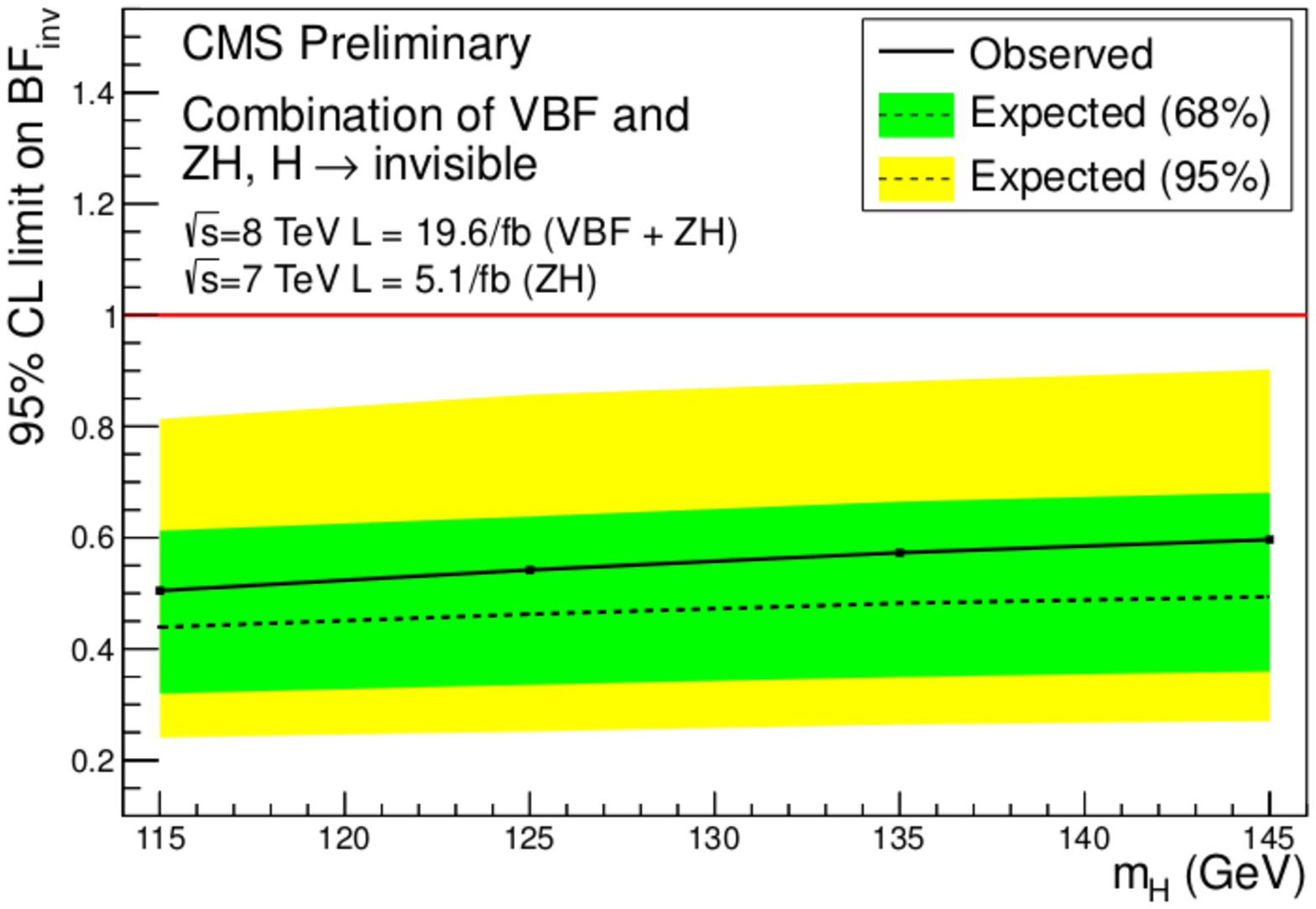}}
%}
\vspace*{-2.1cm}
\end{minipage}
& \hspace*{-11.1cm}
\begin{minipage}{15.5cm}
 \vspace{-1.9cm}
\mbox{
\resizebox{.25\textwidth}{!}{\includegraphics{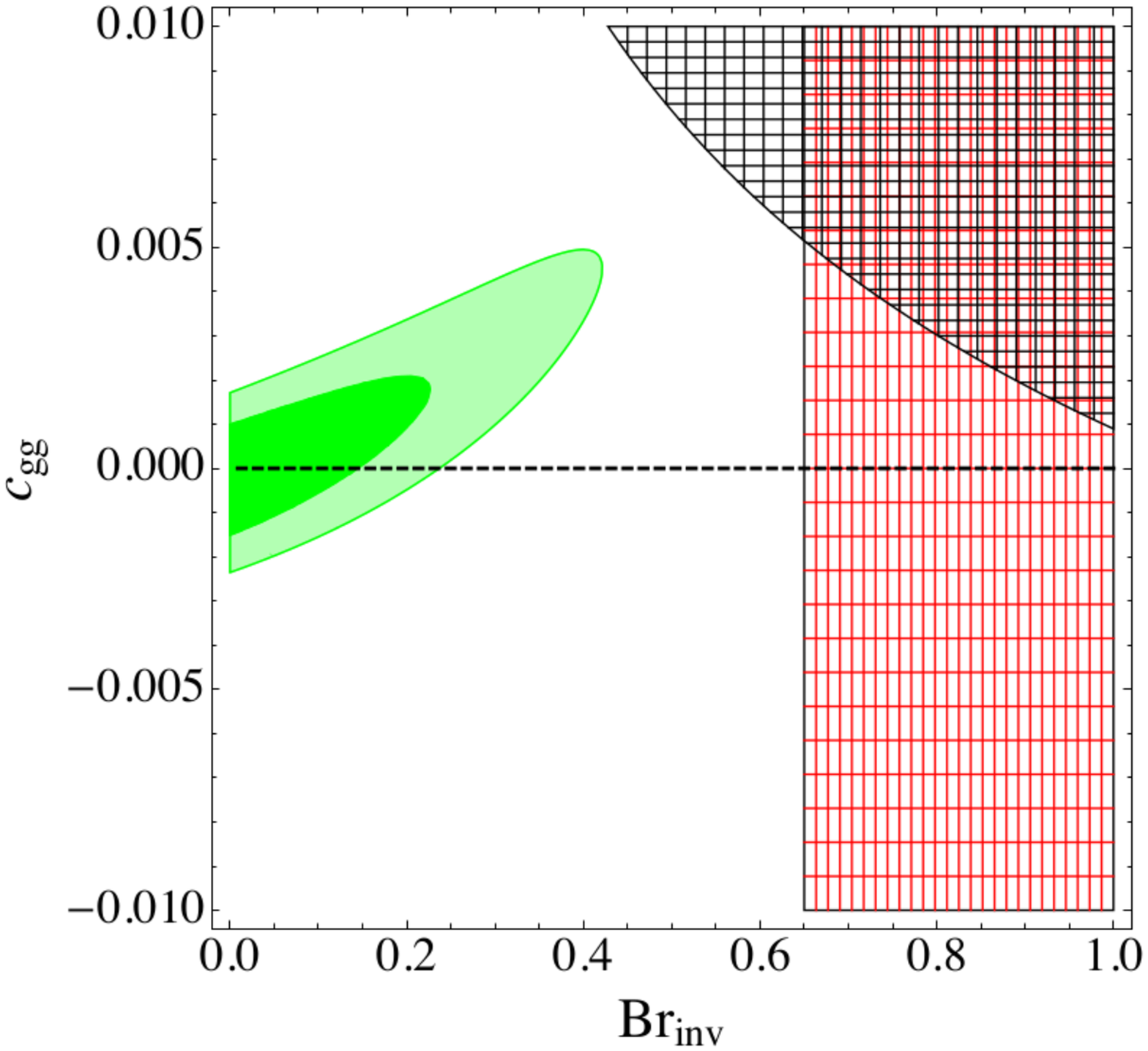}}\hspace*{-1cm}
%\resizebox{.39\textwidth}{!}{\includegraphics{PlotSigma20.ps}}
}
\vspace*{-1.6cm}
\end{minipage}
\end{tabular}
\caption{Left: the Higgs cross section times invisible Higgs decay branching ratio 
normalised to the total SM cross section in the combined $hV$ and VBF channels 
from CMS with the $\approx 20$ fb$^{-1}$ data at 8 TeV \cite{inv-CMS}.
Right: 
$68\%$ CL (light green) and $95\%$ CL (dark green) best fit regions to the 
combined LHC Higgs data. The black region is excluded by the monojet constraints 
while the red region is  excluded by  the ATLAS $Z\!+\!E_T\hspace*{-3mm}\slash$~ search 
\cite{inv-ATLAS}; from Ref.~\cite{monojet}.
\label{fig:Inv-exp}}
%\end{center}
\vspace*{-3mm}
\end{figure}

A more model independent approach would be to perform direct searches for missing
transverse energy. These have been conducted by ATLAS \cite{inv-ATLAS} and CMS 
\cite{inv-CMS} in the $pp\to hV$ process with $V\! \to \! jj, \ell \ell$ and in 
the VBF channel, $qq \to qq
E_T\hspace*{-3mm}\slash$~. As an example, we show in Fig.~\ref{fig:Inv-exp} (left) 
the CMS results for the Higgs cross section times BR$_{\rm inv}$ versus $M_h$  
when the two channels are combined. For $M_h \! \approx \! 125$ GeV a bound 
BR$_{\rm inv} \lsim 50\%$ is obtained at the 95\%CL.

%\begin{figure}[!h]
%\begin{center}
%%\hspace*{-1cm}
%\vspace*{-3cm}
%%\mbox{
%%\resizebox{.3\textwidth}{!}{\includegraphics{ATLAS-inv.ps}}
%\resizebox{.35\textwidth}{!}{\includegraphics{CMS-invisible.ps}}
%%}
%\vspace*{-2.6cm}
%\caption{The Higgs cross section times invisible Higgs decay branching ration 
%normalised to the total SM cross section in the $hV$ channel for ATLAS
%(left) and the VBF channel in CMS (right) with the $\approx 20$ fb$^{-1}$ data
%at 8 TeV.} 
%\label{inv-exp}
%\end{center}
%\vspace*{-6mm}
%\end{figure}

A more promising search for invisible decays is the monojet
channel. In the ggF mode, an additional jet can be emitted at NLO
leading to $gg\!\to\! hj$ final states and, because the QCD corrections 
are large, $\sigma(H\!+\!1j$) is not much smaller than $\sigma(h\!+\!0j$). 
The NNLO corrections besides significantly increasing the $h\!+\!0j$ and 
$h\!+\!1j$ rates, lead to $h\!+\!2j$ events that also occur in VBF and VH. 
Hence, if the Higgs is coupled to invisible particles, it may recoil against 
hard QCD radiation, leading to monojet events. 

In Refs.~\cite{monojet,monojet2}, it has been shown that the monojet signature carries a good 
potential to constrain the invisible decay width of a $\approx 125$ GeV Higgs boson.  
In a model independent fashion,  constraints can be placed on the rates 
\begin{eqnarray}
\label{e.rinv}
R_{\rm inv}^{\rm ggF}  &= & {\sigma (g g \to  h) \times {\rm BR}  
(h \to {\rm inv.})\over \sigma (g g \to h)_{\rm SM} } %\nonumber \\
%R_{\rm inv}^{\rm VBF} &= & {\sigma (q q  \to H q q) \times {\rm BR} 
%(H  \to {\rm inv.})\over \sigma (q q \to H q q)_{SM} }
\end{eqnarray}
Recent monojet searches made by CMS and ATLAS \cite{cms_mono} are sensitive 
to  $R_{\rm inv}$  close to unity. This is shown in  Fig~\ref{fig:Inv-exp} (right) where the 
best fit region to the LHC Higgs data is displayed in the ${\rm Br}_{\rm inv}$--$c_{gg}$ plane, 
where $c_{gg}$ is the deviation of $\sigma(gg\to h)$ from the SM expectation
\cite{monojet}. For the 
SM value $c_{gg}=0$, ${\rm Br}_{\rm inv} \gsim 20\%$ is disfavored at $95\%$ CL while 
for $c_{gg}\! >\! 0$, a larger rate is allowed, up to  ${\rm Br}_{\rm inv}\! \sim \! 50\%$.

The Higgs invisible rate and the dark matter detection rate in direct astrophysical 
searches are correlated in Higgs portal models. Considering the generic cases of 
scalar, fermionic and vectorial dark matter particles $\chi$ that couple only to the 
Higgs, one can translate in each case the LHC constraint  ${\rm BR} (h \to {\rm inv.})$ 
into a constraint on the Higgs couplings to the $\chi$ particles.  It turns out that
these constraints are competitive with those derived from the bounds on
the dark matter scattering cross section on nucleons from the best experiment so far, 
XENON \cite{DM-review}.  

\begin{figure}[!h]
    \begin{center}
    \vspace{-4.6cm}
\resizebox{.5\textwidth}{!}{\includegraphics{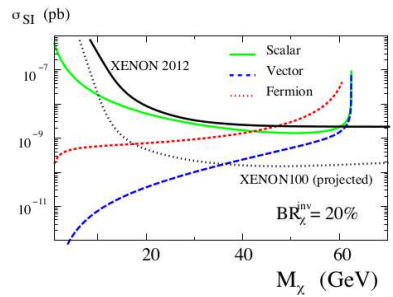}}
\vspace*{-4.9cm}
          \caption{
Bounds on the spin-independent direct detection cross section $\sigma^{\rm
SI}_{\chi p}$ in Higgs portal models derived for an invisible branching fraction 
of 20 \% (colored lines) for a 125 GeV Higgs. These are compared to the  current and 
future direct bounds from the XENON experiment (black lines). From Ref.~\cite{invis-portal}.}
\label{Fig:sigma40}
\end{center}
\vspace*{-10mm}
\end{figure}

This is shown in 
Fig.~\ref{Fig:sigma40} where the maximum allowed values of the scattering cross 
sections are given in the three cases assuming ${\rm BR}^{\rm inv}_\chi \lsim 20\%$.
The obtained spin--independent  rates $\sigma^{\rm SI}_{\chi p}$ are stronger
than the direct limit from the XENON100 experiment in the entire $M_\chi \ll \frac12
M_h$ range.  In other words, the LHC is currently the most sensitive dark matter
detection apparatus, at least in the context of simple Higgs-portal models.

\subsection{Determination of the Higgs parity}

Apart from the measurement of the couplings, one also needs in principle to
establish that the observed Higgs state is indeed a CP even scalar particle 
and hence with ${\rm J^{PC}= 0^{++}}$ quantum numbers\footnote{
To be more general, the spin of the particle needs also to be determined.
The observation of the $h\to \gamma\gamma$ decay channel rules out the 
spin--1 case by virtue of the Landau--Yang theorem \cite{Landau-Yang} and 
leaves only the spin 0 and $\geq 2$ possibilities. The graviton--like 
spin--2 option  is extremely unlikely and, already from the particle signal
strengths in the various channels, it is ruled out in large classes of 
beyond the SM models; see e.g. Ref.~\cite{Ellis}.}.  It is known that the Higgs 
to vector boson ($hVV$) coupling is a possible tool to probe these quantum numbers 
at the LHC \cite{CP-review}. This can be done by studying  various kinematical 
distributions in the 
Higgs decay and production processes. One example is the threshold behavior 
of the $M_{Z^*}$ spectrum in the $h \rightarrow Z Z^* \to 4\ell $ decay channel
and another is the azimuthal distribution between 
the decay planes of the two lepton pairs arising from the $Z, Z^*$ bosons 
from the Higgs  decay. These are sensitive to both the spin and parity of 
the Higgs. 

\begin{figure}[!h]
\vspace*{-1.cm}
\begin{tabular}{ll}
\begin{minipage}{4cm}
\hspace*{-.3cm}
\resizebox{1.1\textwidth}{!}{\includegraphics{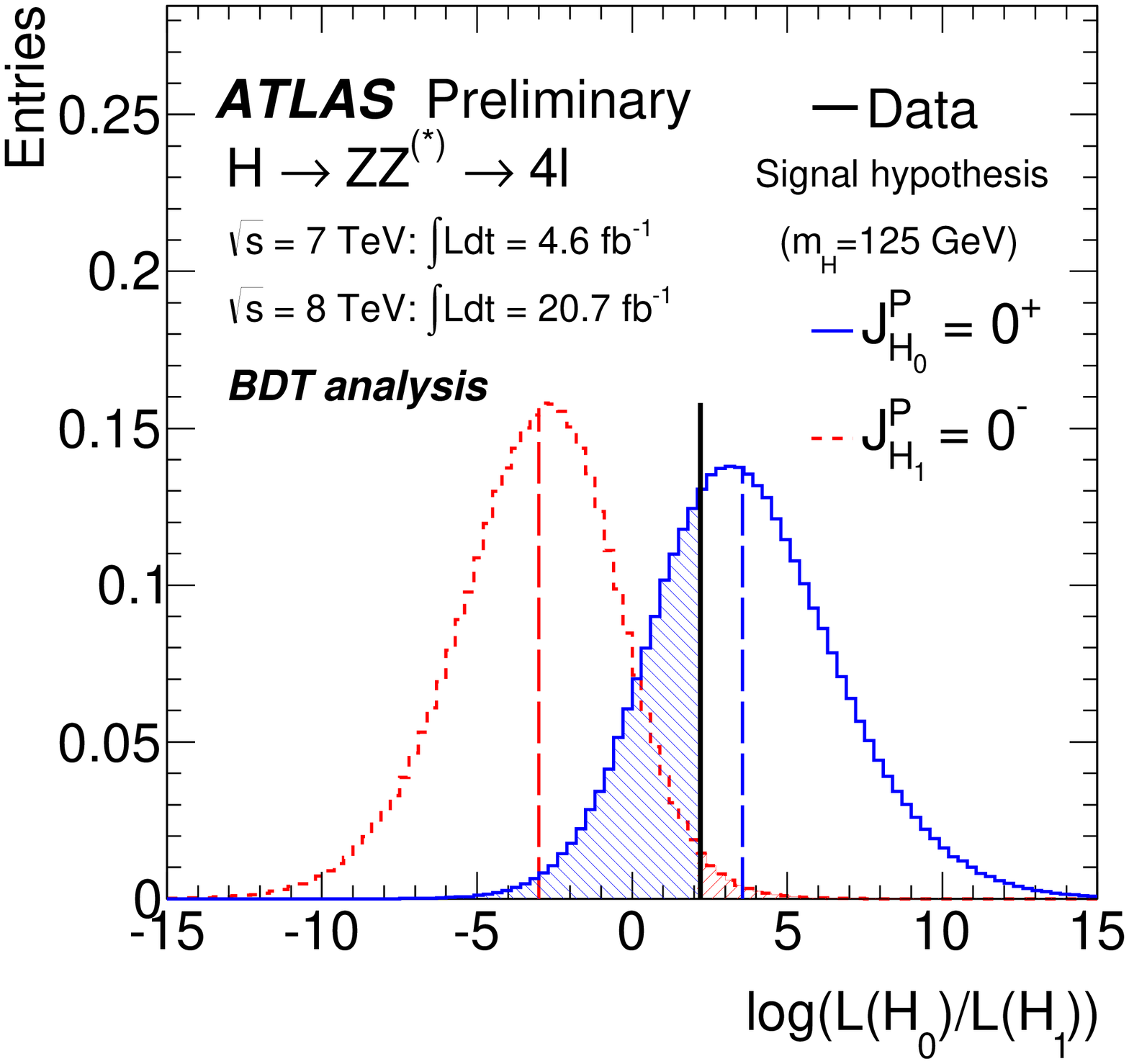}}\hspace*{-.3cm}
\end{minipage}
& \hspace*{-.4cm} 
\begin{minipage}{4cm}
\resizebox{1.1\textwidth}{!}{\includegraphics{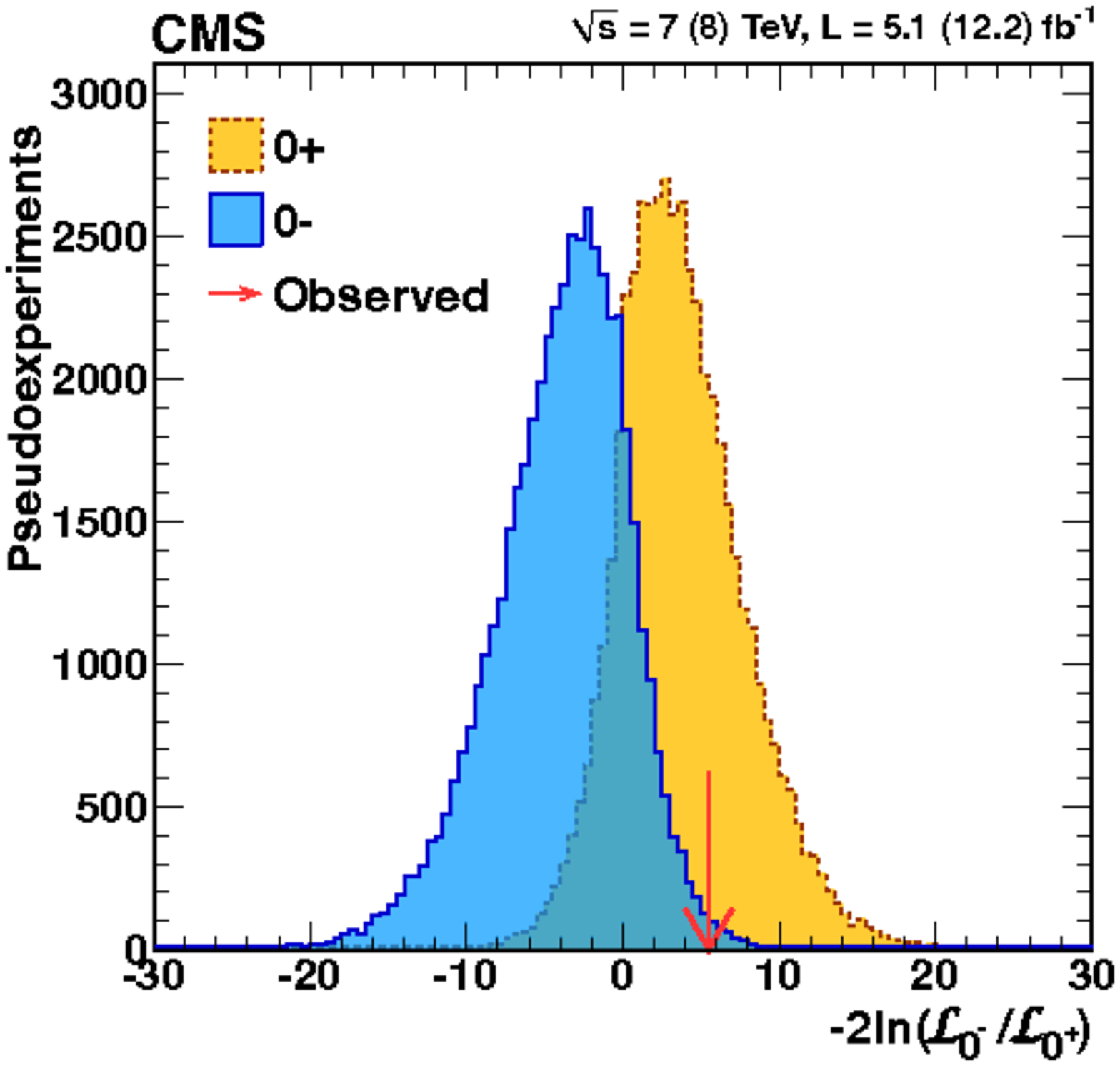}}\hspace*{-.3cm}
\end{minipage}
\end{tabular}
\vspace*{-11mm}
\caption{Discrimination between the $0^+$ and $0^-$ parity hypotheses for the
observed Higgs boson using the kinematics of the $h\to ZZ^* \to 4\ell$ channel
by the ATLAS (left) and CMS (right) collaborations with the  data
collected at 7+8 TeV \cite{CP-exp}.} 
\label{Fig:CP1}
\vspace*{-4mm}
\end{figure}

With the 25 fb$^{-1}$ data collected so far, the ATLAS and CMS collaborations 
performed a matrix-element likelihood analysis which exploits the kinematics
and Lorenz structure of the $h\to ZZ^* \to 4\ell$ channel to see whether the angular 
distributions are more compatible with the $0^+$ or $0^-$ hypothesis (as well
as the spin--2 possibility) \cite{CP-exp}. 
Assuming that it has the same couplings as the SM Higgs boson and that it is
produced mainly from the dominant ggF process, the observed particle 
is found to be compatible with a $0^+$ state and the $0^-$ possibility is excluded at 
the 97.8\% confidence level or higher; see Fig.~\ref{Fig:CP1}.

Other useful diagnostics of the CP nature of the Higgs boson that
also rely on the different tensorial structure of the $hVV$ coupling 
can be made in the VBF process. It was known since a long time that
in this channel, the distribution in the azimuthal angle between the two 
jets produced in  association with the Higgs discriminates a 
CP--even from a CP--odd state \cite{zepp}. This has been extended recently to other
observables in this process, like the rapidity separation between the
two jets \cite{CP-VBF,CP-nous}.

Recently, the VBF channel $pp \to Hjj$ has been reanalyzed in the presence 
of an anomalous $hVV$ vertex that parametrises different spin and CP assignments of 
the produced Higgs boson \cite{CP-nous}. The anomalous $hVV$ coupling is introduced by allowing for  
an effective Lagrangian with higher dimensional operators, that include four momentum 
terms which are absent in the SM.  It was shown that the kinematics of the forward 
tagging jets in this process is highly sensitive to the structure of the anomalous  
coupling and that it can effectively discriminate between different assignments
for the spin (spin-0 versus spin-2) and the parity  (CP--even versus CP--odd) 
of the produced  particle.  In particular, it was found that the correlation between 
the separation in rapidity and the transverse momenta of the scattered quarks, in
addition  to the already discussed distribution of the azimuthal jet
separation,  can be significantly altered compared to the SM expectation.

\begin{figure}[!h]
\vspace*{-6mm}
\begin{center}
\resizebox{0.35\textwidth}{!}{\includegraphics{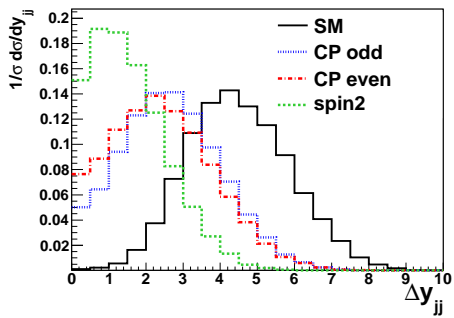} }
\end{center}
\vspace*{-3mm}
\caption[]{Normalized distribution of the difference in rapidity between the 
scattered jets in VBF for  each of the SM and BSM operators (spin--2,  CP--even  
and CP--odd) individually \cite{CP-nous}.}
\label{fig:vbf1Da}
\vspace*{-5mm}
\end{figure}

This is exemplified in Fig.~\ref{fig:vbf1Da} where the difference in rapidity 
between tagging jets ($\Delta y_{jj}$) for each of the higher dimensional operators
in the $hVV$ couplings is displayed. 

%The rapidity differences for each of the operators 
%are shifted to smaller values and, while the $0^+$ and $0^-$ cases display a behavior 
%that is almost identical, for a spin-2 state the peak in the 
%distribution is shifted to smaller values. 

%Also, the transverse momenta of the 
%tagging partons become significantly larger for pure beyond the SM $0^+$ and $0^-$ 
%states. The above mentioned features of the $\Delta y_{jj}$ and $p_T$ distributions 
%are mainly due to the presence of momentum dependent structures in the $hVV$ 
%vertices.  
These kinematical variables define new corners of the phase-space that have not been
explored by the experiments at the LHC to  probe anomalous $hVV$ couplings and
to check the Higgs parity. 
In addition, some of
these observables significantly depend on the c.m. energy  and strong
constraints on anomalous couplings can be obtained by performing measurements at
the LHC with energies of $\sqrt s \!= \!8$ TeV and 14 TeV.  Finally, 
the associated $hV$  production channel can be used as the invariant mass of the $Vh$
system as well as the $p_T$ and rapidities of the $h$ and $V$
bosons are also sensitive to anomalous $hVV$ couplings.

%%%%%%%%%%%%%%%%%%

Nevertheless, there is a caveat in the analyses relying on the $hVV$ couplings. Since a
CP--odd state has no tree--level $VV$ couplings,
%\footnote{The
%$VV$ coupling of a pseudoscalar $A$ boson should be generated through very tiny  loop
%corrections and to be as large as the SM tree--level  $hVV$ one, one needs a very low 
%new physics scale that would spoil the precision
%electroweak data.}, 
all the previous processes project out only the CP--even component
of the $hVV$ coupling \cite{CP-tt} even if the state is a CP--even and odd 
mixture. Thus, in the CP studies above,  one
is simply verifying a posteriori that indeed the CP--even component is
projected out. 

A better way to measure the parity of the Higgs boson is to study the signal strength
in the $h\to VV$ channels \cite{fit,CP-mu}. Indeed, 
the $hVV$ coupling takes the  general form 
$g_{hVV}^{\mu \nu} = -i c_V (M_V^2/v)\; g^{\mu \nu}$
where $c_V$ measures the departure from the SM: $c_V\!=\!1$ for a pure $0^+$ 
state with SM--like couplings and $c_V\approx 0$ for a pure $0^-$ state.   The
measurement of $c_V$ should allow to determine the CP composition of the Higgs
if it is indeed a mixture of $0^+$ and $0^-$ states.  

However, having $c_V\! \neq\! 1$ does not automatically imply a 
CP--odd component: the Higgs sector can be enlarged to contain other 
states $h_i$ with  squared $h_i VV$ couplings $\Sigma_i c_{V_i}^2\, g_{h_iVV}^2$
that  reduce to the SM coupling $g_{hVV}^2$.
This is what occurs in the  MSSM with complex soft 
parameters \cite{CP-review}: one has three neutral states $h_1, h_2$
and $h_3$ with indefinite parity and  their CP--even components share the
SM $hVV$ coupling, $c_{V_1}^2\!+\!c_{V_2}^2\!+\!c_{V_3}^2\!=\!1$. But
in all cases, the quantity  $1\!-\!c_V^2$ gives an {\it upper bound}
on the CP--odd contribution  to the $hVV$ coupling.\vspace*{-.8cm}

\begin{figure}[!h]
\hspace*{-3cm}
\vspace*{-.4cm}
\resizebox{0.7\textwidth}{!}{\includegraphics{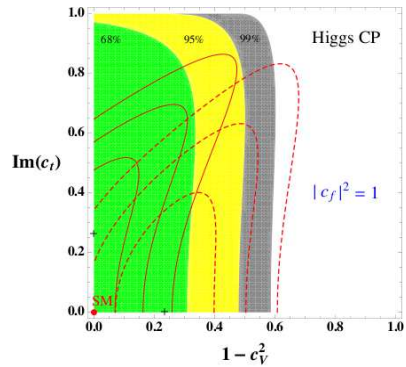}}
\vspace*{-.9cm}
\caption{
Best-fit regions at $68\%$, $95\%$ and $99\%{\rm CL}$ 
in the plane $[1-c^2_V$,Im$(c_t)$ for $\vert c_t \vert^2\!=\!\vert c_f 
\vert^2\!=\!1$. Superimposed are the best-fit regions when including
a theory uncertainty of $\pm 20\%$ \cite{fit}. 
\label{fig:CPfit}}
\vspace*{-.4cm}
\end{figure}

%Hence, the CP composition of the observed Higgs state is less ambiguously 
%probed by measuring the signal strength in the $h\to VV$ decay channels.  

Using $\mu_{VV}$ 
and the ratios $\mu_{\gamma\gamma}/\mu_{VV}$ and $\mu_{\tau\tau}/\mu_{VV}$
as in eq.~(\ref{eq:Chi2R}), it was demonstrated  
that the particle has indeed a large CP component, $\gsim  50\%$ at the 95\%CL, if 
the Higgs couplings to fermions are SM like. This is  shown in
Fig.~\ref{fig:CPfit} where one sees that the pure  CP--odd 
possibility is excluded at the $3 \sigma$ level, irrespective  of the (mixed CP) 
Higgs couplings to fermions provided that $\vert c_f \vert^2\!=\!1$.

%%%%%%%%%%%%%%%%%%%%%%%%%%%%%%%%%%%%%%%%%%%%%%%%%%%%%%%%%%%%%%%%%%%%%%%%%%%%%%%%%%%

\section{Implications from heavy Higgs searches}

We turn now to the constraints on the MSSM Higgs sector that can be obtained
from the search of the heavier $H/A$ and $H^\pm$ states at the LHC and start with 
a  brief summary of their production and decay properties.\vspace*{-5mm} 

\subsection{$\mathbf{H,A, H^\pm}$ decays and production at the LHC}

%\subsection{The high and intermediate $\tb$ regimes}

The production and decay pattern of the MSSM Higgs bosons crucially depend on
$\tb$. In the decoupling regime that is indicated
by the $h$ properties,  the heavier CP--even  $H$ boson has approximately the
same mass as the $A$ state and its interactions are similar. Hence, the MSSM Higgs
spectrum will consist of a SM--like Higgs $h \equiv H_{\rm SM}$  and two pseudoscalar--like
particles, $\Phi\! =\! H/A$. The $H^\pm$ boson will also be mass degenerate 
with the $\Phi$ states and the intensity of its couplings to fermions will be similar.  
In the high $\tb$ regime, the couplings of the non--SM like Higgs bosons to $b$ 
quarks and to $\tau$ leptons are so strongly enhanced,  and the couplings  to top 
quarks and massive gauge bosons suppressed, that the  pattern is rather  simple.

This is first the case for the decays: the $\Phi \! \to \! t\bar t$ channel and all other decay 
modes are suppressed to a level where their branching ratios are negligible and the 
$\Phi$ states   decay almost exclusively into  $\tau^+\tau^-$ and  $b\bar b$ pairs,  
with branching ratios of BR$(\Phi \to \tau \tau) \approx  10\%$ and BR$(\Phi \to 
b \bar b) \approx 90\%$. The $H^\pm$ boson decay into $\tau \nu_{\tau}$ final 
states with a branching fraction of almost 100\% for $H^\pm$ masses  below the $tb$
threshold, $M_{H^\pm} \lsim m_t+m_b$, and a branching ratio of only $\approx
10\%$ for masses above this threshold while the rate for $H^\pm  \to t b$ will
be at the $\approx 90\%$ level in most cases.

Concerning the production,  the strong enhancement of the $b$--quark  couplings at 
high $\tb$ makes that only two processes are relevant in this case:  $gg\!  \to\!  \Phi$ fusion 
with the $b$--loop included and associated production with $b$--quarks, $gg/
q\bar q \! \to \! b\bar b \! +\! \Phi$, which is equivalent to the fusion process  
$b \bar b \to 
\Phi$ when no--additional final $b$--quark is present.  All other processes, in 
particular $V\Phi, t\bar t \Phi$ and VBF have suppressed rates. In both the
$b\bar b$ and $gg$ fusion cases, as $M_\Phi \gg m_b$, chiral symmetry holds and the 
rates are approximately the same for the CP--even $H$ and CP--odd $A$ bosons. While
$\sigma(gg\to \Phi)$ is known up to NLO in QCD \cite{ggH-NLO}, $\sigma(bb\to 
\Phi)$ is instead known up to NNLO  \cite{bbH-NNLO}. 

\begin{figure}[!h] 
\begin{center} 
\vspace*{1mm}
\resizebox{0.24\textwidth}{!}{\includegraphics{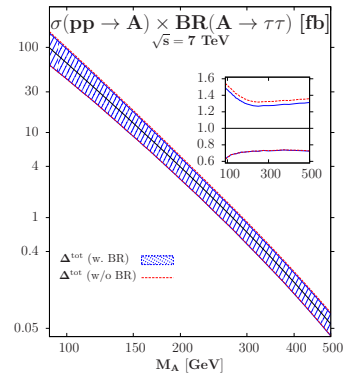}} 
\end{center} 
\vspace*{-4mm}
\caption[]{The combined $\sigma(p p\! \to\! A)\times {\rm BR}(A \!\to\! \tau \tau)$ 
rate with theoretical uncertainties with and without the branching ratio; 
in the inserts, shown are the uncertainties when the rates are normalized to  
the central values. From Ref.~\cite{BD}.}
\vspace*{-4mm}
\label{prod:phi}
\end{figure}

The most powerful search channel for the heavier MSSM Higgs particles at the LHC
is by far the process $pp\! \to \!gg\! +\! b \bar b \! \to \! \Phi \! \to \! \tau^+ \tau^-$.
The precise values of the cross section times branching fraction for this process 
at the LHC have been updated in Refs.~\cite{LHCXS,BD} and an assessment of 
the associated theoretical uncertainties has been made. It turns out that, in
the production cross section, the total
uncertainty from scale variation, the PDFs and $\alpha_s$ as well as from the 
$b$--quark mass are not that small: $\Delta^{\rm TH} \sigma (pp\!\to\! \Phi) \times 
{\rm BR} (\Phi\! \to\! \tau  \tau)  \approx \pm 25\%$ in the entire
$M_\Phi$ range probed at the LHC at $\sqrt s\!=\!8$ 
TeV;  Fig.~\ref{prod:phi}.

Besides the QCD uncertainty, three other features could alter the rate $\sigma
(pp\!\to\! \Phi \! \to\! \tau \tau)$ in the MSSM and they are related to the
impact of the SUSY particle contributions:

$i)$ In the case of $H$ ($A$ does not couple to identical sfermions), there 
are squark (mainly stop) loops that contribute in addition in the $gg\!\to\! H$ process.  
But as they are damped by powers of $\tilde m^2_Q$ for $M_H \lsim 2m^2_Q$,  these
contributions should be small for $\tilde m_Q \gsim 1$ TeV, in particular 
at high $\tb$ where the $b$--contribution is strongly enhanced.

$ii)$ The vertex correction  to the $\Phi b\bar b$ couplings, $\Delta_b$ of 
eq.~(\ref{cb}), grows as $\mu \tb$ and can be very large  in the high  $\tan\beta$
regime.  However, in the full process $p p\! \to\! \Phi\! \to\! \tau^+ \tau^-$, 
this correction appears in both the cross section and the branching fraction
and  largely cancels outs as one obtains, $\sigma\! \times \! {\rm BR} \! \times 
\! (1\! -\! \Delta_b/5)$. A very large contribution $\Delta_b \! \approx \! 1$ 
changes the rate only by 20\%, i.e. less than the QCD uncertainty.

$iii)$ The possibility of light sparticles would lead to the opening of $H/A$
decays into SUSY channels that would reduce BR($\Phi \to \tau \tau$). For $M_\Phi 
\lsim 1$  TeV, the only possibilities are decays into light neutralinos or charginos 
and sleptons. These are in general disfavored at high $\tb$ as the $\Phi \to b\bar 
b+\tau\tau$ modes are strongly enhanced and dominant. 

Thus, only in the unlikely cases where the decay $H\! \to\! \tilde \tau_1 \tilde
\tau_1$ has a branching  rate of the order of 50\%, the squark loop contribution
to the $gg\! \to\! H$ process is of the order 50\%, or the $\Delta_b$ SUSY correction
is larger than 100\%, that one can change the   $pp \to \Phi \to \tau \tau$ rate
by $\approx 25\%$, which is the level  of the QCD uncertainty. One thus
expects $\sigma(pp \to \Phi)\times {\rm BR}(\Phi \to \tau  \tau)$ to be
extremely robust and to depend almost exclusively on $M_A$ and $\tb$.

Finally, for the charged Higgs boson, the dominant search channel is in  $H^\pm \to 
\tau \nu$  final states with the $H^\pm$ bosons produced in top quark decays for 
$M_{H^\pm} \lsim m_t\!-\! m_b \approx 170$ GeV, $pp \to t\bar t$ with $t \to H^+ b 
\to \tau\nu  b$. 
This is parti\-cu\-larly true at high $\tb$ values when BR($t\to H^+b) \propto \bar 
m_b^2 \tan^2\beta$ is significant.  For higher $H^\pm$ masses, one should rely on 
the three--body production  process $pp \to tbH^\pm  \to tb \tau \nu$ but the rates 
are presently rather small. 

\begin{figure}[!h]
\begin{center}
\vspace*{-4.4cm}
\mbox{
\hspace*{-2cm}
\resizebox{0.45\textwidth}{!}{\includegraphics{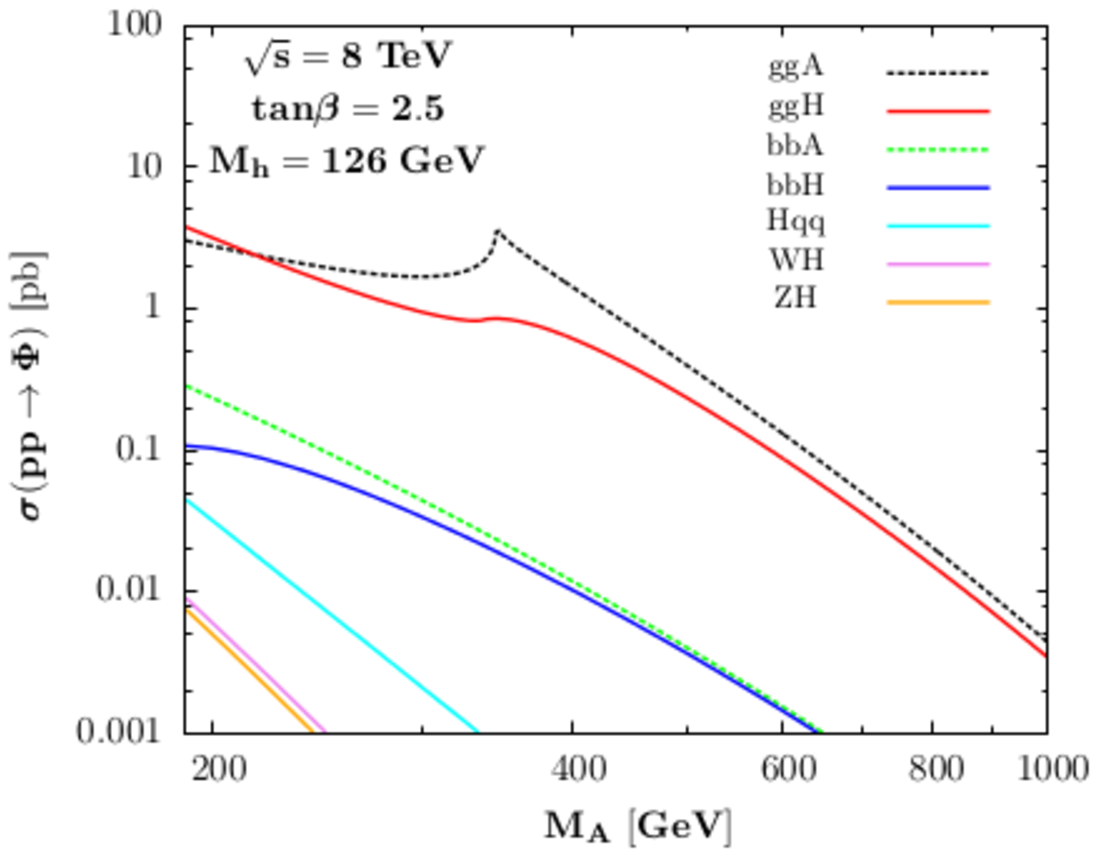}}\hspace*{-4cm} 
\resizebox{0.45\textwidth}{!}{\includegraphics{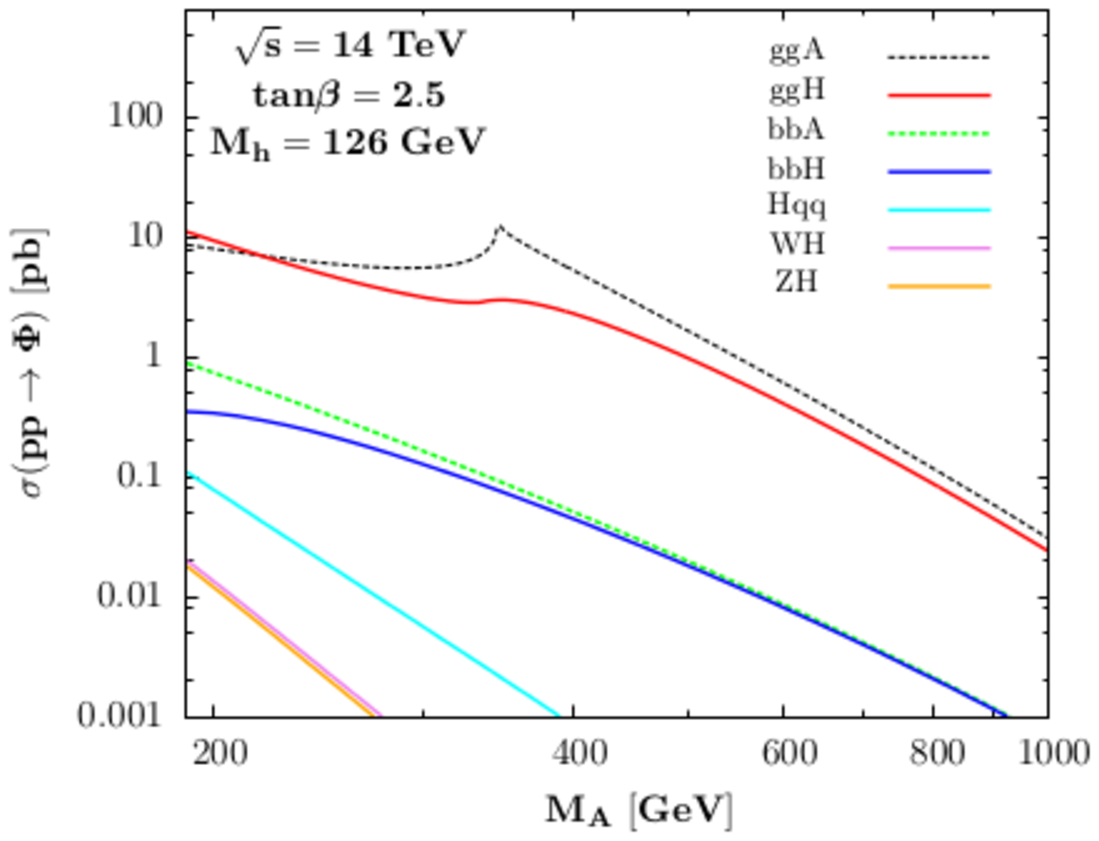}} 
}
\end{center}
\vspace*{-4.4cm}
\caption[]{The production cross sections of the MSSM heavier neutral Higgs bosons at
the LHC at $\sqrt s=8$ for $\tb=2.5$; only the main production channels are 
considered \cite{paper4}.}
\vspace*{-3mm}
\label{Fig:xs}
\end{figure}

In the low $\tb$ regime, $\tb \lsim 5$, the phenomenology of the heavier $A,H,H^\pm$ bosons  
is richer \cite{paper4,Heavy}. Starting with the cross sections, we display in Fig.~\ref{Fig:xs} the rates 
for the relevant production processes at the LHC with $\sqrt s=8$ TeV assuming $\tb=2.5$. 
For smaller $\tb$ values,  the rates except for $pp \to H/A+b\bar b$  are even larger 
as the $H/A\!+\! tt$ and $HVV$ couplings are less suppressed. 

Because of CP invariance which forbids $AVV$ couplings, there is no $AV$ and 
$Aqq$ processes while the rates for associated $t\bar t A$ and $b\bar b A$ 
are small because the $Att\;(Abb)$ couplings are suppressed (not sufficiently enhanced)
compared to the SM Higgs. 
Only the $gg\to A$ process with the dominant $t$ and sub-dominant $b$ contributions 
included  provides large rates. The situation is almost the same for $H$: only 
$gg\to H$ is  significant at  $M_H \gsim 300$ GeV  
and $\tb \lsim 5$; the VBF and HV modes give 
additional small contributions for $\tb \approx 1$. 
For $H^\pm$, the dominant production channel is again top quark decays, $t \to H^+ b$ 
for $M_{H^\pm} \lsim 170$ GeV as for $\tb \lsim 5$, the  $m_t/\tb$ piece of the $H^\pm tb$ coupling becomes large; for higher $H^\pm$ masses,  the main process to be considered is 
$gg/q\bar q  \to H^\pm tb$.

\begin{figure}[!h]
\begin{center}
\vspace*{-4cm}
\mbox{
\hspace*{-2.7cm}
\resizebox{0.4\textwidth}{!}{\includegraphics{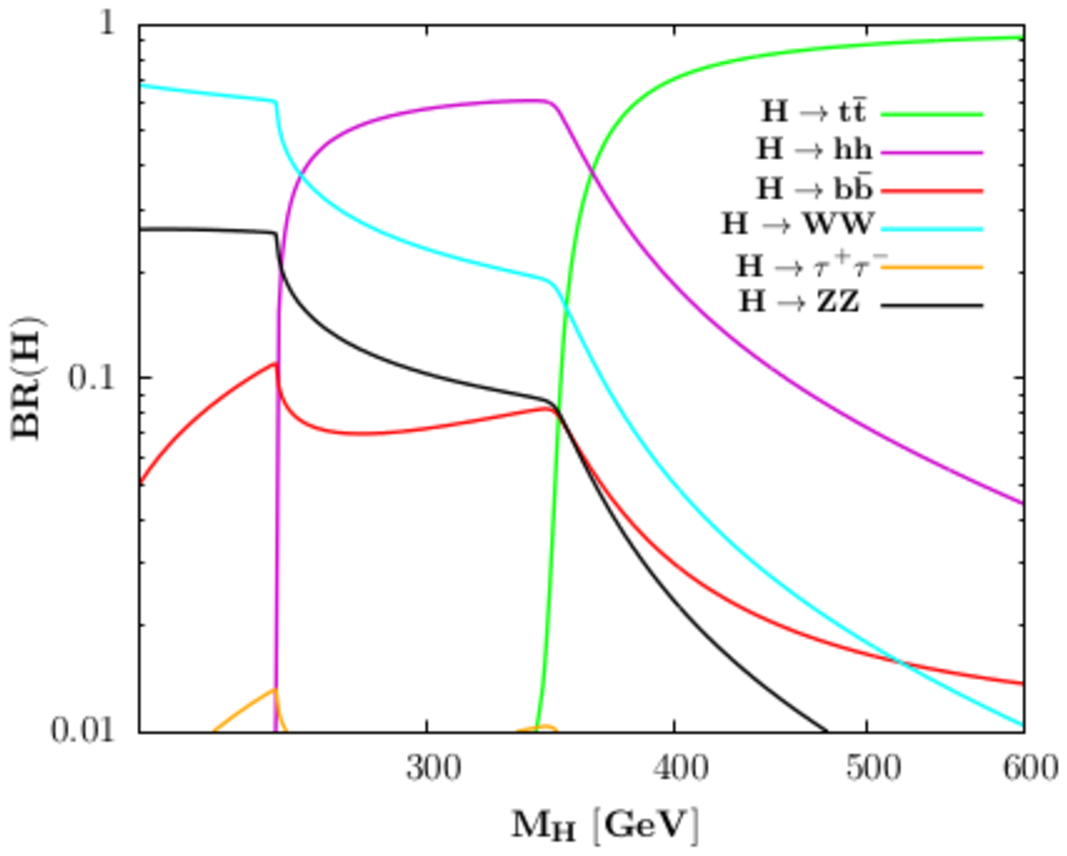}}\hspace*{-3.2cm}
\resizebox{0.4\textwidth}{!}{\includegraphics{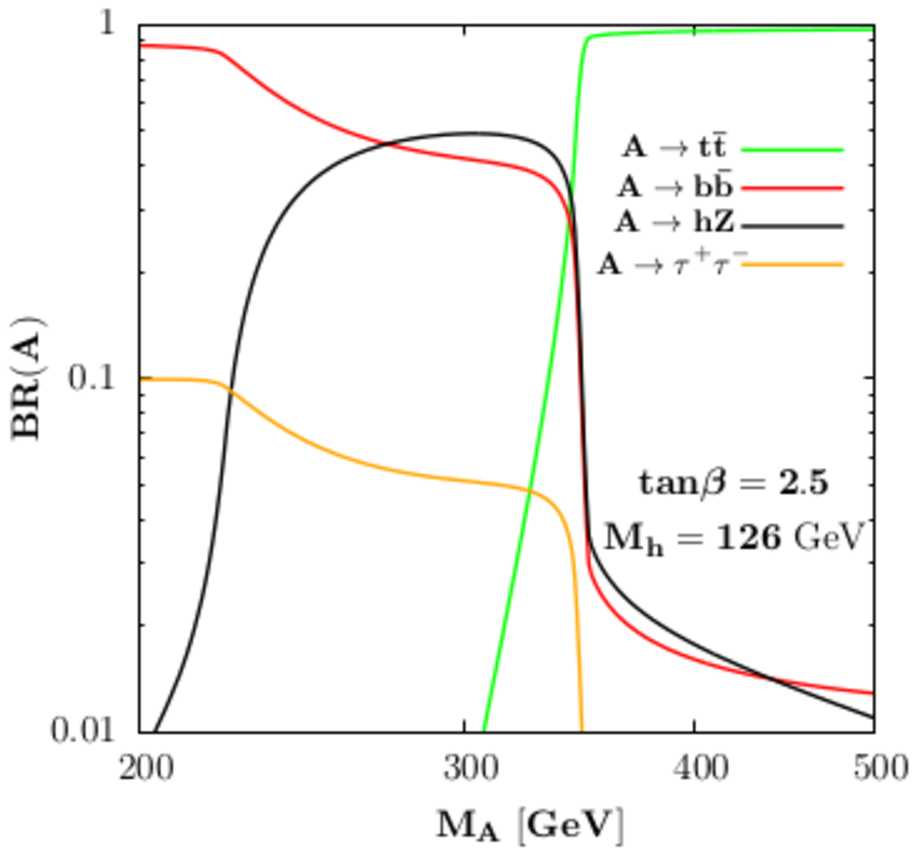}}\hspace*{-3.2cm}}\vspace*{-7.1cm}  
\resizebox{0.4\textwidth}{!}{\includegraphics{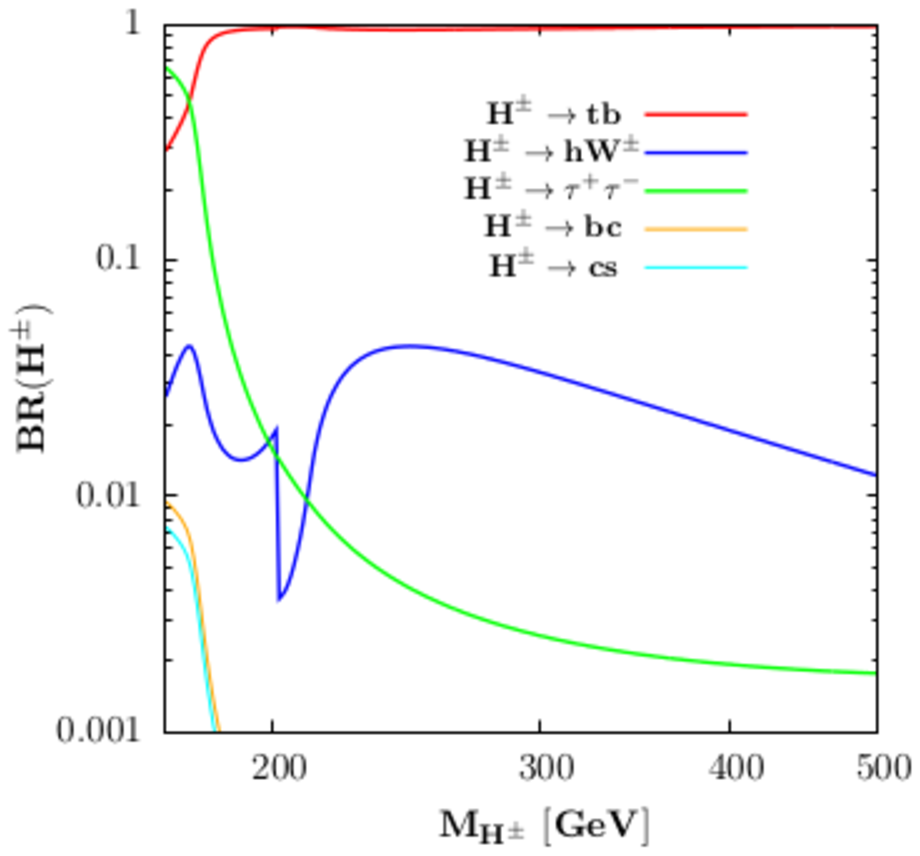}} 
%}
\end{center}
\vspace*{-4cm}
\caption[]{The $H/A/H^\pm$ decay branching ratios as functions of the Higgs masses 
for $\tb=2.5$ \cite{paper4}.}
\vspace*{-5mm}
\label{Fig:br}
\end{figure}

Turning to the $H/A/H^\pm$ decay pattern,  it can be rather involved  at low $\tb$. 
A summary is as follows for $\tb \lsim 3$; see also Fig.~\ref{Fig:br} where the rates
are shown for $\tb=2.5$. 
 
-- Above the $t\bar t\; (tb)$ threshold for $H/A(H^\pm)$,  the decay channels 
$H/ A \rightarrow t\bar{t}$  and $H^+ \to t \bar b$ are by far dominant for 
$\tb \lsim 3$ and do not leave space for any other mode. 

-- Below the $t\bar t$ threshold, the  $H\! \to \! WW,ZZ$ decay rates are still 
significant as $g_{HVV}$ is not completely suppressed.

-- For $2M_h \lsim M_H \lsim 2m_t$, $H\to hh$ is the dominant $H$ decay mode 
as the $Hhh$ self--coupling is large at low $\tb$.

-- For $M_A\! \gsim M_h+M_Z$, $A \to hZ$ decays would occur but the $A \to \tau\tau$ 
channel   is still important with rates  $\gsim\! 5\%$. 

-- In the case of $H^\pm$, the channel  $H^+\! \to \! Wh$ is important for 
$M_{H^\pm}\! \lsim \! 250$ GeV, similarly to the $A \! \to \! hZ$ case.

Hence, many decay and production channels need to be considered in this low $\tb$ regime.

\subsection{Constraints from the LHC Higgs searches}

The most efficient channel to probe the heavier MSSM Higgs bosons is by far 
$pp \! \to \! gg\! +\! bb \! \to \! H/A\! \to \! \tau^+ \tau^-$.  Searches for this 
process have been performed by ATLAS with $\approx 5$ fb$^{-1}$ data at the 7 TeV 
run \cite{ATLAS-tau} and by CMS with $\approx 5+12$ fb$^{-1}$ data at the 7 TeV and
8 TeV runs \cite{CMS-tau}. Upper limits on the production cross section times decay 
branching ratio have been set and they can be turned into constraints 
on the MSSM parameter space. 

\begin{figure}[!h] 
\vspace*{-22mm}
\begin{center} 
\hspace*{-4mm}
\resizebox{0.36\textwidth}{!}{\includegraphics{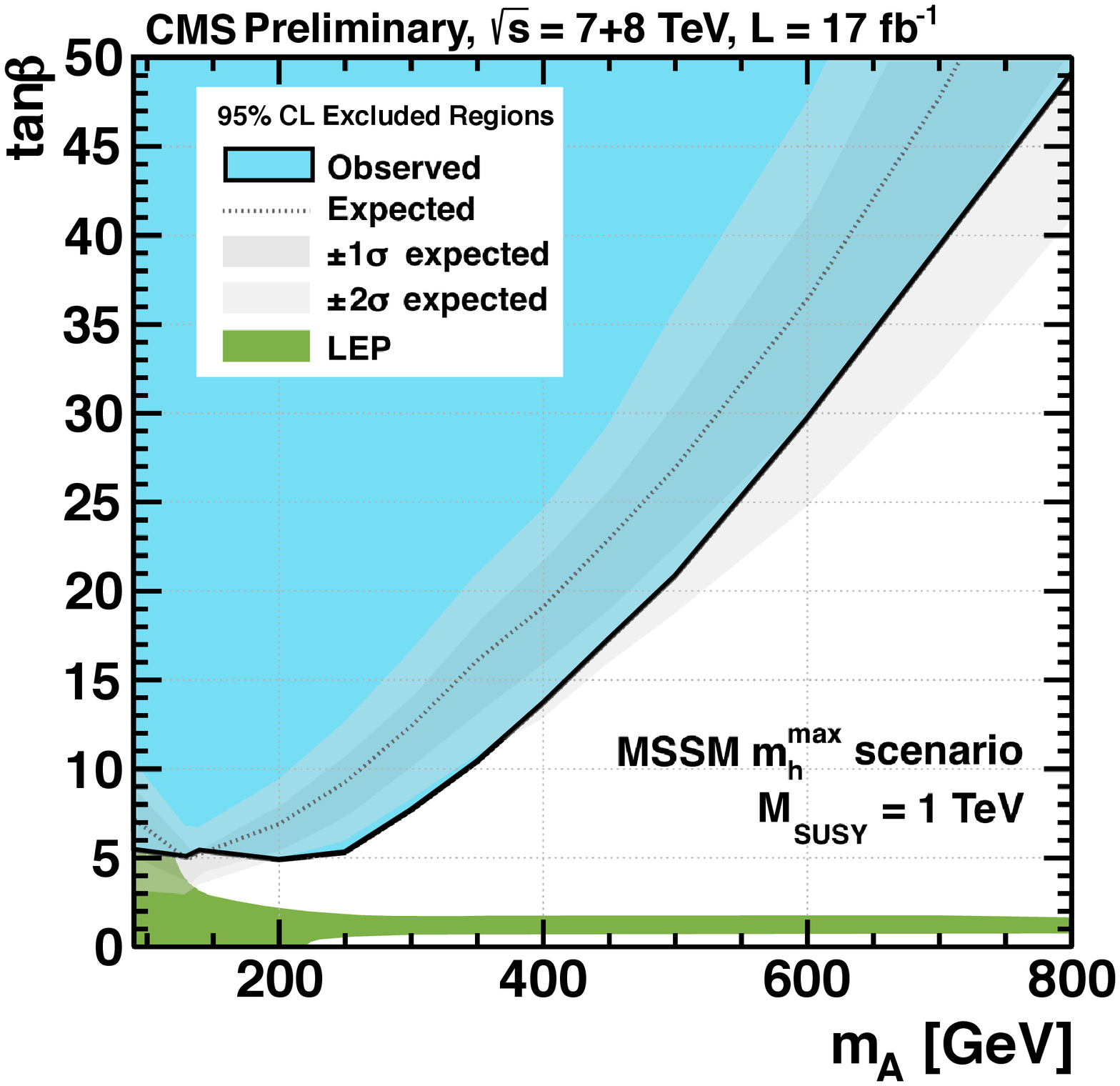}} 
\end{center} 
\vspace*{-20mm}
\caption[]{The expected and observed exclusion limits in the $[\tb, M_A]$
plane in the CMS search of the MSSM neutral Higgs bosons in the channels 
$pp \to h/H/A \to \tau^+\tau^-$  with $\approx 17$ fb$^{-1}$ data collected at 
$\sqrt s=7$+8 TeV \cite{CMS-tau}.}
\vspace*{-3mm}
\label{prod:HA}
\end{figure}

In Fig.~\ref{prod:HA}, displayed is the sensitivity of the CMS $pp\to \Phi \to 
\tau \tau$ analysis with 17 fb$^{-1}$ of data in the $[\tb,M_A]$ plane. The excluded  
region, obtained from the observed limit at the 95\%CL  is drawn in blue. The
dotted line  represents the median expected limit which turns out to be weaker
than the observed limit.  As can be seen,   this constraint is extremely
restrictive and for values $M_A \lsim 250$ GeV, it excludes  almost the entire
intermediate and high  $\tb$ regimes, $\tb \gsim 5$.   The constraint is
less effective for a heavier $A$ boson, but even for $M_A \approx 400$ GeV the high 
$\tb \gsim 10$ region is excluded and one is even sensitive to large values  
$M_A \approx 800$ GeV for  $\tb \gsim 50$. 

There are, however, some caveats to this exclusion limit as discussed previously.  
The first one is that  there is a theoretical uncertainty of order of $\pm 25\%$
that affects the $gg \to \Phi$ and $b\bar b \to \Phi$ production cross sections 
which, when included, will make the constraint slightly weaker as one then needs 
to  consider the lower value predicted for the production rate. A second caveat 
is that SUSY effects, direct corrections to the production and $H/A$ decays into 
sparticles, could alter the rate.  However, as previously argued,  $\sigma(pp 
\! \to \! \Phi) \! \times \! {\rm BR}(\Phi \! \to \!\tau \tau)$ is robust against 
these SUSY effects and the latter will unlikely make a substantial  change of the 
cross section times branching fraction. Finally, the constraint is specifically given 
in the maximal mixing scenario $X_t/M_S= \sqrt 6$ with $M_S=1$ TeV. The robustness
of $\sigma \! \times \! {\rm BR}$ makes that the  exclusion limit  
is actually almost model independent and is valid in far more situations   
than the  ``MSSM $M_h^{\rm max}$ scenario" quoted there, an assumption 
that can be removed without any loss. 

In fact, the exclusion limit can also be extended to the low $\tb$ region which, 
in the chosen scenario with $M_S=1$ TeV, is excluded by the LEP2 limit on 
the lighter $h$ mass (the green area in the figure) but should resurrect if the
SUSY scale is kept as a free parameter. Note also, that $H/A$ bosons
have also been searched for in the channel $gg \to b\bar b \Phi$ with
$\Phi \to b\bar b$ (requiring more than 3--tagged $b$ jets in the final
state) but the constraints are much less severe than the ones derived
from the $\tau\tau$ channel \cite{CMS-bbbb}.

\begin{figure}[!h] 
\vspace*{-3mm}
\begin{center} 
\resizebox{0.32\textwidth}{!}{\includegraphics{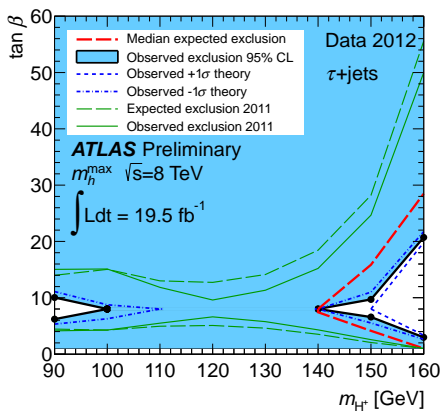}} 
\end{center} 
\vspace*{-3mm}
\caption[]{The $H^\pm$ limits from ATLAS with  $\sqrt 
s=8$ TeV and $\approx 20$ fb$^{-1}$ data in the channel $t\to bH^+ \to b\tau \nu$
\cite{ATLAS-H+}.}
\vspace*{-3mm}
\label{fig:ATLAS-H+}
\end{figure}

Turning to the $H^+$ boson \cite{CMS-H+,ATLAS-H+}, the most recent result has 
been provided by the 
ATLAS collaboration using the full $\approx 20$ fb$^{-1}$ data collected  
at $\sqrt s= 8$ TeV. The $H^\pm$ search as been performed using the $\tau$ 
plus jets channel with a hadronically decaying $\tau$ lepton in the final 
state. For $M_{H^\pm} \lsim 160$ GeV, the results are shown in
Fig.~\ref{fig:ATLAS-H+}. Here, the relevant process is top quark decays, 
$t \to H^+ b$ with the decay $H^+ \to \tau \nu$ having a branching ratio of
almost 100\% at moderate to high $\tb$. For these high values, the $H^+tb$ coupling
has a component $\propto m_b \tan\beta$ which makes BR($t\! \to _!H^+b)$ rather 
large. Almost the entire $\tb \gsim 10$ region   is excluded by the ATLAS analysis.  

In addition, the branching fraction for the decay $t \to bH^+$ is also significant at
low $\tb$ values, when the component of the  coupling $g_{tbH^+} \propto \bar m_t /\tb$ 
becomes dominant. On the other hand, the
branching fraction for the decay $H^\pm \to \tau \nu$ does not become very small as
it has competition only from $H^+ \to c\bar s$ which, even for $\tb \approx 1$,
does not dominate. Hence, the rates for $pp \! \to \! t\bar t$ with $t \! \to \! bH^+ \!\to 
\! b\tau 
\nu$ are comparable for $\tb \approx 3$ and $\tb \approx 30$ and the processes can 
also probe the low $\tb$ region. This is exemplified in Fig.~\ref{fig:ATLAS-H+}
where one can see that the entire area below $\tb \approx 5$ is also excluded.
Remains then, for $H^\pm$ masses close to 90 GeV (where the detection efficiency is
lower) and 160 GeV (where one is limited by the phase-space), the intermediate
area with $\tb \approx 5$--10 where the $H^\pm tb$ coupling is not strongly enhanced

This ATLAS search has been extended to larger values of $M_{H^\pm}$ where the charged Higgs
is produced in association with top quarks, $gb \to tH^+$, but the constraints
are poor (only the region $\tb\! \gsim \! 50$ is excluded for $M_{H^\pm}\! =\!200$--300 GeV)
as the cross section for this process is low.

The reopening of the low $\tb$  region allows  to consider  a plethora of very 
interesting channels for the heavier Higgs bosons to be also investigated at the 
LHC: heavier CP--even $H$ decays into massive gauge  bosons $H\to WW,ZZ$ and 
Higgs bosons $H\to hh$, CP--odd Higgs decays into a vector and a Higgs boson, 
$A \to hZ$, CP--even and CP--odd Higgs decays  into top quarks, $H/A \to t \bar 
t$, and even the charged Higgs decay $H^\pm \to Wh$. These final states have 
been searched for in the context of a heavy SM Higgs boson or for new resonances in 
some non--SUSY beyond the SM scenarios and the analyses  
can be adapted  to the case of the heavier MSSM Higgs bosons. They would then allow 
to cover a  larger part of the parameter space of the MSSM Higgs sector in 
a model--independent way, i.e. without using the information on the scale 
$M_S$ and more generally on the SUSY particle spectrum that appear in the 
radiative  corrections. 

\begin{figure}[!h]
\begin{center}
\vspace{-7.cm}
\mbox{\hspace*{-2.5cm}
\resizebox{0.75\textwidth}{!}{\includegraphics{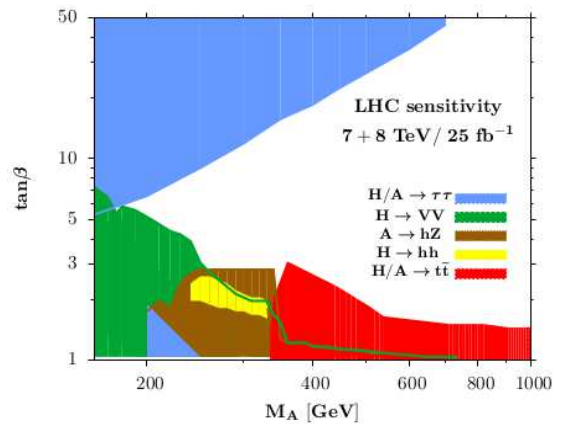}}  
}
\end{center}
\vspace*{-7.2cm}
\caption[]{The estimated sensitivities in the various search channels for the 
heavier MSSM Higgs bosons in the $[\tb,M_A]$ plane: $H/A \! \to \! \tau \tau$, 
$H \! \to \! WW\! +\! ZZ$, $H/A \! \to \! t\bar t$, $A\! \to \! hZ$  
and $H \! \to \! hh$ \cite{paper4}. The projection is made for the LHC with 7+8 TeV and the full 
25 fb$^{-1}$ of data collected so far. The radiative corrections are such that 
the $h$ mass is $M_h=126$ GeV.}
\label{Fig:sensitivity}
\vspace*{-2mm}
\end{figure}

In Ref.~\cite{paper4} a preliminary analysis of these channels has been 
performed using current information given by the ATLAS and CMS collaborations
in the context of searches for the SM Higgs boson or other heavy resonances
(in particular  new $Z'$ or Kaluza--Klein gauge bosons that decay into $t\bar 
t$ pairs). 
The results are shown in Fig.~\ref{Fig:sensitivity} with an extrapolation to the
full 25 fb$^{-1}$ data of the 7+8 TeV LHC run (it has been assumed  that the
sensitivity scales simply as the  square root of the number of events). The
sensitivities from the usual $H/A \to \tau^+\tau^-$  and $t \to bH^+ \to b \tau
\nu$ channels are also shown. The green and red areas correspond to the domains
where the $H\to VV$ and $H/A \to t\bar t$ channels become   constraining. The 
sensitivities in the $H\to hh$ and $A\to hZ$ modes are  given by, respectively, 
the yellow and brown areas which peak in the mass range $M_A=250$--350 GeV that
is visible at low $\tb$ values. 

The outcome is impressive. These channels, in particular the $H \to VV$ and $H/A
\to t \bar t$ processes, are very constraining as they cover the entire low
$\tb$ area that was previously excluded by the LEP2 bound up to $M_A \approx
500$ GeV. Even  $A \to hZ$ and $H \to hh$ would be visible at the current LHC 
in small portions of the parameter space.

\subsection{Could the observed state be the heavier H boson?}

Let us briefly discuss the possibility, raised with the early LHC 
data, that the observed particle is the heavier MSSM $H$ boson, as advocated 
for instance in Refs.~\cite{H=observed,benchmarks}. The possibility $M_H\! \approx \! 
125$ GeV with $H$ couplings close to those of the SM Higgs, occurs for low 
values of $M_A$, $\approx 100$--120 GeV, and moderate values of $\tan \beta$, 
$\approx 10$. In this case, $H$ has approximately SM--like properties, while $h$ 
has a mass of order 100~GeV or below and suppressed  couplings to vector bosons. 
A dedicated scan for this region of parameter space has been performed 
\cite{paper2} and the results were confronted with the measured Higgs mass
$M_h\! =\!123$--129 GeV and couplings that comply with the LHC $\approx 10$ fb$^{-1}$ data 
collected at $\sqrt s\! =\! 7\!+\!8$ TeV. Both the signal strengths in the various 
search channels of the observed Higgs boson and the limits from the $pp \to \tau^+ 
\tau^-$ channel obtained by the CMS  collaboration have been considered. 

It was found that 
among the large flat scan with $10^8$ points, only $\approx 2 \times 10^{-5}$ of the 
generated points would remain after imposing these LHC constraints. These points 
were then excluded by applying the constraints from flavour  physics \cite{B-physics}
(see also
Ref.~\cite{eviction}), mainly
the radiative decay $b\to s\gamma$,  and dark matter constraints 
\cite{DM-review} (as they do 
not satisfy the constraint  of $10^{-4} <  \Omega h^2 < 0.155$ 
when  accounting 
for all uncertainties). The updated $pp\to \tau^+\tau^-$
search performed by CMS  with 17 fb$^{-1}$ data, which excludes all 
values $\tb \gsim 5$ for $M_A \lsim 250$ GeV as shown in Fig.~\ref{prod:HA}, now
definitely rules out this scenario.  

\begin{figure}[h!]
\begin{center}
\vspace*{-33mm}
\resizebox{0.4\textwidth}{!}{\includegraphics{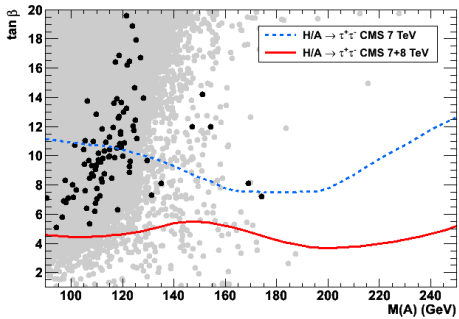}}
\end{center}
\vspace*{-32mm}
\caption{The parameter space $[M_A, \tan \beta]$ with points for the heavier $H$ boson
to be observed with a mass in the range 123--129 GeV (light grey points) and after 
flavour and dark matter relic density  constraints (black points) \cite{paper2}. 
The CMS excluded 
regions from the 2011 and 2012 $\Phi \! \to \! \tau^+ \tau^-$ searches are shown by 
the dashed blue and continuous red lines, respectively.}
\label{fig:MA}
\vspace*{-5mm}
\end{figure}

This is exemplified in Fig.~\ref{fig:MA} where we zoom in the $[M_A, \tan \beta]$ 
plane for low values of the inputs and apply the constraints listed above.  
The small region in which the $H$ boson was allowed to be the observed state  (black
points) by the previous  $H/A \to \tau^+ \tau^-$ CMS search (dashed blue line), is 
excluded by the new data (in red). In fact, the latest ATLAS limits from $H^\pm$
searches given in Fig.~\ref{fig:ATLAS-H+} also exclude now the possibility $M_A\approx 
100$--120 GeV and, hence, the scenario where $H$ is the observed Higgs state\footnote{Note 
that the recent $pp\to \tau \tau$ and $H^+\to \tau \nu$  limits also
exclude the so--called ``intense  coupling regime" \cite{intense}, in which the
three neutral Higgs bosons could be light and  close in mass.}.

\subsection{Higgs production with SUSY particles}

Finally, let us comment on the possibility of the Higgs bosons being
produced in processes involving sparticles. First of all, there
is the option of Higgs decays into SUSY particles. In the case
of the lighter $h$ boson, the only possibility when the LEP2 constraints
are taken into account is the decay $h \to \chi_1^0 \chi_1^0$ which
has been discussed in the context of invisible Higgs decays in section 3.4.  
In view of the strong LHC limits on squark masses, the only SUSY channels
of the heavier $H/A/H^\pm$ states that might be kinematically open 
would be the decays into chargino, neutralinos and sleptons. For $H/A$,
these decays have been discussed in the context of the $\tau$ 
searches as they might reduce the $H/A \to \tau\tau$ branching fractions
but no specific search for these SUSY final states has been performed
so far.
   
Turning to associated Higgs production with sparticles, the most important 
process was expected to be $pp \! \to \! \tilde t_1 \tilde t_1+$ Higgs
which could benefit from the possibly large Higgs--stop coupling \cite{Hstop}.
The large value of $M_S$ and hence the lightest stop mass from current 
constraint makes this process unlikely. Another possibility would be
associated production with stau's where the phase-space could be
more favorable but the rates are in general much smaller.   

The only channel which could lead to a detectable signal with the data
collected so far would be Higgs particles from decays of charginos and 
neutralinos. In particular the decays $\chi_2^0  \to \chi_1^0 h$, with 
$\chi_2^0$ directly produced in association with $\chi_1^\pm$ in the
process $pp \to \chi_2^0 \chi_1^\pm$ leading a lepton, a Higgs (decaying
either into $b\bar b$ or into multi-leptons via $h\to ZZ^*, WW^*$)
and missing energy \cite{Hcascade}.

\begin{figure}[h!]
\begin{center}
\vspace*{-28mm}
\resizebox{0.38\textwidth}{!}{\includegraphics{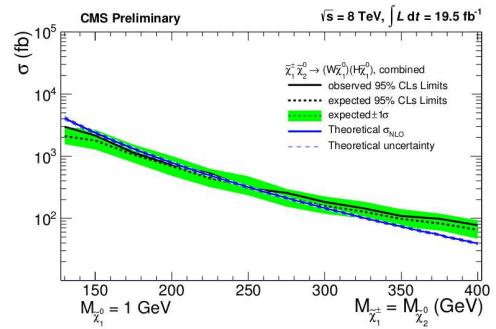}}
\end{center}
\vspace*{-30mm}
\caption{
The interpretations of the CMS results from the combination of all lepton and 
$E_T\hspace*{-3mm}\slash$~ searches with $\approx 20$ fb$^{-1}$ data collected at 
$\sqrt s=8$ TeV. The expected and observed limits on the $pp\to \chi_2^0 \chi_1^\pm$ 
cross section times the $\chi_2^0 \chi_1^\pm \to Wh \chi_1^0 \chi_1^0$ branching fraction 
(with the green band is for experimental uncertainties) is compared to the 
theoretical prediction \cite{CMS-hchi}.}
\label{fig:hchi}
\vspace*{-5mm}
\end{figure}

The CMS collaboration has reported the results for searches of leptons and missing energy 
with a luminosity of $\approx 20$ fb$^{-1}$ data collected at $\sqrt s=8$ TeV \cite{CMS-hchi}. They set a  
 limit on the cross section times branching ratio for the possible SUSY process
$pp\to \chi_2^0 \chi_1^\pm$ with $\chi_2^0 \to \chi_1^0 h$ and $\chi_1^\pm \to W \chi_1^0$.
As can be observed from Fig.~\ref{fig:hchi} where the cross section times branching
ratio is displayed as a function of the masses $m_{\chi_1^\pm}=m_{\chi_2^0}$ (with the
assumption that the LSP neutralino is very light, $m_{\chi_1^0}=1$ GeV),  the data show
no excess over the SM backgrounds. 

\section{What next?}

The last two years were extremely rich and exciting for particle physics. With the 
historical discovery at the LHC of a Higgs boson by the ATLAS and CMS collaboration 
crowned by a Nobel price this fall, and the first probe of its basic properties,
they witnessed a giant step in the unraveling of the mechanism that breaks the 
electroweak symmetry and generates the fundamental particle masses. They promoted 
the SM as the appropriate theory, up to at least the Fermi energy scale, to describe 
three of Nature's interactions, the electromagnetic, weak and strong forces, 

However, it is clear that these two years have also led to some frustration as no 
signal of physics beyond the SM has emerged from the LHC data. The hope of observing 
some signs of the new physics models that were put forward to address the 
hierarchy problem, that is deeply rooted in the Higgs mechanisms, with  Supersymmetric 
theories being the most attractive ones, did not materialize. 

The discovery of the Higgs boson and the non--observa\-tion of
new particles has nevertheless far reaching consequences for supersymmetric theories 
and, in particular, for their simplest low energy formulation, the MSSM. The mass of 
approximately 125 GeV of the observed Higgs boson implies that the scale of 
SUSY--breaking is rather high, at least ${\cal O}$(TeV). This is backed up by
the limits on the masses of strongly interacting SUSY particles set by  
the ATLAS and CMS searches, which in most cases exceed the TeV range
\cite{LHC-SUSY,Craig}.  This
implies that if SUSY is indeed behind the stabilization the Higgs mass against 
very high scales that enter via quantum corrections, it is either
fine--tuned at the permille level at least or its low energy manifestation
is more complicated than expected. 
 
The production and decay rates of the observed Higgs particles, as well as its
spin and parity quantum numbers, as measured by the ATLAS and CMS collaborations 
with the $\approx 25$ fb$^{-1}$ data collected at $\sqrt s\!=\!7$ and 8 TeV, 
indicate that its couplings to fermions and gauge bosons are approximately SM--like. 
In the context of the MSSM, this implies that we seem to be in the decoupling regime 
and this 125 GeV particle can be only identified with the lightest $h$ boson, 
while the other $H/A/H^\pm$ states must be heavier than approximately the Fermi 
scale. This last feature is also backed up by the constraints from direct searches of 
these heavier Higgs states at the LHC. 

This drives up to the question that is now very often asked in particle physics
(and elsewhere): what to do next? The answer is, for me,  obvious: we are only in 
the beginning of a new era\footnote{One can rightfully use here the words of Winston 
Churchill in November 1942 after the battle of El Alamein (which in Arabic 
literally means 
``the two flags" but could also mean ``the two worlds" or even ``the two scientists"!): 
``Now, this is not the end; it is not even the beginning to the end;
but it is, perhaps, the end of the beginning".}. Indeed, it was expected 
since a long time that the probing of the EWSB mechanism 
will be at least a two chapters story. The first one is the search and the 
observation of a Higgs--like particle that will confirm the scenario of the SM and 
most of its extensions, that is, a spontaneous symmetry breaking by a scalar field 
that develops a non--zero vacuum expectation value. This long chapter has just 
been closed by the ATLAS and CMS collaborations with the spectacular observation 
of a Higgs boson. This observation opens a second and equally important chapter: 
the precise determination of the Higgs profile and the unraveling of the EWSB 
mechanism itself. 

A more accurate measurement of the Higgs couplings to fermions and gauge bosons 
will be mandatory to establish the exact nature of the mechanism 
and, eventually, to pin down effects of new physics if additional ingredients beyond 
those of the SM are involved. This is particularly true in weakly interacting theories
 such as SUSY in which the quantum effects are expected to be small.
These measurements could be performed at the upgraded LHC with an energy close 
to $\sqrt s\!=\!14$ TeV, in particular if a very high luminosity, a few ab$^{-1}$, 
is achieved \cite{H-LHC,Snowmass}. 

At this upgrade, besides improving the measurements performed so far, rare but important 
channels such as associated Higgs production with top quarks, $pp\!\to\! t\bar t h$, 
and Higgs decays into $\mu^+ \mu^-$ and $Z\gamma$ states could be probed. Above all,  
a determination of the self--Higgs coupling could be made by searching for double Higgs 
production e.g. in the gluon fusion channel $gg\to hh$ \cite{HHH}; this would
be a first step towards the reconstruction of the scalar potential that is responsible 
of EWSB. A proton collider with an energy $\sqrt s\!=\!30$ to 100 TeV could 
do a similar job \cite{Snowmass}.

In a less near future, a high--energy lepton collider, which is nowadays discussed in various 
options (ILC, TLEP, CLIC, $\mu$--collider)  would lead to a more accurate probing of the Higgs properties \cite{ILC}, promoting the scalar sector of the theory to 
the high--precision level of the gauge and fermionic sectors achieved by LEP and SLC
\cite{PDG}. 

Besides the high precision study of the already obser\-ved Higgs, one
should also continue to search for the heavy states that are predicted by
SUSY, not only the superparticles but also the heavier Higgs bosons. 
The energy upgrade to  $\approx \!14$ TeV (and eventually beyond) and the planed 
order of magnitude (or more) increase in luminosity will allow to probe much higher
mass scales than presently.  

In conclusion, it is not yet time to give up on Supersymmetry and on New Physics
in general but, rather, to work harder to be fully prepared for the more precise 
and larger data that will be delivered by the upgraded LHC. It will be soon
enough to ``philosophize" in two years from now, when the physics landscape will 
become more clear.\bigskip 

\noindent {\bf Acknowledgements:}\smallskip 

\noindent This review relies heavily on work performed in the last two years
in collaboration with S. Alekhin, A. Arbey, J. Baglio, M. Battaglia, A. Falkowski, 
R. Godbole, R. Grober, O. Lebedev, A. Lenz, N. Mahmoudi, L. Maiani, Y. Mambrini, 
B. Mellado,  K. Mohan, G. Moreau, M. Muhlleitner, J. Quevillon, A. Polosa, V. 
Riquer and M. Spira. I thank them all for their input and for having made these
two last years very fruitful and extremely exciting. Discussions with members
of ATLAS and CMS (that I congratulate in passing) 
are also acknowledged. I thank the CERN Theory Unit for its hospitality during
this period. This work is supported by the ERC Advanced Grant Higgs@LHC.

\end{document}